\documentclass[twocolumn,aps,prl,superscriptaddress]{revtex4}

\usepackage{amsmath}
\usepackage{graphicx}
\usepackage{MnSymbol}

\begin{document}

\title{Viscoelastic crack propagation: review of theories and applications}

\author{N. Rodriguez}
\affiliation{BD Medical-Pharmaceutical Systems, Prefillable Systems, 
1 Becton Drive, MC 427, 07417, Franklin Lakes, NJ, USA}
\author{P. Mangiagalli}
\affiliation{Sanofi, 13, quai Jules Guesde-BP 14-94403 VITRY SUR SEINE Cedex, France}
\author{B.N.J. Persson}
\affiliation{PGI-1, FZ J\"ulich, Germany, EU}
\affiliation{www.MultiscaleConsulting.com}

\begin{abstract}
We review a theory of crack propagation in viscoelastic solids.
We consider both cracks in infinite systems and in finite sized systems.
As applications of the theory we consider two adhesion problems, namely
pressure sensitive adhesives and the ball-flat adhesion problem. We also study 
crack propagation in the pig skin dermis, which is of medical relevance, and rubber
wear in the context of tires.
\end{abstract}

\maketitle

\setcounter{page}{1}
\pagenumbering{arabic}




\vskip 0.3cm
{\bf 1 Introduction}

The cohesive strength of solids usually depend on crack-like defects, and the energy to propagate cracks
in the material. Similarly, the strength of the adhesive bond between two solids is usually determined by the energy to propagate interfacial 
cracks. Here we are interested in crack propagation in viscoelastic materials, such as rubber\cite{Knaus2,Sch,Knaus1,adhesion,Kramer1,Gent,Gennes,Brener,Crack1,CP,CP1,Creton,Gong,HuiX,Green1,Green2,Green3}. 
This topic is of great importance, e.g., the wear of tires or wiper blades resulting from the removal of small rubber particles by crack propagation\cite{wear}. 

In this article we will review a theory for crack propagation in viscoelastic solids. 
We will consider crack propagation in both infinite sized solids and 
in finite sized solids. The latter is also relevant for rubber wear where small particles 
(often micrometer sized) are removed from the rubber surface
by the high tensile  stresses which exist in the asperity contact regions during sliding. 
We will also consider interfacial crack propagation which is important
for adhesion. As applications of the theory we consider: 
(a) Adhesion for the sphere-flat contact problem, and for pressure sensitive adhesives. 
(b) Crack propagation in the skin dermis as may be relevant for intradermal fluid injection. 
(c) Rubber wear for the case of a tread block sliding on a road surface.

\begin{figure}[tbp]
\includegraphics[width=0.4\textwidth,angle=0]{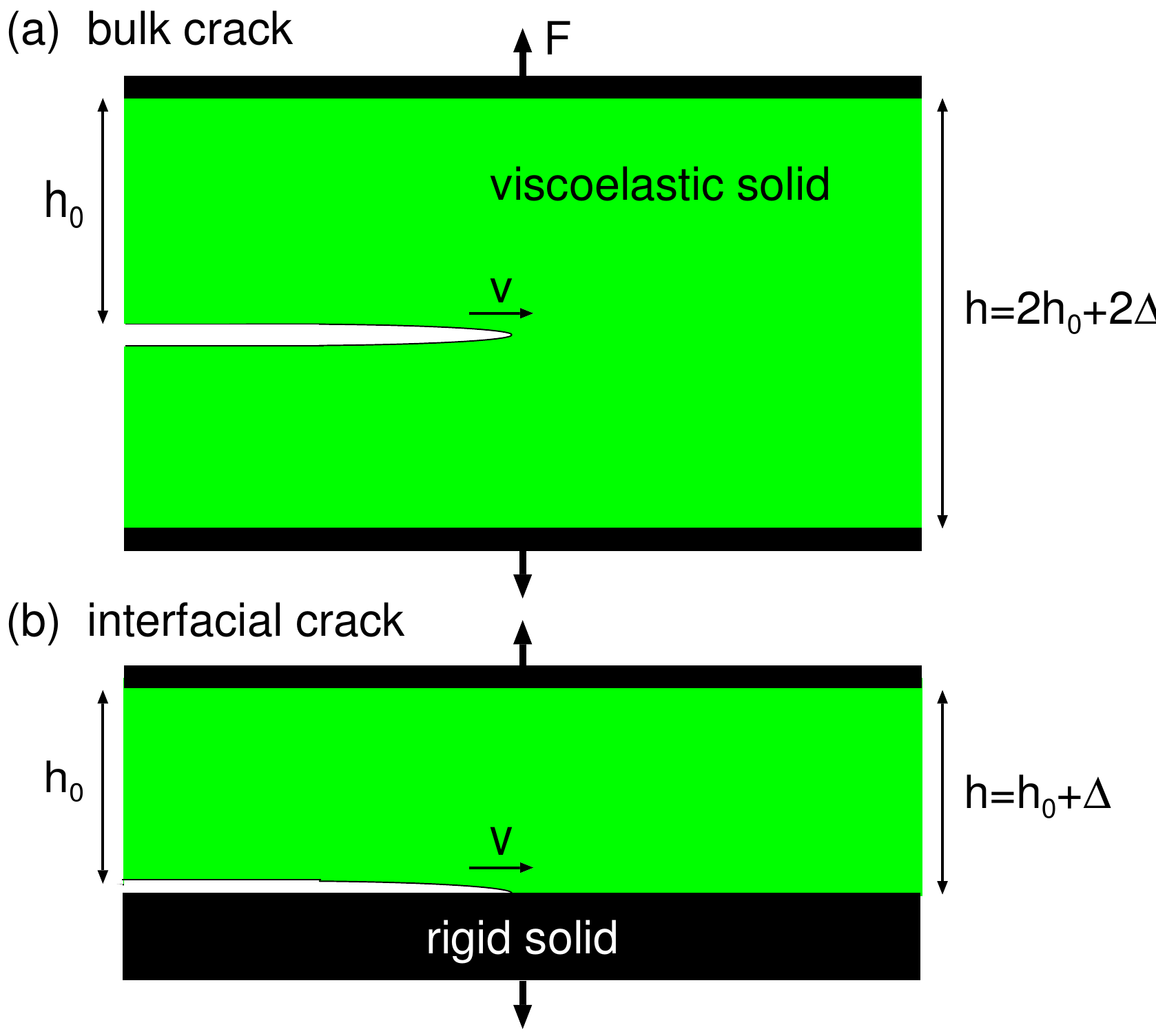}
\caption{
(a) Crack propagation in the bulk of a viscoelastic solid (cohesive crack propagation), and (b) at the interface
between a viscoelastic solid and a countersurface (adhesive crack propagation).
}
\label{crackpicslab.pdf}
\end{figure}

\vskip 0.3cm
{\bf 2 Theory of crack propagation in viscoelastic solids}

Rubber wear usually involves crack propagation in the bulk of the material (see Fig. \ref{crackpicslab.pdf}(a)).
For a bulk crack the stress and strain are usually very high close to the crack tip
and nonlinear effects, involving the breaking strong covalent bonds, chain pull-out and and cavity formation, will occur close to the
crack tip. This region of space is denoted the crack-tip process zone. The detailed nature of 
the crack-tip process zone is still a research topic, specially in cases involving heterogeneous media.

Another important set of applications involves interfacial crack propagation, e.g., between rubber materials and a hard
counter surface (see Fig. \ref{crackpicslab.pdf}(b)). In this case the strain and stresses at a crack 
tip can be much smaller, in particular if the interaction
at the interface is dominated by the weak van der Waals interaction. In this case nonlinear viscoelastic effects 
may occur only in a very small region close to the crack tip where the bond breaking occurs.

\begin{figure}[tbp]
\includegraphics[width=0.20\textwidth,angle=0]{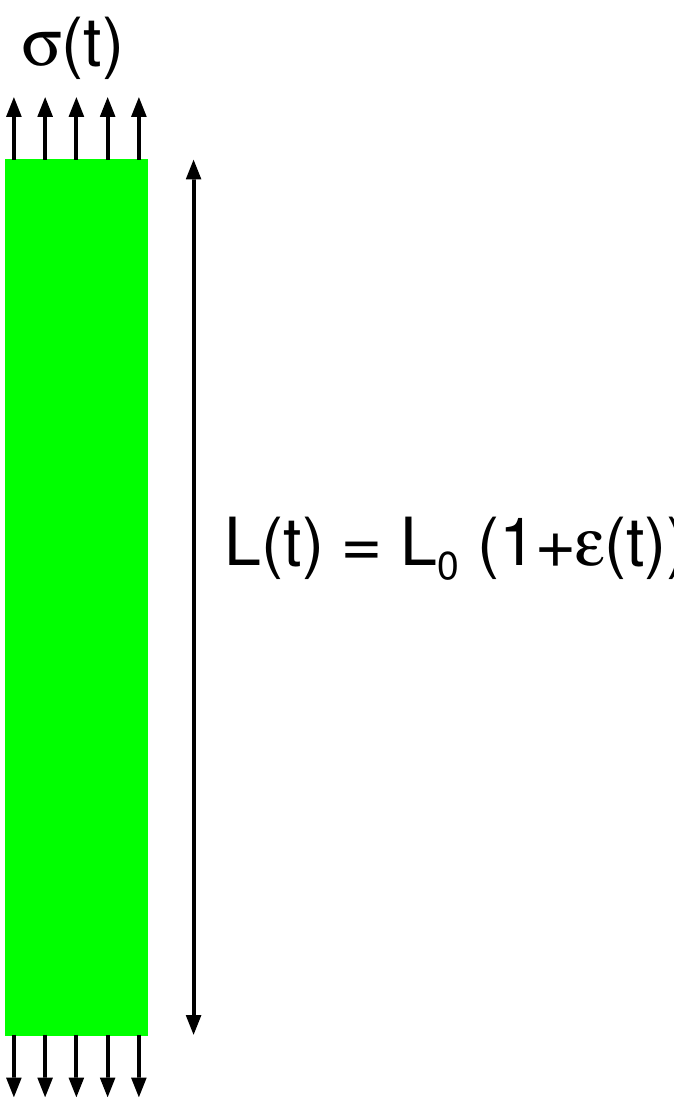}
\caption{
The length of a strip of rubber exposed to the stress $\sigma (t)$ will fluctuate
as $L(t)=L_0 (1+\epsilon(t))$, where $L_0$ is the unperturbed length and $\epsilon(t)$ the strain
at time $t$.
}
\label{RubberStrip.pdf}
\end{figure}

\begin{figure}[tbp]
\includegraphics[width=0.45\textwidth,angle=0]{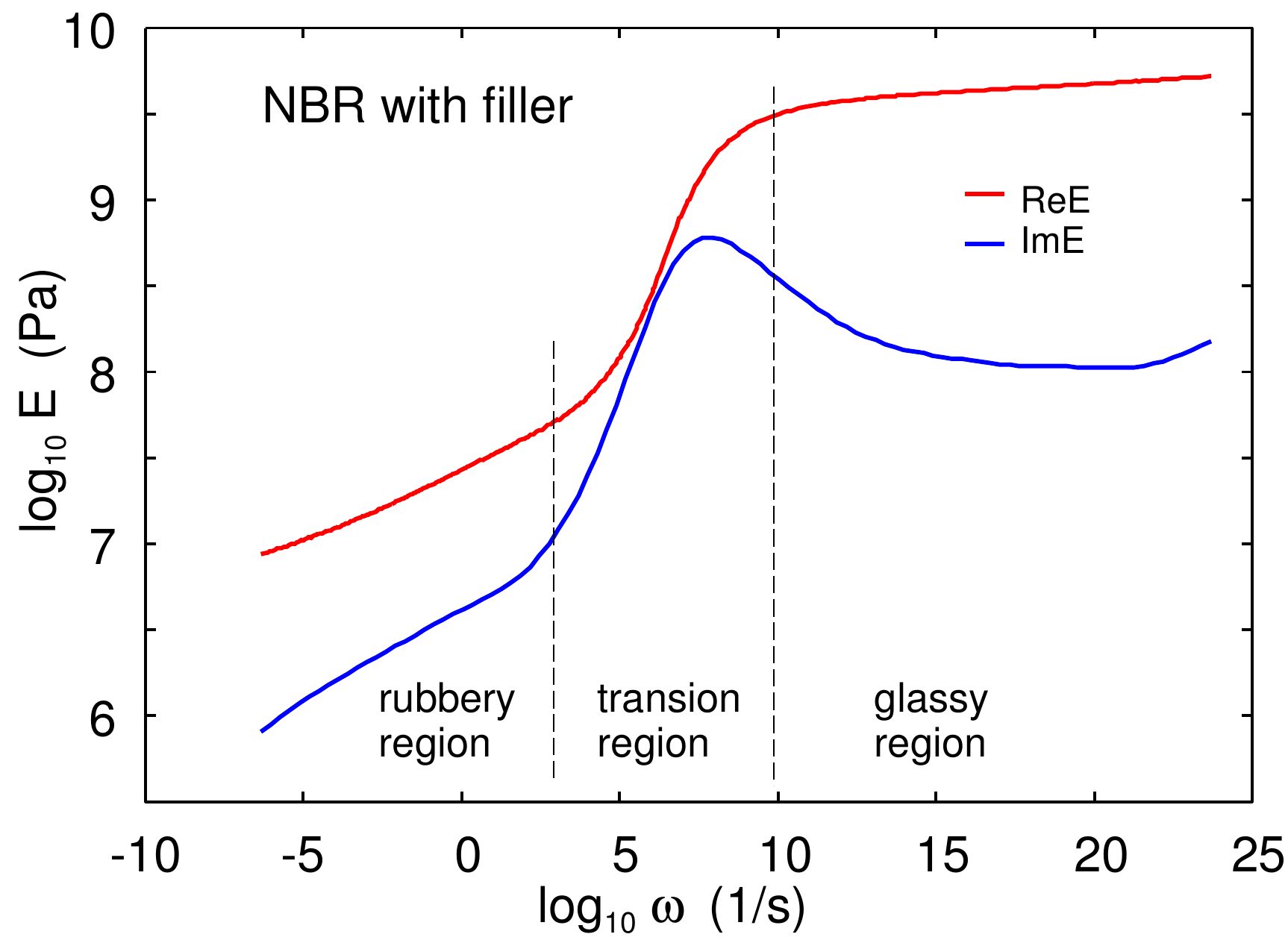}
\caption{
The real and the imaginary part of the viscoelastic modulus as a function of frequency $\omega$ (log-log scale).
For a NBR rubber compound with filler at $T=20^\circ {\rm C}$.}
\label{1logOmega.2logE.pdf}
\end{figure}

\begin{figure}[tbp]
\includegraphics[width=0.45\textwidth,angle=0]{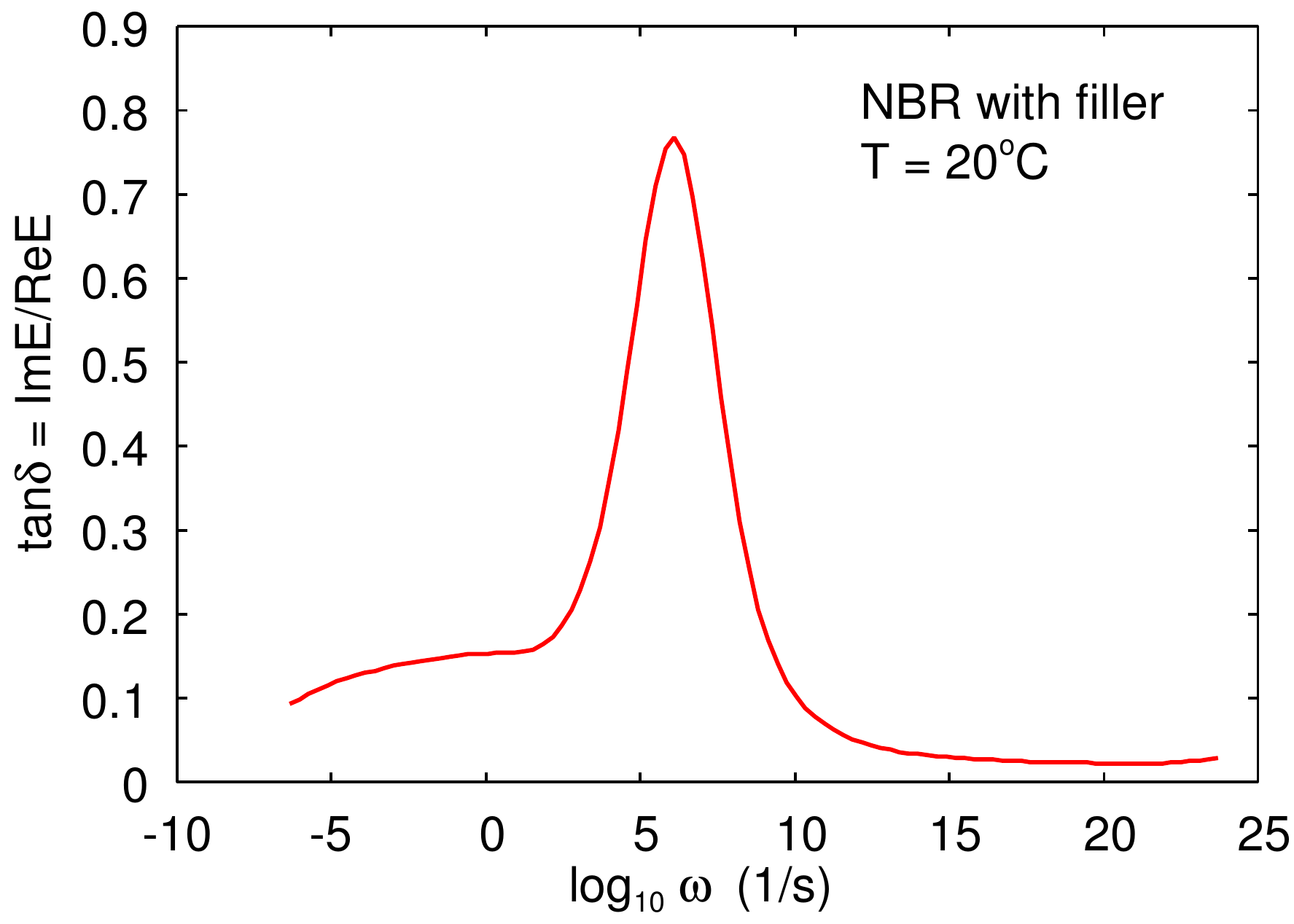}
\caption{
The loss tangent ${\rm tan}\delta = {\rm Im}E/{\rm Re}E$ as a function of the logarithm of the frequency $\omega$.
For a NBR rubber compound shown in Fig. \ref{1logOmega.2logE.pdf} at $T=20^\circ {\rm C}$.}
\label{1logOmega.2logLosstangent.pdf}
\end{figure}

\begin{figure}[tbp]
\includegraphics[width=0.45\textwidth,angle=0]{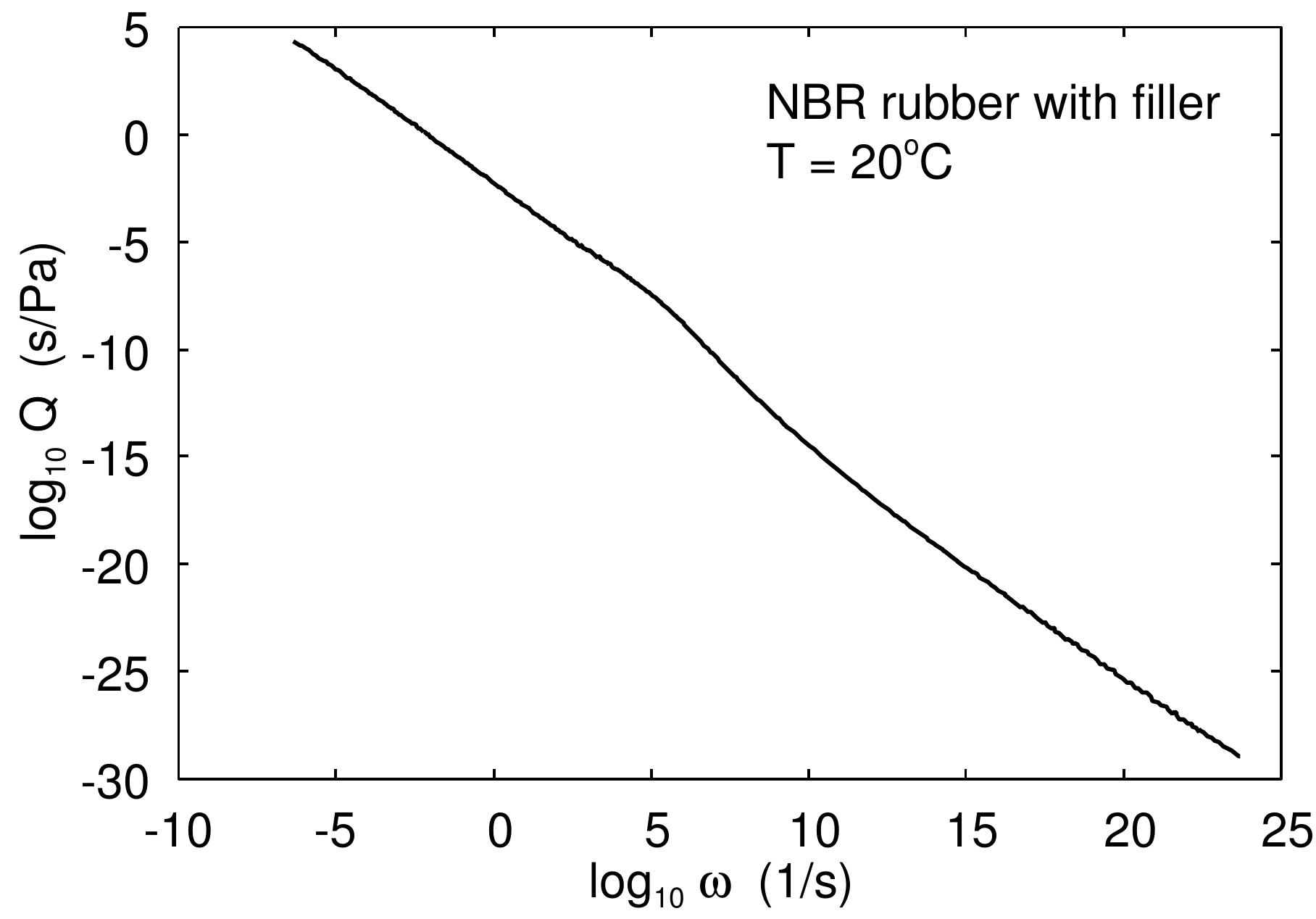}
\caption{
The crack loss function $Q(\omega)=(1/\omega){\rm Im}[1/E(\omega)]$ as a function of the logarithm of the frequency $\omega$.
For a NBR rubber compound shown in Fig. \ref{1logOmega.2logE.pdf} at $T=20^\circ {\rm C}$.}
\label{1logOmega.2logCrackLossFactor.pdf}
\end{figure}

\vskip 0.2cm
{\bf 2.1 Viscoelastic modulus}
 
Assume that a rectangular block of a linear viscoelastic material is exposed to the stress $\sigma (t)$. This will result in a strain
$\epsilon(t)$ (see Fig. \ref{RubberStrip.pdf}). If we write
$$\sigma(t)=\int_{-\infty}^\infty d\omega \ \sigma(\omega ) e^{-i\omega t}$$
$$\epsilon(t)=\int_{-\infty}^\infty d\omega \ \epsilon(\omega ) e^{-i\omega t}$$
then
$$\sigma(\omega )=E(\omega) \epsilon (\omega)\eqno(1)$$
For viscoelastic materials like rubber the viscoelastic modulus $E(\omega)$ is a complex quantity, where the imaginary part is related
to energy dissipation (transfer of mechanical energy into the random thermal motion). In a typical case $E(\omega)$ 
depends on the frequency as indicated in Fig. \ref{1logOmega.2logE.pdf} (log-log scale). 
For low frequencies (or high temperatures) the rubber respond as a
soft elastic body (rubbery region) with a viscoelastic modulus $E(\omega)$ of order $\approx 1 \ {\rm MPa}$ for the rubber used in tires,
or for the human skin dermis, or
$\approx 1 \ {\rm kPa}$ for the weakly crosslinked rubber used in pressure sensitive adhesives. At very high frequencies
(or low temperatures) is behaves as a stiff elastic solid (glassy region) with the viscoelastic modulus 
$E(\omega)$ of order $\approx 1 \ {\rm GPa}$. In the transition region it exhibits strong internal damping
and this is the region important for energy loss processes, e.g., involved in rubber friction. 
In this context the loss
tangent ${\rm Im}E(\omega) /{\rm Re}E(\omega)$ is very important and is shown in Fig. \ref{1logOmega.2logLosstangent.pdf}.

The viscoelastic modulus $E(\omega )$ is a causal linear response function. This imply that the real and the imaginary part of 
$E(\omega )$ are not independent functions but given one of them one can calculate 
the other one using a Kramers-Kronig equation\cite{Kramers}.
One can also derive sum-rules, and the most important in the present context is
$${1 \over E_0} -{1 \over E_1}= {2\over \pi} \int_{0}^\infty d\omega {1 \over \omega} {\rm Im} {1\over  E( \omega )},\eqno(2)$$
where $E_0=E(0)$ is the static ($\omega =0$) modulus, and $E_1=E(\infty)$ the modulus for infinite high frequency $\omega = \infty$.
The function 
$$Q(\omega)={1\over \omega} {\rm Im} {1\over E(\omega)}\eqno(3)$$
occurring in the integral in (1) is very important in viscoelastic crack propagation, and we will denote it as the the crack loss-function.
It is shown in Fig. \ref{1logOmega.2logCrackLossFactor.pdf} for the same rubber compound (acrylonitrile butadiene (NBR) with filler particles)
as in Fig. \ref{1logOmega.2logE.pdf} and \ref{1logOmega.2logLosstangent.pdf}. Note that $Q(\omega )$ decays monotonically with increasing 
frequencies, and is hence largest in the rubbery region in spite of the small magnitude of the damping in this frequency region.
This has important implications for the finite-size effect in rubber crack propagation (see below).

\vskip 0.2cm
{\bf 2.2 Opening crack in infinite solid}
 
We consider first an opening crack in an infinite viscoelastic solid characterized by the viscoelastic modulus $E(\omega)$
which depends on the frequency $\omega$. Consider a crack loaded in tension (mode I) (see Fig. \ref{crackpicslab.pdf}). 
The energy dissipated per unit time and unit length of the
crack line, $P$, is given by
$$P=\int d^2x \ \dot \epsilon_{ij} \sigma_{ij}\eqno(4)$$
where $\dot \epsilon_{ij}$ is the strain rate tensor and $\sigma_{ij}$ the stress 
tensor (summation over repeated indices is implicitly understood). For an opening crack
the stress field close to the crack tip has the universal form (also for a viscoelastic solid)
$$\sigma ({\bf x},t) \approx  {K\over (2\pi |{\bf x}-{\bf v}t|)^{1/2}}\eqno(5)$$
where $K$ is the stress intensity factor, and
where ${\bf v}$ is the velocity of the crack tip. Using (4) and (5) and the relation (1) between stress and strain 
one can calculate\cite{Brener}
$$P=v K^2 {2\over \pi} \int_0^{\omega_{\rm c}} d\omega F(\omega) Q(\omega)\eqno(6)$$
where we have introduced a high-frequency cut-off $\omega_{\rm c} = 2 \pi v/a$, where $a$ is the radius of the crack tip. 
The function
$$F(\omega ) = \left [ 1-\left ({\omega \over \omega_{\rm c}} \right )^2\right ]^{1/2}\eqno(7)$$
Now, let us consider the energy conservation condition relevant to the crack propagation. 
In the present case, the elastic energy stored in the solid in front of the crack tip is dissipated at the crack tip. 
The flow of elastic energy into the crack is given by $vG$ (where $G$ is the crack propagation energy per unit surface area), 
which must equal the fracture energy term 
$vG_0 $ (the energy dissipated in the crack tip process zone)
plus the bulk viscoelastic dissipation term $P$ given by (6). Energy conservation gives
$$v G = vG_0 +P\eqno(8)$$
Using (6) and (8) gives
$$G=G_0+K^2 {2\over \pi} \int_0^{\omega_{\rm c}} d\omega F(\omega) Q(\omega)$$
Using the standard relation $G=K^2/E_0$ from the theory of cracks\cite{CRACK} we obtain
$$G={G_0\over 1-E_0 {2\over \pi} \int_0^{\omega_{\rm c}} d\omega F(\omega) Q(\omega)}\eqno(9)$$

Equation (9) depends on the cutoff length $a$, and (9) is of limited practical importance unless we have a way of 
determining this length. 
Experiments have shown that the crack-tip radius in polymers 
increases with increasing speed of the crack tip\cite{Jagota}. We choose $a$ equal to the radius of the crack tip, 
which we determine as follows. The stress at the crack tip must be equal to the stress necessary to break the atomic 
bonds at the tip in order for the tip to propagate. 
If $\sigma_{\rm c}$ denotes this stress, which is a characteristic property of the material in question, we obtain, 
from (5)
$$\sigma_{\rm c} = {K\over (2\pi a)^{1/2}}\eqno(10)$$
where $a$ depends on the crack tip velocity. Combining this with $G=K^2/E_0$ gives
$$G= {2 \pi a \sigma_{\rm c}^2\over E_0}\eqno(11)$$
Combining (9) and (11) gives
$${a_0\over a} = 1-E_0 {2\over \pi} \int_0^{\omega_{\rm c}} d\omega F(\omega) Q(\omega)\eqno(12)$$
where $\omega_{\rm c} = 2 \pi v/a$ and where $a_0=E_0G_0/(2\pi \sigma_{\rm c}^2)$.
Since  $\omega_{\rm c}$ depends on $a$ this is an implicit equation for $a=a(v)$. 
Thus the theory gives both the (velocity-dependent) radius of the crack tip, 
$a(v)$, and the crack propagation energy $G(v)=G_0 a(v)/a_0$.

For large crack tip velocities $G(v)\approx G_0 E_1/E_0$ or $a(v) \approx a_0 E_1/E_0$.
The ratio between the high frequency and low frequency modulus, $E_1/E_0$, is typically very 
large, e.g., $\sim 1000$ for the rubber in Fig. \ref{1logOmega.2logE.pdf}. 
Hence for large crack tip velocity the denominator in (9) will almost vanish.
Thus any small error in the evaluation of the integral
will result big numerical error for $G(v)$ and $a(v)$. For numerical
accuracy reason it is therefore useful to rewrite (12) using the relation (2). 
If we eliminate $E_0$ in (12) using (2) we get
$${a_0\over a} = 1- 
{E_1 {2 \over \pi}  \int_{0}^{\omega_a} d\omega {1
        \over \omega} F(\omega ) {\rm Im} {1\over E(\omega )} 
\over 1+ E_1 {2\over \pi} \int_{0}^\infty d\omega {1 \over \omega} 
{\rm Im} {1\over E( \omega )}}.\eqno(13)$$
Since $E(\omega )$ typically varies with $\omega$ over very many decades in frequency, 
for the numerical evaluation of the integrals in (13) 
it is convenient to write (see Ref. \cite{Crack1}) $\omega = \omega_0 e^\xi$, so that if $\omega$ varies over $\sim 30$
decades, $\xi$ varies only by a factor $\sim 100$.

\begin{figure}[tbp]
\includegraphics[width=0.25\textwidth,angle=0]{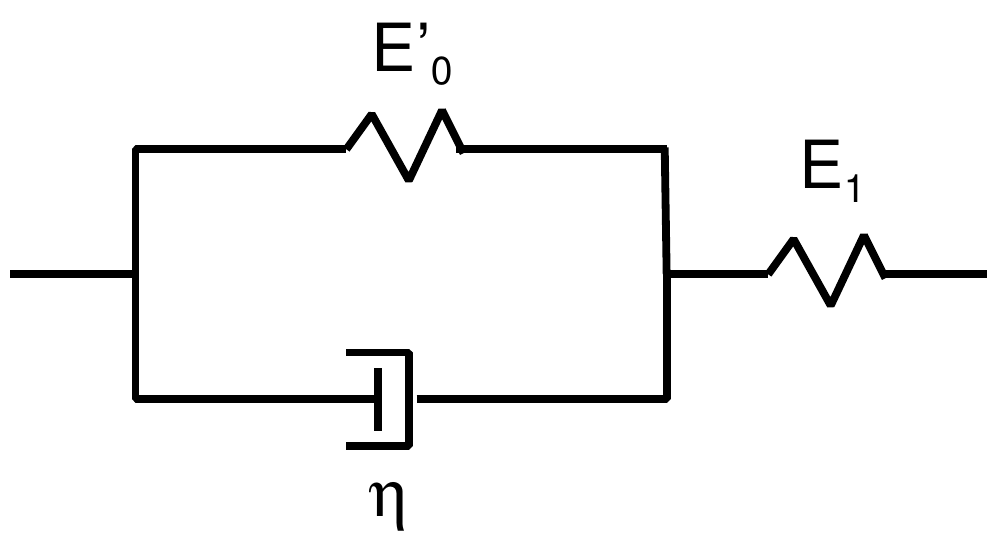}
\caption{
Three element viscoelastic model 
used in model calculation of the crack propagation energy $G(v)$.
The low frequency modulus $E(0)=E_0=E_0'E_1/(E_0'+E_1)$ and the high frequency modulus $E(\infty)=E_1$
and the viscosity $\eta$ are indicated.
}
\label{rheologypic.pdf}
\end{figure}

\begin{figure}[tbp]
\includegraphics[width=0.45\textwidth,angle=0]{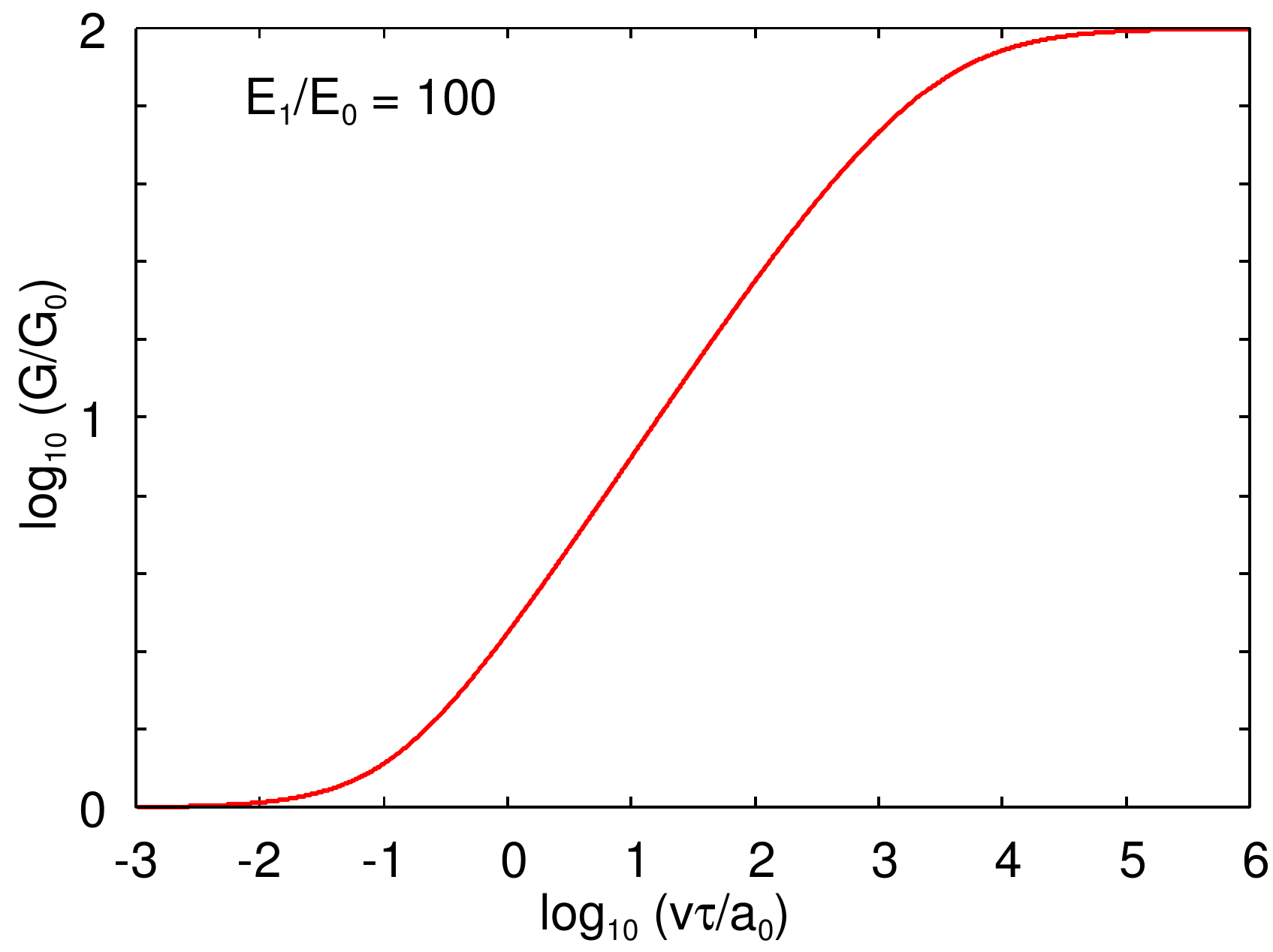}
\caption{
The crack propagation energy $G$ (in units of adiabatic value $G_0$) 
as a function of the crack tip speed $v$ (in units of $a_0/\tau$) (log-log scale)
for the three element viscoelastic model
shown in Fig. \ref{rheologypic.pdf}.}
\label{1logv.2logG.ideal.pdf}
\end{figure}

\begin{figure}[tbp]
\includegraphics[width=0.45\textwidth,angle=0]{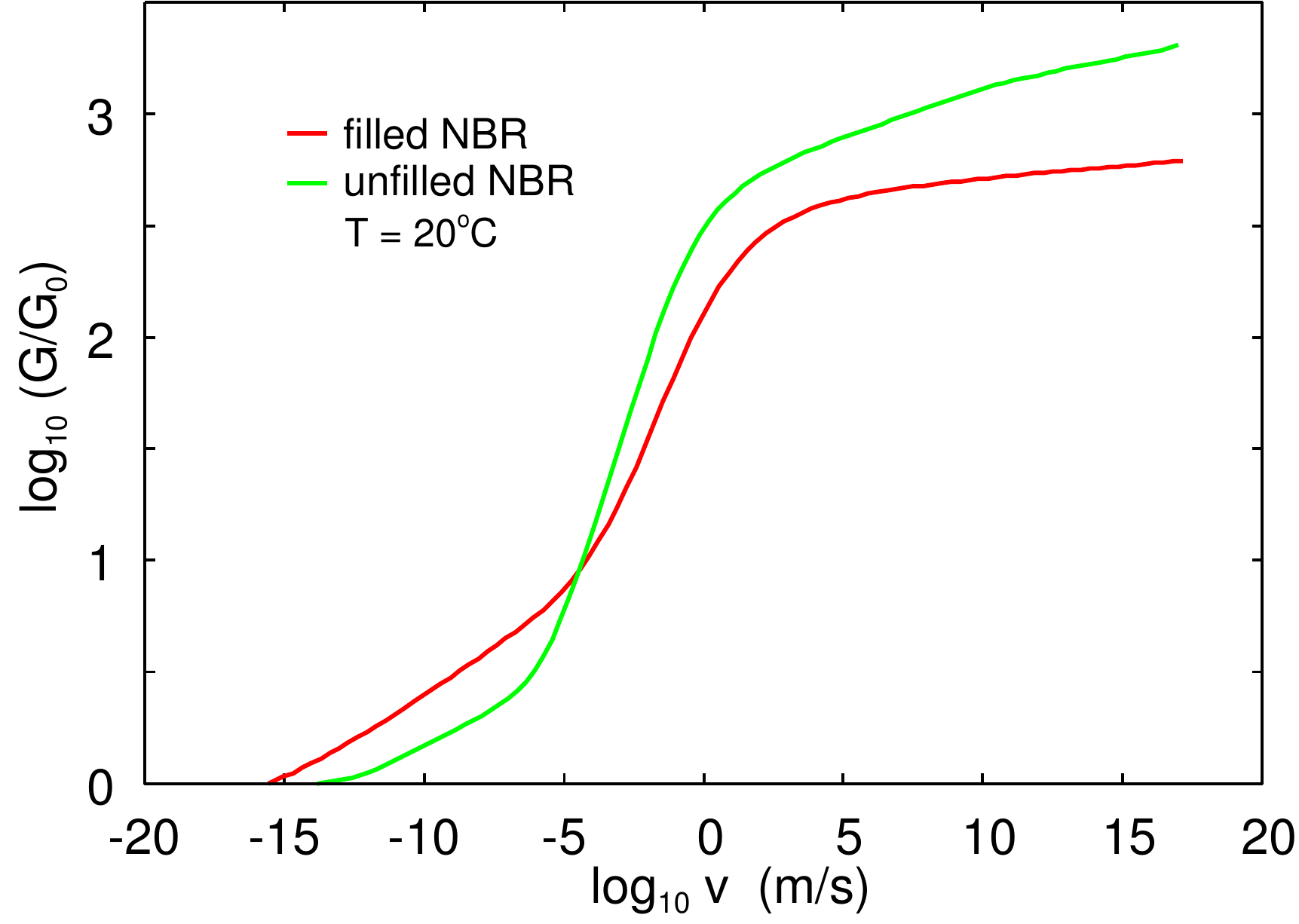}
\caption{
The crack propagation energy $G$ (in units of adiabatic value $G_0$) 
as a function of the crack tip speed $v$ (log-log scale) for filled and unfilled
NBR rubber at $T=20^\circ {\rm C}$. In the calculations we have used the measured
low-strain viscoelastic modulus (shown in Fig. \ref{1logOmega.2logE.pdf} for the filled NBR compound).
}
\label{1logv.2logG.filled.unfilled.NBR.pdf}
\end{figure}

\vskip 0.2cm
{\bf 2.3 Numerical results}
 
We now present some numerical results for the dependency of the crack propagation energy $G(v)$
on the crack tip velocity.
We first consider the highly idealized three element viscoelastic model shown in Fig. \ref{rheologypic.pdf}.
The low frequency modulus $E(0)=E_0$ and the high frequency modulus $E(\infty)=E_1$
and the viscosity $\eta$ are indicated in the figure.
Real rubber materials have a very wide range of relaxation times while the present model is characterized by a single
relaxation time $\tau$. However, this model has been used in most model studies so far, and is therefore a good test
case.

Fig. \ref{1logv.2logG.ideal.pdf} shows the
crack propagation energy $G$ (in units of adiabatic value $G_0$) 
as a function of the crack tip speed $v$ (in units of $a_0/\tau$) (log-log scale)
for the three element viscoelastic model
shown in Fig. \ref{rheologypic.pdf}. 
In the calculation we have assumed $E_1/E_0 = 100$ and that is the reason for why $G/G_0$ increases from 1 to 100
with increasing crack tip speed.

The results presented in Fig. \ref{1logv.2logG.ideal.pdf} are virtually identical
to the numerical results obtained by Greenwood using the Barenblatt process zone model\cite{Baren}. This shows that 
the detailed nature of the process zone is not very important
as the present study use a completely different description of the process zone (just a cut-off radius $a(v)$) then in the Barenblatt model
where a linearly extended process zone is used. The advantage of the present approach is that it can trivially
be applied to real materials using the measured viscoelastic modulus $E(\omega )$.

To illustrate this, in Fig. \ref{1logv.2logG.filled.unfilled.NBR.pdf}
we show the crack propagation energy $G$ (in units of adiabatic value $G_0$), 
as a function of the crack tip speed $v$ (log-log scale) for filled and unfilled
NBR rubber at $T=20^\circ {\rm C}$. In the calculations we have used the measured
low-strain viscoelastic modulus (shown in Fig. \ref{1logOmega.2logE.pdf} for the filled NBR compound).
For the unfilled compound $E_1/E_0 > 1000$ and this explain the large increase in $G(v)$ with
increasing crack speed.

\vskip 0.2cm
{\bf 2.4 Opening crack in finite solid}

Most theories of cracks in viscoelastic solids assumes an infinite large 
system\cite{adhesion,Kramer1,Brener,Crack1,CP,CP1,Saul}.
A few studies exist for the slab geometry\cite{Knaus,Langer}, where the solid is infinite in the $x$-direction but 
finite in the perpendicular $z$-direction, say with thickness $h_0$ (see Fig. \ref{crackpicslab.pdf}). 
If the surfaces $z=0$ and $z=h_0$ are clamped the stress field from the crack tip in a slab is screened by the solid walls, 
in which case the slab geometry is similar to a finite
solid with linear size $L \approx h_0$, and the results presented below are approximately valid also for the slab geometry. 

The viscoelastic energy dissipation is determined by an integral over the relevant frequencies of
the crack dissipation function $Q(\omega)$. We show $Q(\omega)$ for a typical case in Fig. \ref{1logOmega.2logCrackLossFactor.pdf}.
The biggest contribution to the integral over frequencies of $Q(\omega)$ will be from the lowest frequency region
in spite of the fact that in this region the loss tangent ${\rm tan} \delta$ is very small. Hence, any effect which
influence the low-frequency part of the viscoelastic modulus can have a big impact on the crack propagation energy.
One such influence is finite-size effects. 

The theory described in Sec. 2.2 (see also Ref. \cite{Brener,Crack1,CP,CP1}) is for an opening crack in an infinite viscoelastic media.
The theory predicts that as the crack tip velocity $v\rightarrow \infty$, the crack propagation energy $G\rightarrow (E_1/E_0) G_0$,
where $G_0$ is the crack propagation energy as $v \rightarrow 0$, where no viscoelastic energy dissipation takes place. The
high and low frequency modulus, $E_1$ and $E_0$, respectively, are both real, and can be obtained from the complex viscoelastic
modulus $E(\omega )$ as the frequency $\omega \rightarrow \infty$ and  $\omega \rightarrow 0$, respectively.

For an infinite solid there will always be a region far enough from the crack tip where the solid can be considered
as purely elastic and characterized by the static (or low frequency) modulus $E_0 = E(\omega = 0)$. This follows
from the observation that if the crack tip propagate with the velocity $v$ the time-dependent deformations of the
rubber a distance $r$ from the crack tip are characterized by the frequency $\omega = v/r$. Thus, as $r\rightarrow \infty$
we get $\omega \rightarrow 0$. However, all solids have a finite extent, say with linear dimension $L$. In this case
$r < L$ and hence $\omega > v/L$. It follows that for high crack-tip speed, the frequency $\omega$ will be very large {\it everywhere}, 
and the rubber will be in the glassy, purely elastic, state everywhere in the solid. Hence in this limiting case there is no viscoelastic
energy dissipation i.e. $G(v) \approx G_0$ for large enough $v$. This is not the case for infinite solids where
$G(v) = (E_1 /E_0) G_0$ for large enough $v$.

Here we will study how the finite-size of real solid objects influence the crack propagation energy.
For example, consider the pull-off of a rubber ball from a flat surface. 
This can be considered as a circular opening crack propagating towards the center of the circular contact region.
Let $L$ be the linear size of the contact area. 
The region where
the viscoelastic crack propagation theory is valid is limited to distances
from the crack tip $a < r < L$, where $a$ is the crack tip radius. 
Some time-dependent deformations of the rubber will occur also for $r> L$ in this case, but only for
$r<L$ the stress field (as a function of $r$) has the inverse square-root singular nature characteristic of
crack-like defects. 
When
the crack tip moves with the velocity $v$ the viscoelastic spectra will
be probed in the frequency range $2 \pi v /L < \omega < 2 \pi v /a$. We denote $\omega_L = 2 \pi v /L$ and $\omega_a = 2 \pi v/a$.

We can (approximately) use the theory for viscoelastic crack propagation in an infinite medium also for a
finite system of linear size $L$ by using the following procedure: We replace the measured viscoelastic modulus $E(\omega)$ with another modulus
$\tilde E(\omega)$ where ${\rm Im} \tilde E^{-1}(\omega) = {\rm Im} E^{-1}(\omega )$ for $\omega > \omega_L$ and ${\rm Im} \tilde E^{-1}(\omega) = 0$ for
$\omega < \omega_L$. This imply that viscoelastic energy dissipation will only occur for distances from the crack tip $r < L$. 
Given ${\rm Im} \tilde E^{-1}(\omega )$ we obtain ${\rm Re} \tilde E^{-1}(\omega )$ using a Kramers-Kronig relation\cite{Kramers}, which holds for all causal linear
response functions. We can choose the high-frequency modulus (which is real) $\tilde E_1 = E_1$ but the static (or low frequency modulus) 
$\tilde E_0 > E_0$, which is expected to be of order $\tilde E_0 \approx {\rm Re} E(\omega_L)$. 

To obtain $G(v)$ for a finite size solid we replace $E(\omega)$ in (13) with $\tilde E(\omega)$ defined so that 
${\rm Im} \tilde E^{-1}(\omega) = {\rm Im} E^{-1}(\omega )$ 
for $\omega > \omega_L$ and ${\rm Im} \tilde E^{-1}(\omega) = 0$ for
$\omega < \omega_L$. We get 
$${a_0\over a} = 1- 
{E_1 {2 \over \pi}  \int_{\omega_L}^{\omega_a} d\omega {1
        \over \omega} F(\omega ) {\rm Im} {1\over E( \omega )} 
\over 1+ E_1 {2\over \pi} \int_{\omega_L}^\infty d\omega {1 \over \omega} 
{\rm Im} {1\over E( \omega )}}.\eqno(14)$$
Note that as $v\rightarrow 0$ and $v\rightarrow \infty$, $a \rightarrow a_0$ and $G\rightarrow G_0$.

\begin{figure}[tbp]
\includegraphics[width=0.45\textwidth,angle=0]{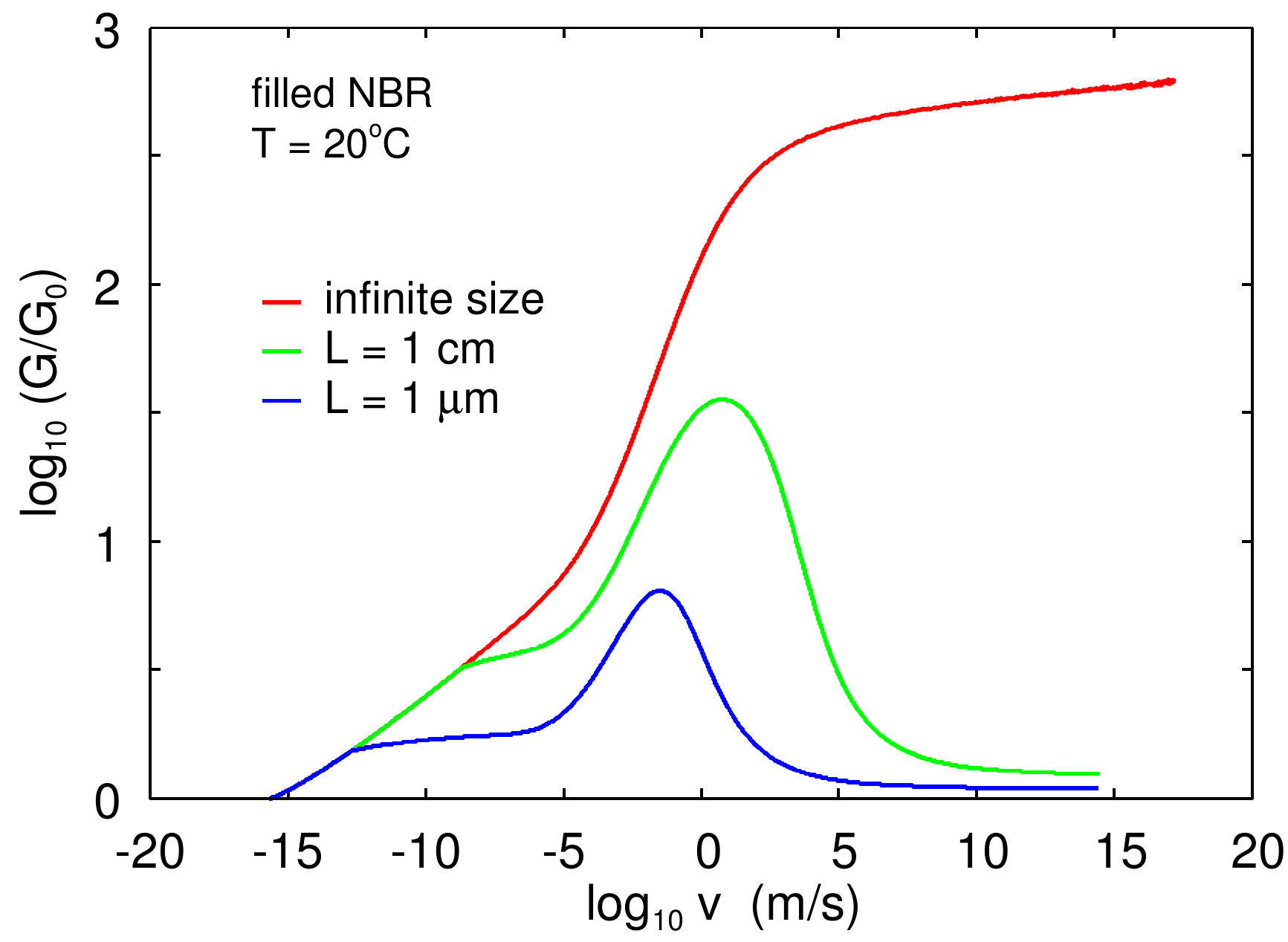}
\caption{
The viscoelastic enhancement factor $G_{\rm open} /G_0$ as a function of the crack-tip speed
for filled NBR at $T=20^\circ {\rm C}$. Results are shown for the system sizes $L=1 \ {\rm \mu m}$, $10 \ {\rm cm}$,
and for infinite system.
}
\label{1logv.2logG.1mum.1cm.infinite.NBR.filled.pdf}
\end{figure}

Fig. \ref{1logv.2logG.1mum.1cm.infinite.NBR.filled.pdf} shows the viscoelastic enhancement 
factor $G_{\rm open} /G_0$ as a function of the crack-tip speed
for filled NBR at $T=20^\circ {\rm C}$. Results are shown for the system sizes $L=1 \ {\rm \mu m}$, $10 \ {\rm cm}$,
and for infinite system.
The system size $L=10 \ {\rm cm}$ is typical for the sample size used
in studies of the crack propagation in macroscopic rubber samples, and is clearly not equivalent to the infinite sample size.
The reason for the strong finite size effects is that the function $Q(\omega)$ decreases monotonically with increasing frequencies,
making the integrals in (14) very sensitive to the lower cut-off frequency $\omega_L$ determined by the size of the system.
Note that for the finite sized sample there is a maximum in the $G(v)$-curve, 
corresponding to an instability in the crack-tip motion.

\vskip 0.2cm
{\bf 2.5 Role of temperature}

The crack propagation energy $G(v,T)=G_0 (1+f(v,T))$ depends strongly on the temperature. The viscoelastic factor $1+f(v,T)$
depends on temperature via the temperature-frequency shift factor $a_T$ since $f(v,T)=f(va_T,T_0)$ where $T_0$ is a reference
temperature with $a_{T_0}=1$. Thus, increasing the temperature shift the factor $1+f(v,T)$ to higher sliding speeds.
The factor $G_0=G_0(v,T)$ depends also on the crack tip speed and the temperature because breaking the bonds
in the crack tip process zone is a thermally activated, stress aided process\cite{Chaud1,Pcrack2}. 
This temperature effect is particular important for low-energy bonds, e.g., for
weak adhesive bonds\cite{Chaud1,Pcrack2,Pcrack0,Pcrack1}.

In a recent study\cite{PRX} using fluorogenic mechanochemistry with 
quantitative  confocal  microscopy  mapping,  it was found how  many  and  where  covalent 
bonds are broken as an elastomer fractures. The measurements reveal that  bond scission 
near the crack plane can be delocalized over up to hundreds of micrometers and increase 
$G_0$ by  a  factor  of  $\approx 100$  depending  on  temperature  and  stretch  rate,  pointing  to  an  intricated 
coupling   between   strain   rate   dependent   viscous   dissipation   and   strain   dependent 
irreversible network  scission. These findings shows that energy dissipated by covalent bond scission accounts for a much larger 
fraction of the total fracture energy than previously believed.  

At low crack tip speed the temperature will
everywhere be close to surrounding (background) temperature, but for a fast moving crack tip the energy dissipated close to
the crack tip will not have time to diffuse away resulting in a higher temperature close to the crack tip. Including this
temperature increase in the theory is a complex topic addressed in Ref. \cite{CP,CP1,addC}. 

We note that it is possible to reformulate (14) as an integral over temperature rather than frequency
which is useful if the viscoelastic modulus has been measured only as a function of temperature for one frequency.
Assume that the viscoelastic modulus has been measured as a function of temperature for the frequency $\omega=\omega_1$,
i.e. $E(\omega_1,T)$ is known.
Let $T_1$ denote the temperature of interest so that (14) takes the form:
$${a_0\over a} = 1- 
{E_1 {2 \over \pi}  \int_{\omega_L}^{\omega_a} d\omega {1
        \over \omega} F(\omega ) {\rm Im} {1\over E(a_{T_1} \omega )} 
\over 1+ E_1 {2\over \pi} \int_{\omega_L}^\infty d\omega {1 \over \omega} 
{\rm Im} {1\over E(a_{T_1} \omega )}}.\eqno(15)$$
where $E(a_{T_1} \omega)=E(\omega,T_1)=E(a_{T_1}\omega,T_0)$.
Next, let us write $\omega = \omega_1 a_T / a_{T_1}$. We consider $T$ as the new integration variable
and get
$$d\omega {1\over \omega} = dT ({\rm ln} a_T)'$$
where $({\rm ln} a_T)' = d({\rm ln} a_T)/dT$.
Denoting the solution to $\omega_a = \omega_1 a_T/a_{T_1}$ as $T_a$
and to $\omega_L = \omega_1 a_T/a_{T_1}$  as $T_L$ we can write (15) as
$${a_0\over a} = 1- 
{E_1 {2 \over \pi}  \int_{T_a}^{T_L} dT \ (-{\rm ln}a_T)'  \ F(T) {\rm Im} {1\over E(T)} 
\over 1+ E_1 {2\over \pi} \int_{0}^{T_L} dT \ (-{\rm ln}a_T)'  \ {\rm Im} {1\over E(T)}}.\eqno(16)$$
where $E(T) = E(a_{T_1}\omega,T_0)=E(a_{T_1}[\omega_1 a_T / a_{T_1}],T_0)=E(a_T \omega_1,T_0)=E(\omega_1,T)$ and where
$$F(T ) = \left [1-\left ( {\omega_1\over \omega_a}{a_T\over a_{T_1}}\right )^2\right ]^{1/2}\eqno(17)$$

\begin{figure}[tbp]
\includegraphics[width=0.45\textwidth,angle=0]{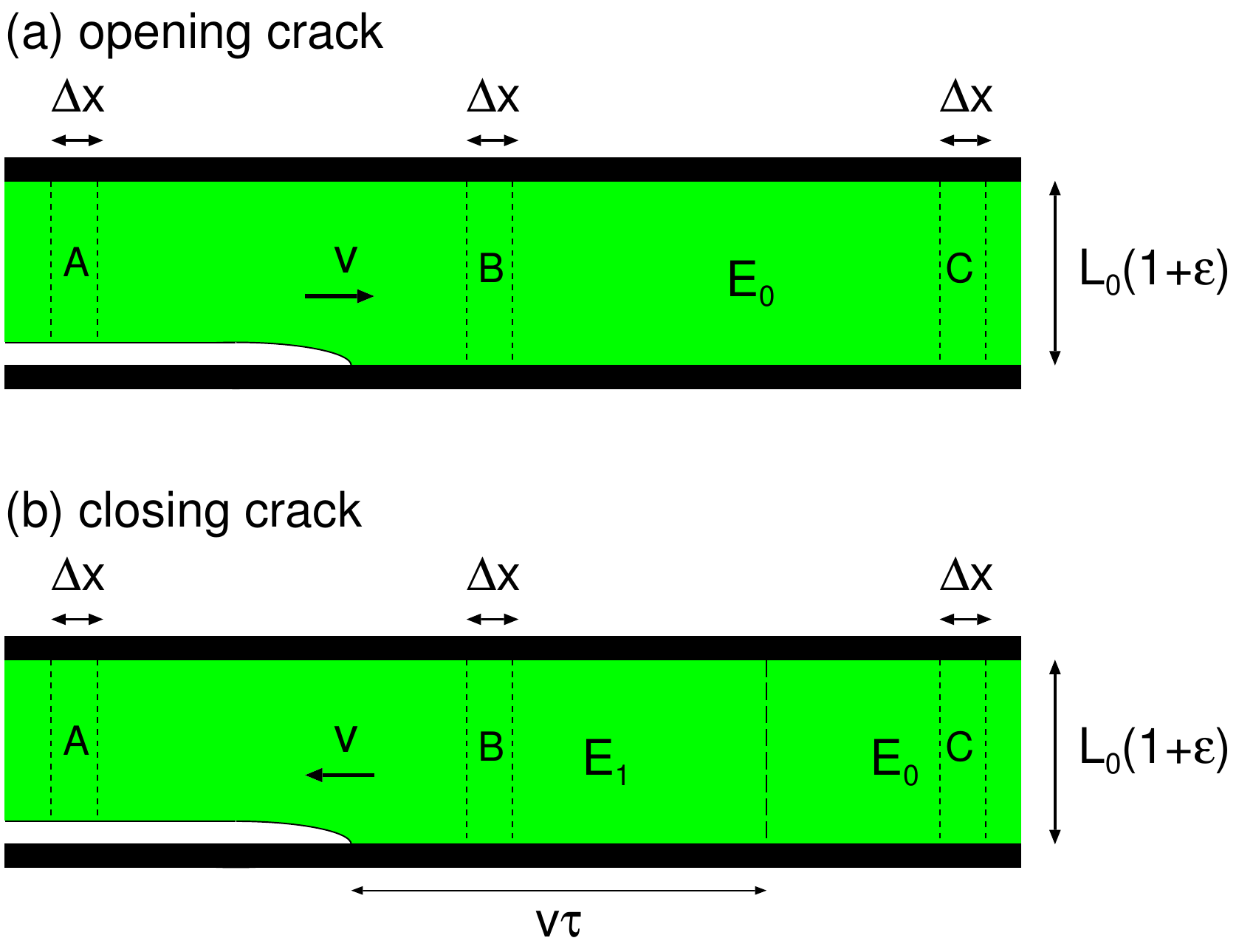}
\caption{
Fast moving opening crack (a) and closing crack (b) in thin viscoelastic slab under tension. In (a) negligible viscoelastic energy dissipation
occur and the crack propagation energy is given by the energy to break the interfacial bonds, $G\approx G_0$. In (b) viscoelastic energy 
dissipation occur and the effective crack propagation energy $G\approx (E_0/E_1)G_0$ is reduced by a factor $E_0/E_1$ (see text for details). 
}
\label{OpenClosePic.pdf}
\end{figure}

\vskip 0.2cm
{\bf 2.6 Closing crack}

When an opening crack propagates in the bulk of a viscoelastic solid the breaking of the bonds in the crack tip process zone is
usually an irreversible process: the broken (dangling) bonds formed during the crack opening 
react quickly with molecules from the atmosphere, or with mobile molecules
in the solid. Hence if the external crack driving force is removed no closing crack propagation involving the reformation of the original
bonds will occur. However, for interfacial crack propagation the situation may be very different. Thus, in many cases rubber bind to
a countersurface mainly with the weak and long-ranged van der Waals bonds. In this case the bonds broken during  
crack opening and the bonds formed during crack closing may be very similar. 

For an opening crack in a viscoelastic solid energy conservation require that $vG=vG_0+P$ i.e. $G>G_0$.
For a closing crack the energy conservation condition gives
$$vG=vG_0-P$$
so that $G<G_0$. Physically, the energy gained by the binding of the solids at the crack interface is 
in part lost as viscoelastic energy dissipation inside the solid. For an opening crack, as the crack speed $v\rightarrow \infty$
we have $G/G_0 \rightarrow E_1/E_0$ but for a closing crack $G/G_0 \rightarrow E_0/E_1$. The latter result is most
easily understood by considering 
the simple crack problem shown in Fig. \ref{OpenClosePic.pdf}.

Fig. \ref{OpenClosePic.pdf} shows a fast moving opening crack (a) and closing crack (b) in a thin viscoelastic slab under tension. 
In case (a) the slab is elongated by $L_0 \epsilon$, 
and we wait until a fully relaxed state is formed before inserting the crack. 
Thus the elastic energy stored in the strip C of width $\Delta x$ is
$\sigma \epsilon L_0 \Delta x/2=E_0 \epsilon^2 L_0 \Delta x/2$. For a fast moving crack, in the present finite-size set up 
($L_0$ is finite), there will be negligible 
viscoelastic energy dissipation in the solid and $G\approx G_0$ is determined by the energy conservation condition 
$G_0 \Delta x = E_0 \epsilon^2 L_0 \Delta x/2$ or $G_0=E_0 \epsilon^2 L_0/2 = \sigma_0^2 L_0/(2E_0)$. The fact that $G\approx G_0$ in this case
is a finite-size effect (for an infinite system we would instead get $G=(E_1/E_0)G_0$). 

For the closing crack (case (b)) the
situation is different: For a fast moving crack the strip A is quickly elongated when it approach the crack tip, which require a large stress
$\sigma=E_1 \epsilon$ determined by the high frequency modulus $E_1$. Since the crack moves very fast the stress in the strip will remain at
this large value even when the crack tip has moves far away from the strip as in position B. However, due to viscoelastic relaxation
the stress will finally arrive at the relaxed value $\sigma=E_0 \epsilon$ as at position C. The time this takes depends on the nature of the viscoelastic
relaxation process, e.g., for a process characterized by a single relaxation time $\tau$, 
a time $t> \tau$ (and distance $s>v\tau$) would be needed
to reach the relaxed state. During this relaxation mechanical energy is converted into heat. Since the crack tip is far away from
the region where this relaxation process takes place, it does not know about it, and the interfacial binding energy is converted into 
elastic energy in the rapid stretching of the strip in the process going from strip position A to B. Thus
$G_0 \Delta x = E_1 \epsilon^2 L_0 \Delta x/2$. However, the crack propagation energy $G$ refer to the relaxed state configuration
so that $G \Delta x = E_0 \epsilon^2 L_0 \Delta x/2$. Thus $G = E_0 \epsilon^2 L_0/2 = (E_0/E_1) E_1 \epsilon^2 L_0/2 = (E_0/E_1)G_0$.

Using the Barenblatt description of the crack tip process zone, Greenwood has shown that for an infinite
sized system $G_{\rm open} G_{\rm close} \approx G_0^2$.
Thus if we write the opening crack tip propagation energy as
$$G_{\rm open} = [1+f(v,T)]G_0\eqno(18)$$
then
$$G_{\rm close} \approx {G_0 \over 1+f(v,T)}\eqno(19)$$
Thus we can use the theory presented in Sec. 2 to predict the crack propagation energy also for closing cracks.
However, the theory for closing cracks in viscoelastic solids is still not fully understood, e.g., for a fast moving crack
a region of compressible stress occur close to the crack tip for which no physical explenation exist\cite{HuiX,Green1}.  

The results presented here are crucial for adhesion involving viscoelastic solids, e.g., rubber materials. Thus in a typical case
no adhesion can be detected when two solids are squeezed in contact (closing crack propagation), but strong adhesion is observed 
during separation (opening crack propagation). One well known case is the contact involving adhesive tape: when the tape is pushed in contact 
no adhesion can be detected but during separation a strong adhesion force prevail. In general these are several reasons for 
contact hysteresis (e.g. related to roughness or chain interdiffusion) but in many cases the most important effect is viscoelasticity.

\begin{figure}[tbp]
\includegraphics[width=0.45\textwidth,angle=0]{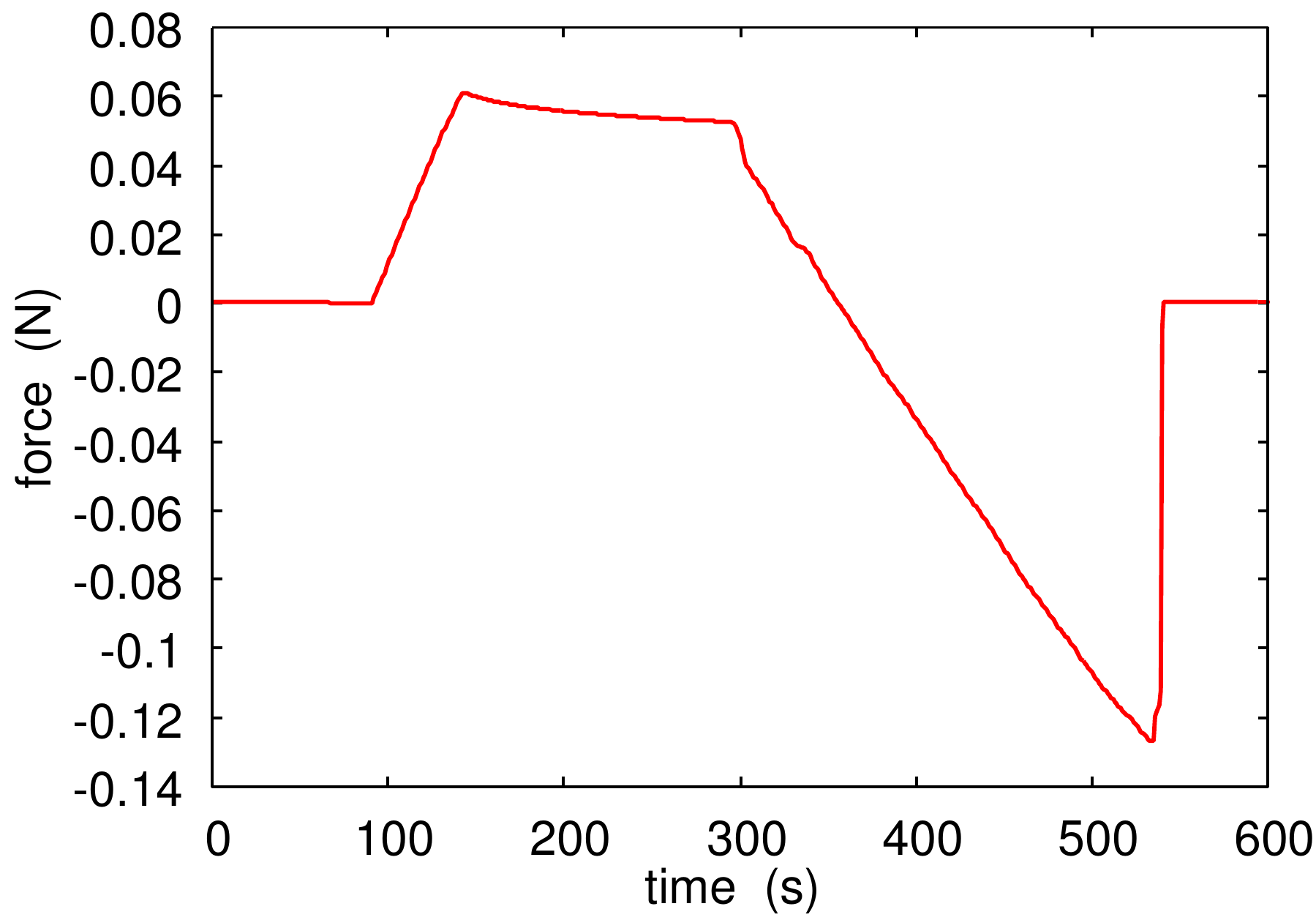}
\caption{
The interaction force between a glass ball (diameter $2R=2.5 \ {\rm cm}$) moved in contact with a pressure sensitive adhesive film 
(double sided adhesive tape attached to a smooth glass surface) and then removed. The approach and retraction speed is 
$36 \ {\rm \mu m/s}$. Note the strong adhesion hysteresis: no adhesion is observed during approach but a strong adhesion
(corresponding to the work of adhesion $G\approx 2 \ {\rm J/m^2}$) is observed during pull-off.
}
\label{SkotchTape.1time.2force.pdf}
\end{figure}

As an example illustrating contact hysteresis effects, in Fig. \ref{SkotchTape.1time.2force.pdf} we show 
the interaction force between a glass ball (diameter $2R=2.5 \ {\rm cm}$) moved in contact with a pressure sensitive adhesive film 
(double sided adhesive tape attached to a smooth glass surface) and then removed. The approach and retraction speed is 
$36 \ {\rm \mu m/s}$. Note the strong adhesion hysteresis: no adhesion is observed during approach but a strong adhesion
(corresponding to the work of adhesion $G\approx 2 \ {\rm J/m^2}$) is observed during pull-off.

\begin{figure}[tbp]
\includegraphics[width=0.3\textwidth,angle=0]{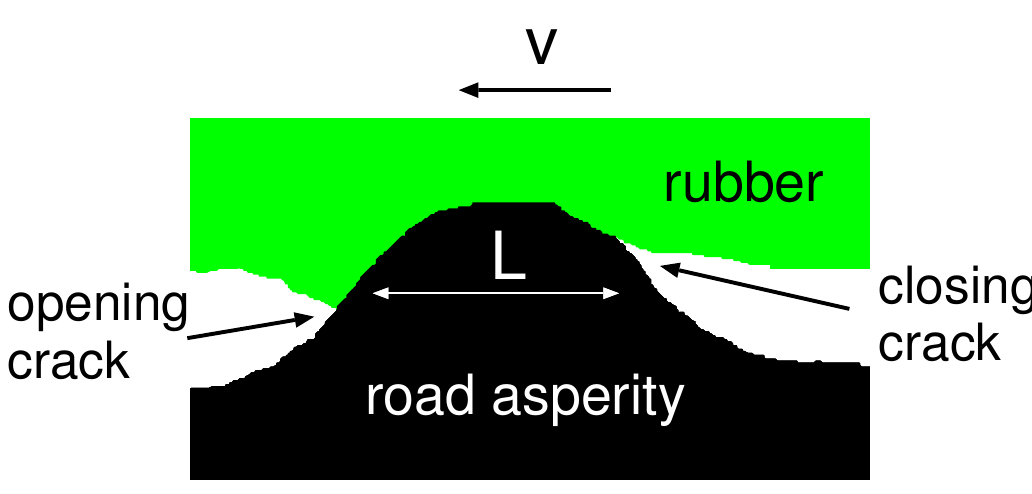}
\caption{A rubber block in contact with a road asperity. We assume the contact region is circular with the
diameter $L=2r_0$. During sliding (speed $v$) a closing crack is observed on
the entrance side and an opening crack on the exit side. During sliding the distance $\Delta x$ the dissipated energy
in an asperity contact region $(G_{\rm open}-G_{\rm close}) L \Delta x $
The dissipated energy can also be written as $\tau_{\rm f} \pi r_0^2 \Delta x$, where $\tau_{\rm f}$ is an effective frictional
shear stress, giving $\tau_{\rm f} \approx (4/\pi) (G_{\rm open}-G_{\rm close})/ L$. 
}
\label{rubberroadcont.pdf}
\end{figure}

\vskip 0.2cm
{\bf 2.7 Implications for sliding friction}

When a rubber block slides on hard and rough substrate surface, such as an asphalt or concrete road surface, the 
rubber-road contact will in general not be complete, but it will consist of many small asperity contact regions.
The contact area is usually a very small fraction of the nominal contact area, e.g., 
for a tire it may be only $\sim 1 \ {\rm cm}^2$. A very important contribution
to the friction force is derived from the interaction between the rubber molecules and the road surface in the area
of real contact. For clean surfaces two different (adhesive) contributions to the frictional force 
have been considered, namely from the opening crack on the exit-side 
of the asperity contact region (see Fig. \ref{rubberroadcont.pdf})\cite{Klupp0,Klupp}, and from
bonding-stretching-debonding process within the area of real contact\cite{Schall,PV}.
If the typical diameter of a contact region is $L$ one can show that the contribution
from the opening cracks gives a contribution to the frictional shear stress given by (see Fig. \ref{rubberroadcont.pdf}) 
$\tau_{\rm f} \approx  (G_{\rm open}-G_{\rm close})/L$. 

For a rubber tread block sliding on an asphalt road surface, contact mechanics calculations 
(including adhesion)\cite{SSR,Persson3} show that the lateral size of a typical
asperity contact region is of order $L\approx 1 \ {\rm \mu m}$. For this case 
it was shown in Ref. \cite{WithGH} that the maximum of
the frictional shear stress is about $10-20$ times smaller than the adhesive contribution to the friction 
needed to explain measured friction data. We conclude that the contribution to the friction from the opening crack propagation cannot 
explain the observed magnitude (or velocity dependency) of the shear stress acting in the area of real contact. This 
suggest another origin for the main contribution to the friction from
the area of real contact. In Ref. \cite{Nam,Per1} it was proposed that molecular bonding-stretching-debonding process\cite{Schall,PV}
in the area of real contact can explain the observed magnitude (and velocity dependency) of the contribution
to the friction from the area of real contact. 

If the asperity contact regions would be much smaller than $\sim 1 \ {\rm \mu m}$, the crack opening contribution to the friction
could be much more important and may dominate the adhesive contribution. Furthermore, the adhesive contribution to 
rolling friction on a smooth rubber surface, and the friction associated with Schallamach waves, 
are both determined by the crack opening (and closing) contribution\cite{RobTom}.

\begin{figure}[tbp]
\includegraphics[width=0.45\textwidth,angle=0]{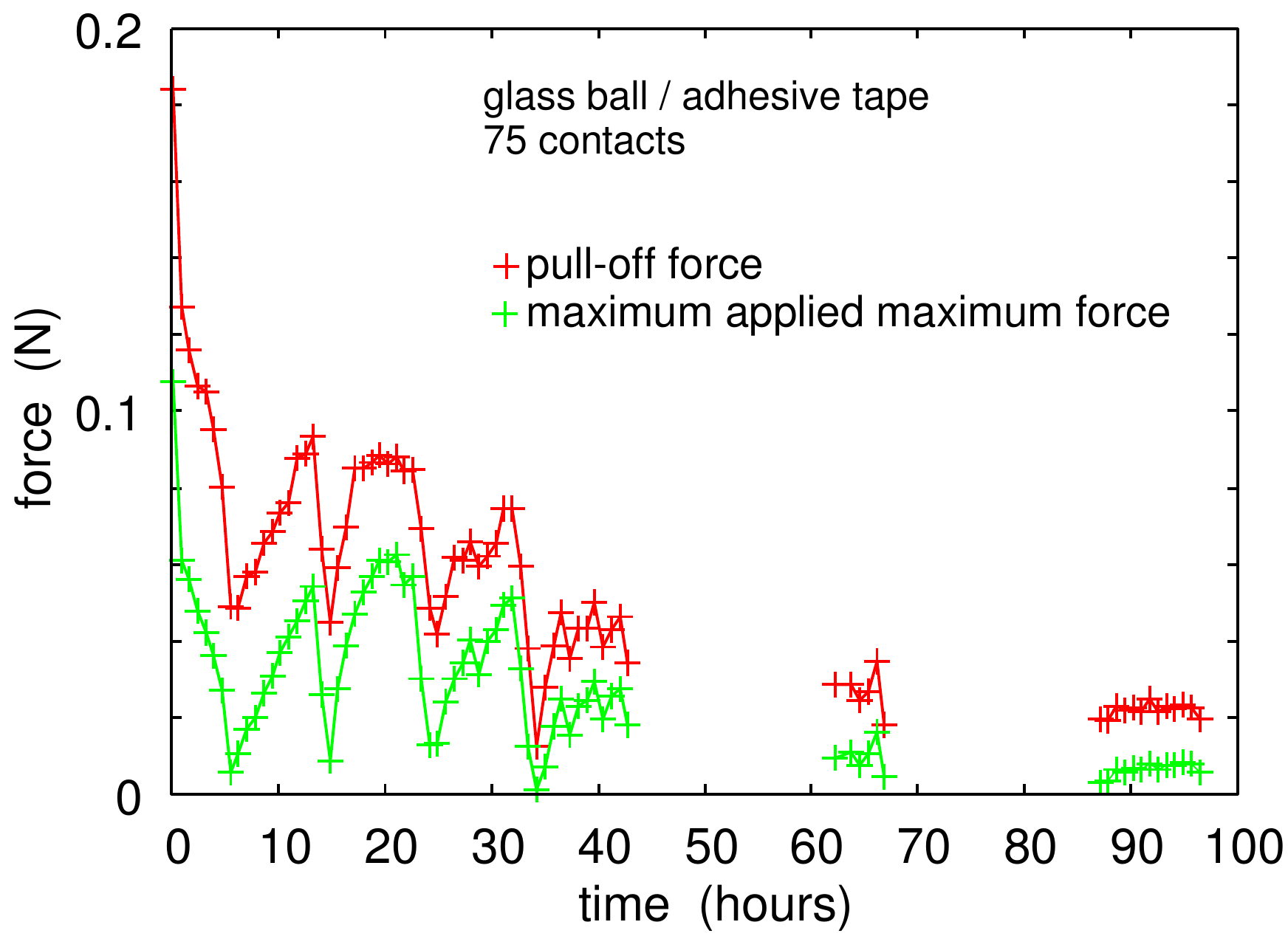}
\caption{The pull-off force (red) and the maximum applied squeezing force (green) as a function of time during
repeated contacts (75 contacts) between a glass ball (diameter 2.5 cm) and an adhesive tape attached to
a smooth flat glass plate. Based on force-time curves such as shown in Fig. \ref{SkotchTape.1time.2force.pdf}.
Note that the pull-off force is proportional to the applied squeezing force 
(see Fig. \ref{1Fn.2Fpulloff.SkotchTape.pdf}).
}
\label{1time.hours.2FmaxFpulloff.pdf}
\end{figure}

\begin{figure}[tbp]
\includegraphics[width=0.45\textwidth,angle=0]{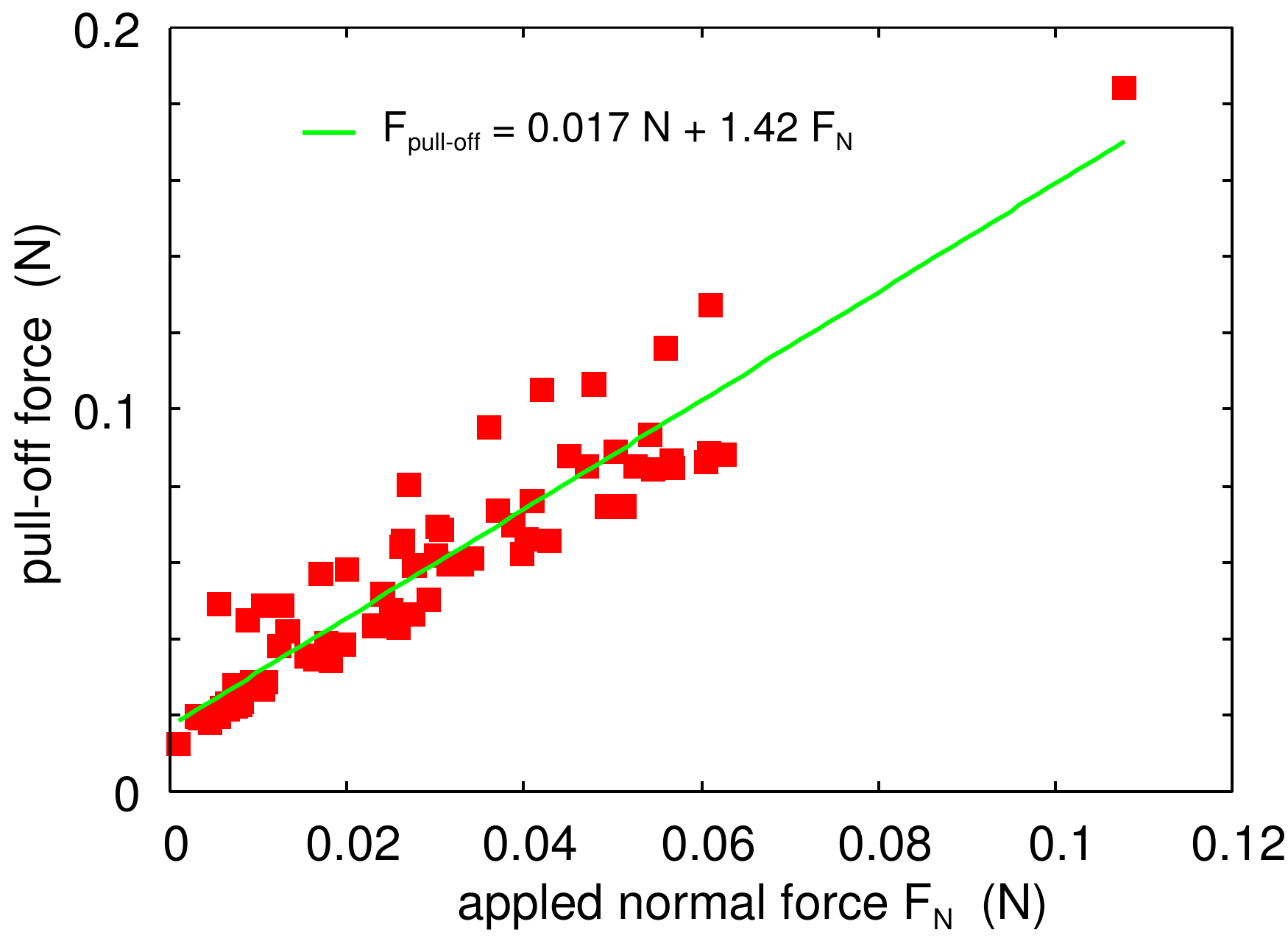}
\caption{The pull-off force between a glass ball and an adhesive tape as a function of the maximum applied squeezing force. 
Based on the data shown in Fig. \ref{1time.hours.2FmaxFpulloff.pdf}.
}
\label{1Fn.2Fpulloff.SkotchTape.pdf}
\end{figure}

\begin{figure}[tbp]
\includegraphics[width=0.45\textwidth,angle=0]{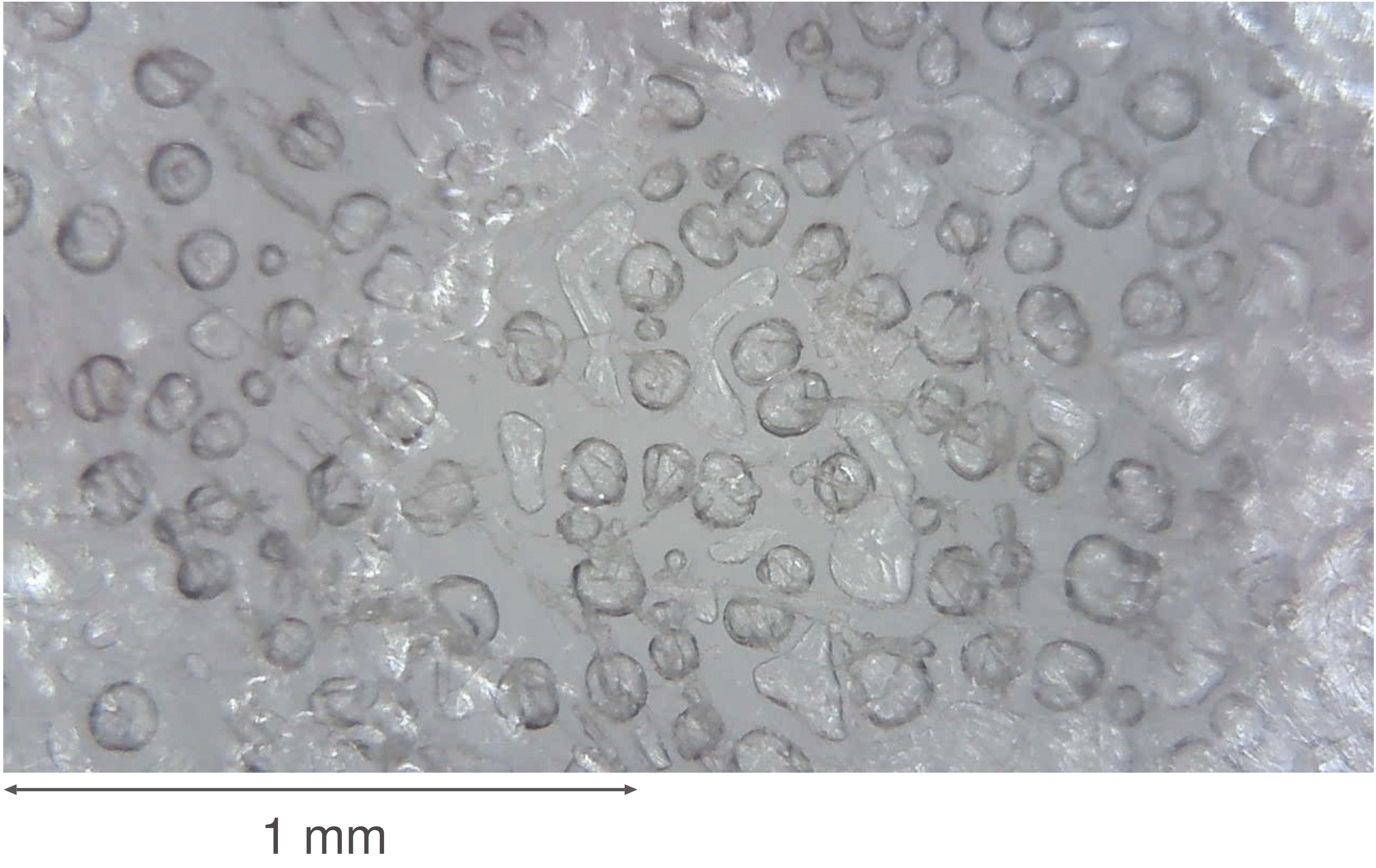}
\caption{Optical picture of the double sided adhesive film (tesa 5338) (attached to a smooth glass plate)
used in the adhesion experiments. 
Note the surface roughness. Single sided adhesive films gives very smooth surface if pulled rapidly (but a rough surface
if pulled very slowly), while the present film gives a rough surface independent of the pull-off speed.
}
\label{TesaDoubleSided.pdf}
\end{figure}

\vskip 0.3cm
{\bf 2.8 Role of surface roughness}

Surface roughness has a big influence on interfacial crack propagation. 
For very soft rubber compounds, like pressure sensitive adhesives,
the pull-off force is proportional to the relative area of real contact $A/A_0$. We illustrate this in Fig. \ref{1time.hours.2FmaxFpulloff.pdf} 
which shows the squeeze-together force and the pull-off force between a pressure sensitive adhesive film, 
attached to a smooth glass plate, and a glass ball. 
Note that the pull-off force is proportional to the applied normal force (see Fig. \ref{1Fn.2Fpulloff.SkotchTape.pdf})
which we attribute to the fact that the area of real contact is proportional to the normal force.
Thus due to surface roughness the adhesive film (see Fig. \ref{TesaDoubleSided.pdf}) 
makes only partial contact with the glass ball in the nominal contact area,
and the effective crack propagation energy for opening crack
$$G \approx (A/A_0) G_{\rm open} (v),$$
where $A/A_0$ is the relative contact area at the rim of the nominal contact area at the point of snap-off
(where the opening crack speed is $v$). The crack propagation energy (also denoted the work of adhesion) 
$G_{\rm open} (v)$ is the interfacial 
crack propagation energy for smooth surfaces.

For a thick rubber film the pull-off force is given by the Johnson-Kendall-Roberts theory
$$F_{\rm pull-off} = {3\pi \over 2} R G$$
If the elastic modulus of the rubber compound is high enough the effective crack propagation energy
$G$ must be corrected for the elastic energy stored when the rubber when the rubber surface is bent to make contact with the substrate
$$G\approx [(A/A_0)G_0 - U_{\rm el}](1+f(v,T)),$$ 
where $U_{\rm el}$ is the elastic energy per unit surface area due to the surface roughness.
Thus if the roughness is big enough, $(A/A_0)G_0 \approx U_{\rm el}$, the pull-off force will vanish.

\vskip 0.3cm
{\bf 3 Applications}

\begin{figure}[tbp]
\includegraphics[width=0.45\textwidth,angle=0]{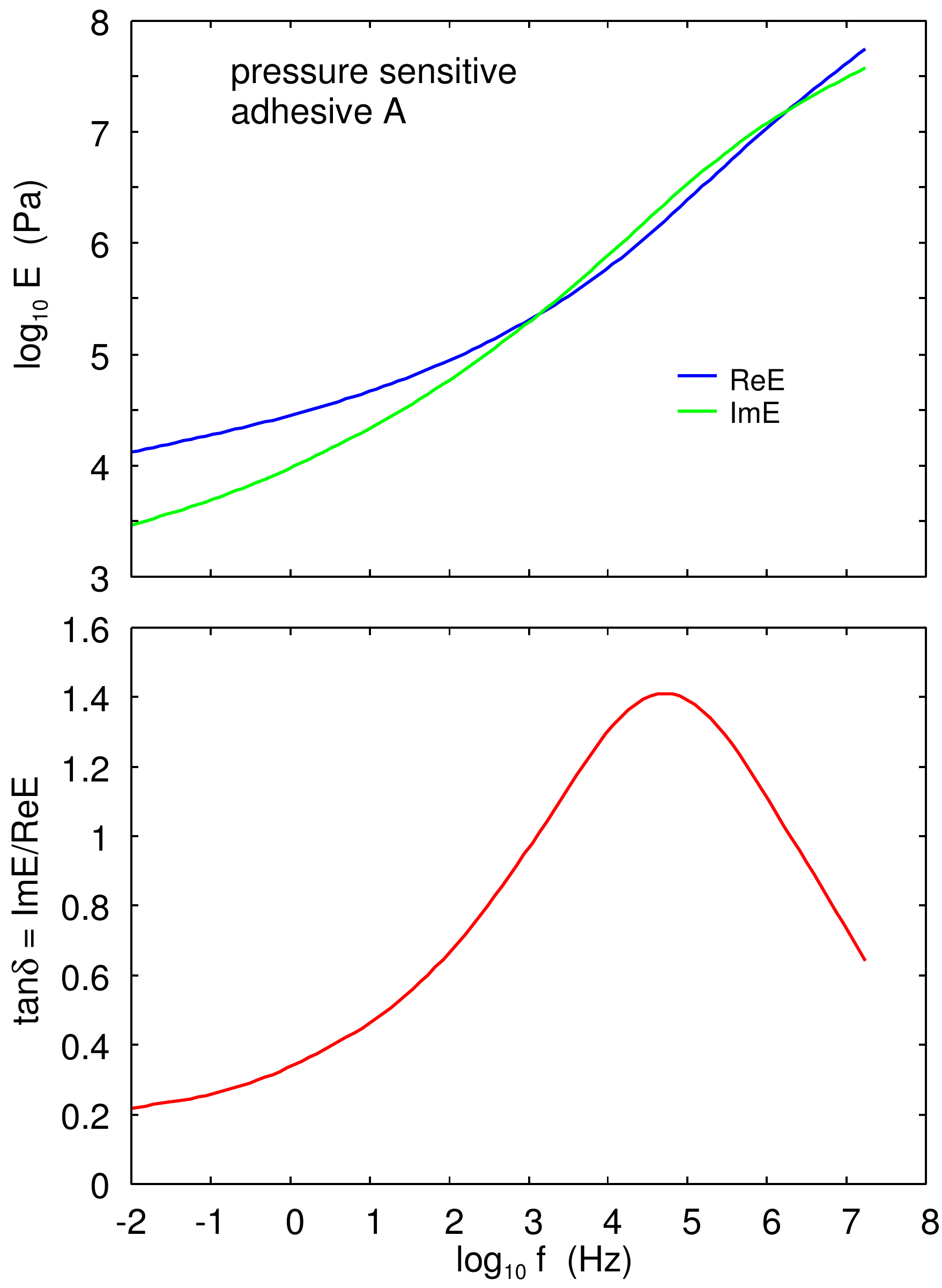}
\caption{
The viscoelastic modulus as a function of the logarithm of the frequency for the pressure sensitive 
adhesive A used in Ref. \cite{Cre1}.
(a) shows the real and imaginary part of $E$ (log-scale) and (b) ${\rm Im}E/{\rm Re }E = {\rm tan} \delta$. 
}
\label{1logf.2tand.pdf}
\end{figure}

\begin{figure}[tbp]
\includegraphics[width=0.45\textwidth,angle=0]{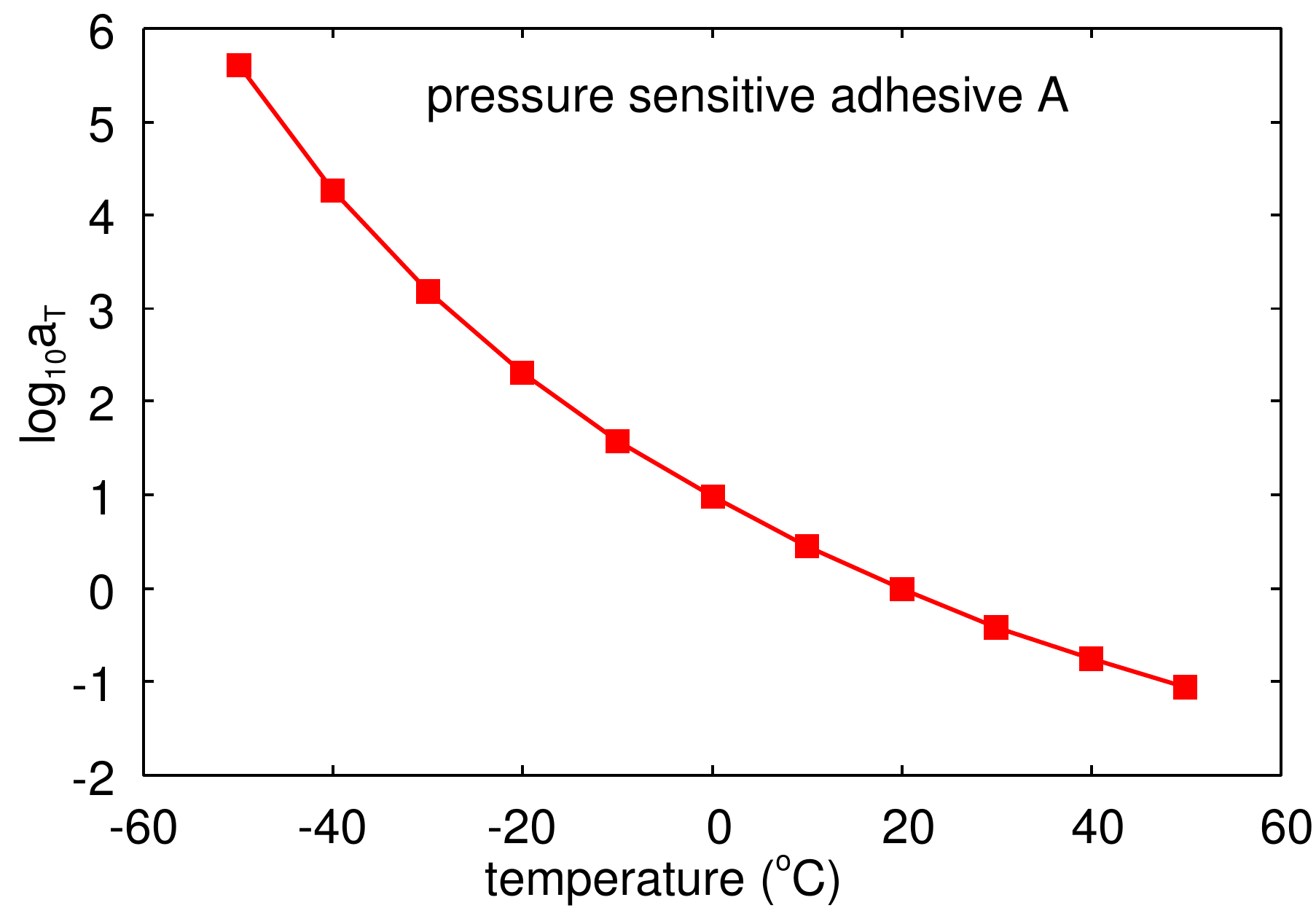}
\caption{
The logarithm of the (horizontal) frequency-temperature shift factor $a_T$ as a function of the 
temperature for the pressure sensitive adhesive A.
}
\label{1Temp.2logaT.pdf}
\end{figure}

\begin{figure}[tbp]
\includegraphics[width=0.45\textwidth,angle=0]{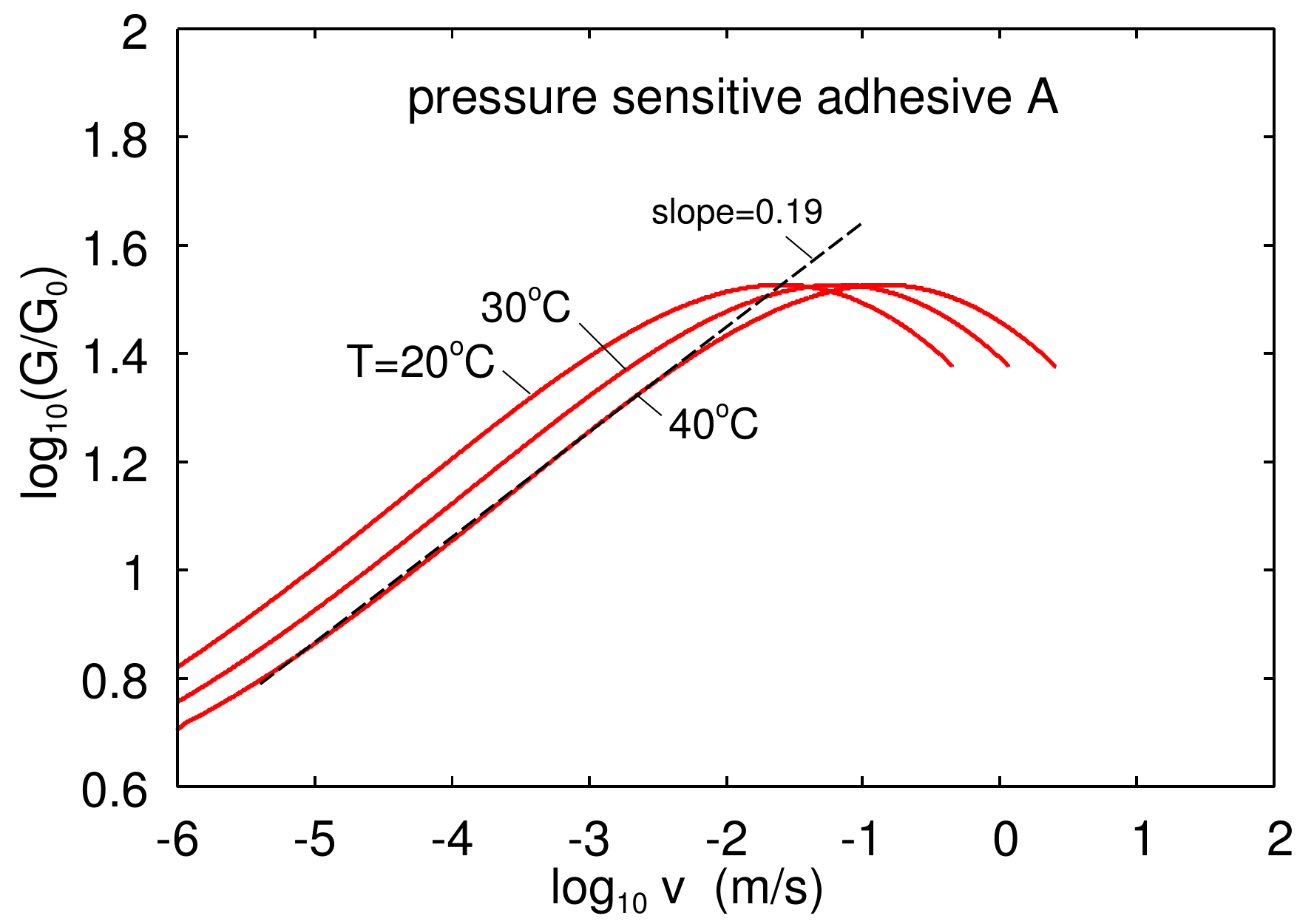}
\caption{
The viscoelastic enhancement factor $G_{\rm open} /G_0$ as a function of the crack-tip speed
for the pressure sensitive adhesive A, 
for the temperatures $T=20^\circ {\rm C}$ (red lines), $30^\circ {\rm C}$ (green lines) and $40^\circ {\rm C}$ (blue lines). 
The results are shown for the system sizes (film thickness) $L=20 \ {\rm \mu m}$.
}
\label{1logv.2G.pressure.sensitive.adhesive.pdf}
\end{figure}

\vskip 0.2cm
{\bf 3.1 Pulling adhesive tape}

A pressure sensitive adhesive typically consist of a soft (weakly crosslinked) rubber film (with tacky additives) 
on a stiffer polymer film, e.g. of polyester type.
In recent studies of peeling of adhesive tapes\cite{Bar,Dalbe,Cret4,Affe} 
the crack propagation energy $G(v)$ was measured as a function of the peeling velocity $v$.
Thus, for example, peeling of the 3M Scotch 600 tape, which consist of a polymer film covered by a thin
$d\approx 20 \ {\rm \mu m}$ (acrylic) adhesive film, resulted in a $G(v)$ function very similar
in form to what is shown in Fig. \ref{1logv.2logG.1mum.1cm.infinite.NBR.filled.pdf}, 
with a maximum around $v_{\rm m} \approx 0.1 \ {\rm m/s}$.
For peeling velocities $v< v_{\rm m}$ the crack-tip process zone is very complex
involving cavitation and stringing, and $G_0(v)$ is likely to depend on the crack tip speed. 
Thus for $v<v_{\rm m}$ the velocity dependency of $G(v)$ 
will depend not only on the bulk viscoelastic energy dissipation, but also on $G_0(v)$, which was considered as a constant above.

The complex processes occurring close to the crack tip for low peeling velocities result in a 
very rough rubber surface which appear white due to light scattering from the surface inhomogeneities\cite{WithGorb}.
However, high peeling speeds result in a much smoother (and transparent) rubber film. This indicate a much simpler
crack-tip process zone for $v>v_{\rm m}$. Thus, the theory developed above may be directly applied to $v> v_{\rm m}$.
In this velocity region the decrease in $G(v)$ may result from the finite thickness of the adhesive film as predicted
by the theory above. This is expected for a thin film, but not for an infinite system where $G(v)$ increases monotonically
with the crack tip speed (see Fig. \ref{1logv.2logG.ideal.pdf} and \ref{1logv.2logG.filled.unfilled.NBR.pdf}). 
This origin of a maximum in the $G(v)$ curve was already suggested by de Gennes\cite{DeG}.

Let us present some numerical results for a pressure sensitive (acrylic) rubber compound used in an earlier study (see Ref. \cite{Cre1}).
Fig. \ref{1logf.2tand.pdf} shows the viscoelastic modulus as a function of frequency (log-log scale) for a pressure sensitive 
adhesive denoted A in Ref. \cite{Cre1}.
(a) shows the real and imaginary part of $E$ and (b) ${\rm Im}E/{\rm Re }E = {\rm tan} \delta$. 
Fig. \ref{1Temp.2logaT.pdf} shows the logarithm of the (horizontal) frequency-temperature shift factor $a_T$ as a function of the temperature for the 
same compound.

Using the viscoelastic modulus in Fig. \ref{1logf.2tand.pdf} and assuming a  $L=20 \ {\rm \mu m}$ thick rubber film,
in Fig. \ref{1logv.2G.pressure.sensitive.adhesive.pdf}
we show the calculated viscoelastic enhancement factor $G_{\rm open} /G_0$ as a function of the crack-tip speed
for the pressure sensitive adhesive A, 
for the temperatures $T=20^\circ {\rm C}$ (red lines), $30^\circ {\rm C}$ (green lines) and $40^\circ {\rm C}$ (blue lines). 
The slope of the curve and the velocity where $G(v)$ is maximal is close to what is observed in Ref. \cite{Dalbe,Cret4}.

The maximum of the $G(v)$ curve in the experiments presented in Ref. \cite{Dalbe,Cret4} is about $100 \ {\rm J/m^2}$.
We have found that at the maximum $G/G_0 \approx 30$. So this imply $G_0 \approx 3 \ {\rm J/m^2}$. This is much bigger than
the adiabatic work of adhesion $\Delta \gamma = \gamma_1+\gamma_2-\gamma_{12}$, which probably is around
$0.05 \ {\rm J/m^2}$ (because of the inert backing of the tape). 
So $G_0$ is increased by a factor of $\sim 60$ or so compared to the adiabatic case.
We attribute this to the cavitation and stringing  in the crack tip process zone.

\begin{figure}[tbp]
\includegraphics[width=0.45\textwidth,angle=0]{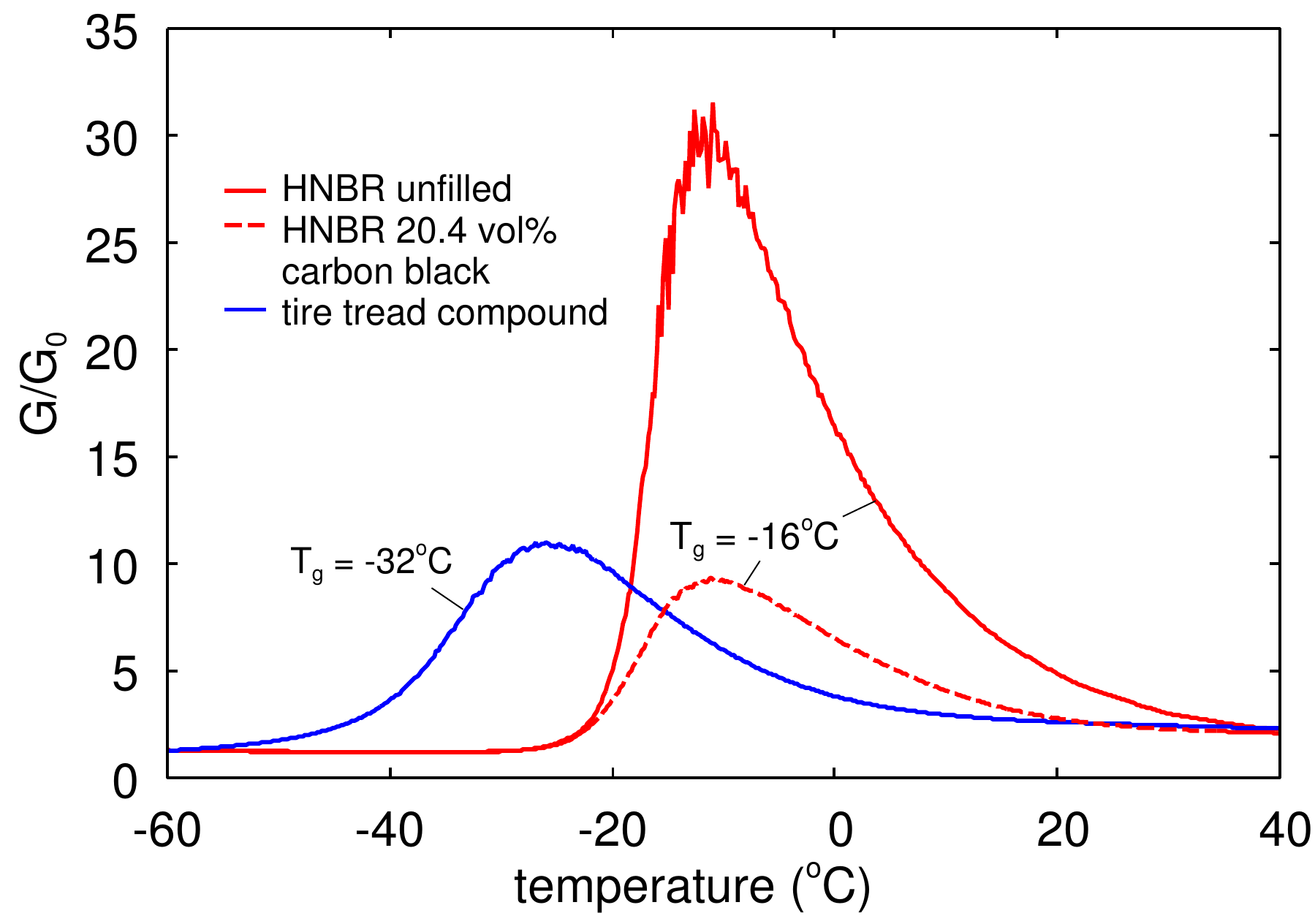}
\caption{
The viscoelastic enhancement factor $G_{\rm open}/G_0$ as a function of the temperature for the crack-tip
speed $v=10 \ {\rm \mu m/s}$. The system size $L=100 \ {\rm \mu m}$ as is the typical diameter of the contact between a ball
and a flat surface in JKR\cite{JKR} adhesion experiments. Results are shown for unfilled and filled ($20.4 \ {\rm Vol.}\%$ carbon black) HNBR 
(red solid and dashed lines, respectively) and for a tread compound (blue curve). The glass transition temperature of the
HNBR and tread compound are $-16^\circ {\rm C}$ and $-32^\circ {\rm C}$, respectively.
}
\label{1Temp.2G.over.G0.pdf}
\end{figure}

\begin{figure}[tbp]
\includegraphics[width=0.45\textwidth,angle=0]{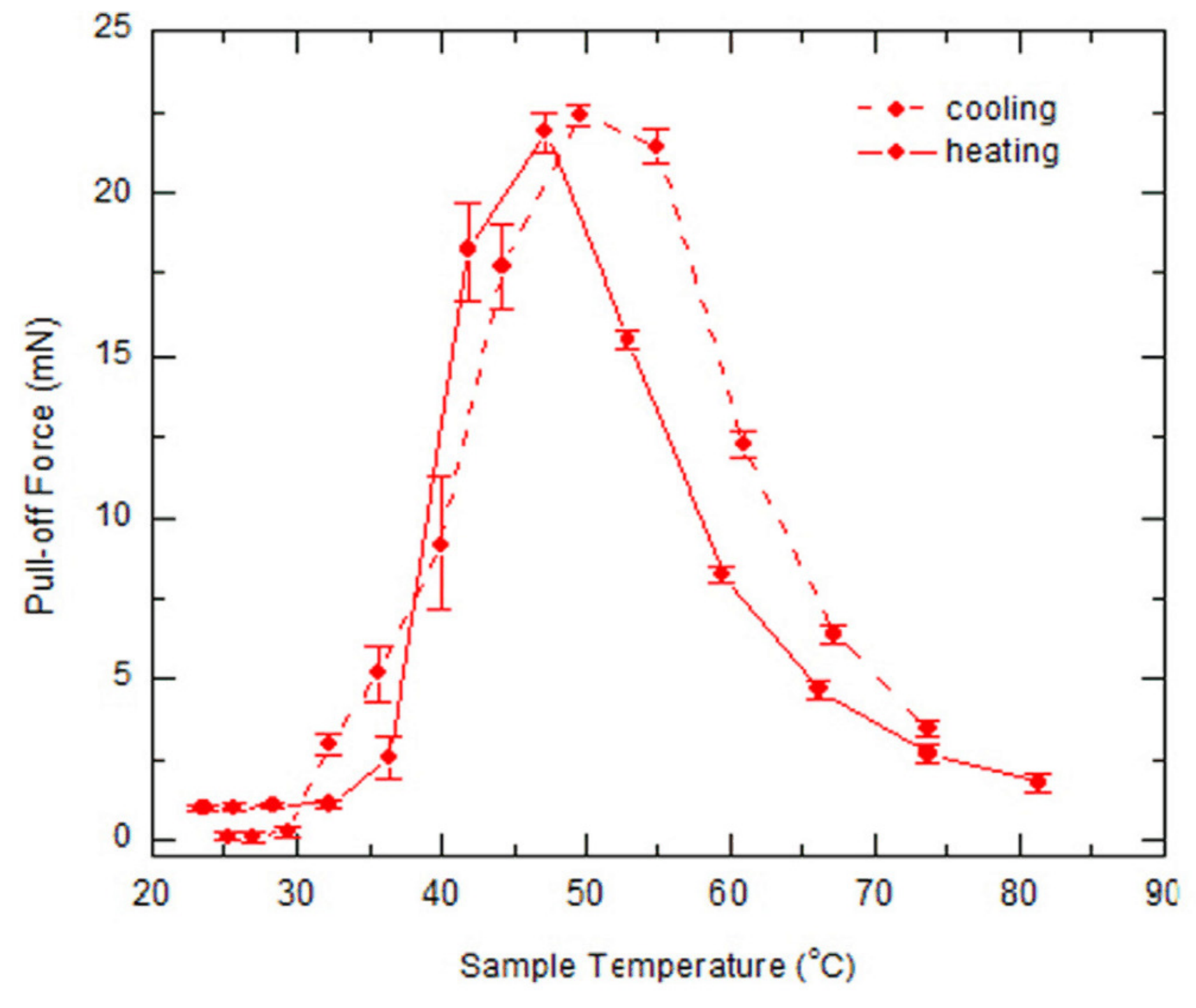}
\caption{
The pull-off force as a function of temperature for a glass ball with radius $R=1.5 \ {\rm mm}$ 
first squeezed in contact with a flat surface of a (photopolymerizable) acrylic polymer ($T_{\rm g} = 53^\circ {\rm C}$),
and then pulled-off with the speed $v_z \approx 10 \ {\rm \mu m/s}$. From Ref. \cite{Elmar}.
}
\label{ExperimentAdhesion.pdf}
\end{figure}

\vskip 0.2cm
{\bf 3.2 Ball-flat adhesion: role of finite size effects}

Here we compare the 
theory prediction with a ball-flat pull-off
(adhesion) experiment\cite{Kendal,softM,Lorenz,Saw,Creton1,Kramer2,JKR,Elmar,dor1,dor2,dor3,dor4}. 
Adhesion experiments are typically performed by moving a spherical
ball (radius $R$) in and out of contact with a substrate. This type of experiments can be analyzed using the Johnson-Kendall-Roberts (JKR) theory, which predict
the pull-off force $F_{\rm pull-off} = (3\pi/2) G R$, where $G$ is the work of adhesion. 
The work of adhesion is the energy per unit surface area to propagate an interfacial (opening or closing) crack. 
Hence, for viscoelastic solids such as rubber, there will be a viscoelastic contribution to $G$ given by the theory above: $G/G_0 = 1+f(v,T)$,
where $G_0$ the the work of adhesion in the adiabatic limit (crack speed $v \rightarrow 0$).

Here we are interested in a hard ball in contact with a flat rubber surface. 
In a typical adhesion experiment the ball radius $R$ is of order a few mm, and the diameter of the area of contact when the pull-off instability occur,
of order $\sim 0.1 \ {\rm mm}$. The linear size $L \sim 0.1 \ {\rm mm}$ of the contact region (at the point of the onset of snap-off)
determines the size of the region in space where the stress-field exhibit the inverse-square-root (singular) behavior expected 
as a function of the distance away from the tip of crack-like defects. 

Fig. \ref{1Temp.2G.over.G0.pdf} shows the calculated [from (6)] viscoelastic enhancement 
factor $G_{\rm open}/G_0$ as a function of the temperature for the crack-tip
speed $v=10 \ {\rm \mu m/s}$. The system size $L=0.1 \ {\rm mm}$.
Results are shown for unfilled and filled ($20.4 \ {\rm Vol.}\%$ carbon black) Hydrogenated Nitrile Butadiene Rubber (HNBR) rubber
(red solid and dashed lines, respectively) and for a tread compound (blue curve). The glass transition temperature of the
HNBR and tread compounds are $-16^\circ {\rm C}$ and $-32^\circ {\rm C}$, respectively.

Let us compare the results in Fig. \ref{1Temp.2G.over.G0.pdf} with the experimental results obtained in Ref. \cite{Elmar}
for a silica-glass ball with radius $R=1.5 \ {\rm mm}$ 
first squeezed in contact with a flat surface of an acrylic polymer,
and then pulled-off with the speed $v_z \approx 10 \ {\rm \mu m/s}$. 
We will assume that both surfaces are perfectly smooth. The acrylic polymer surfaces where produced by photopolymerization 
with the polymer confined between two smooth glass plates. 
The glass ball is also expected to be very smooth, but no information about the surface roughness was
given in Ref. \cite{Elmar}.

The acrylic polymer used in Fig. \ref{ExperimentAdhesion.pdf} has a much higher glass transition temperature
than the HNBR rubber used in the calculations ($T_{\rm g} = 53^\circ {\rm C}$ compared to $-16^\circ {\rm C}$ for HNBR),
which will result in a shift of the adhesion curve along temperature axis, but the 
temperature dependency of the pull-off force (which is proportional to $G$) for the acrylic polymer is very similar
to the temperature dependency on the work of adhesion for unfilled HNBR. 
In particular, both the full-width-at-half-maximum (FWHM) (about $20^\circ {\rm C}$), and the asymmetry of the peak in the pull-off force and the work of adhesion,
are nearly the same. This is indeed expected because the change in the viscoelastic
modulus from the rubbery region to the glassy region is nearly the same for both polymers (from $\approx 3 \ {\rm MPa}$ to $\approx 2 \ {\rm GPa}$).
We also note that the viscoelastic enhancement in the pull-off force 
observed in the experiment (roughly $15-30$) is very similar to what the theory predict.
Unfortunately, the (complex) frequency-dependent modulus $E(\omega )$ for the acrylic polymer was not reported on in
Ref. \cite{Elmar} so no detailed comparison between theory and experiment is possible.

\begin{figure}
\includegraphics[width=0.4\textwidth]{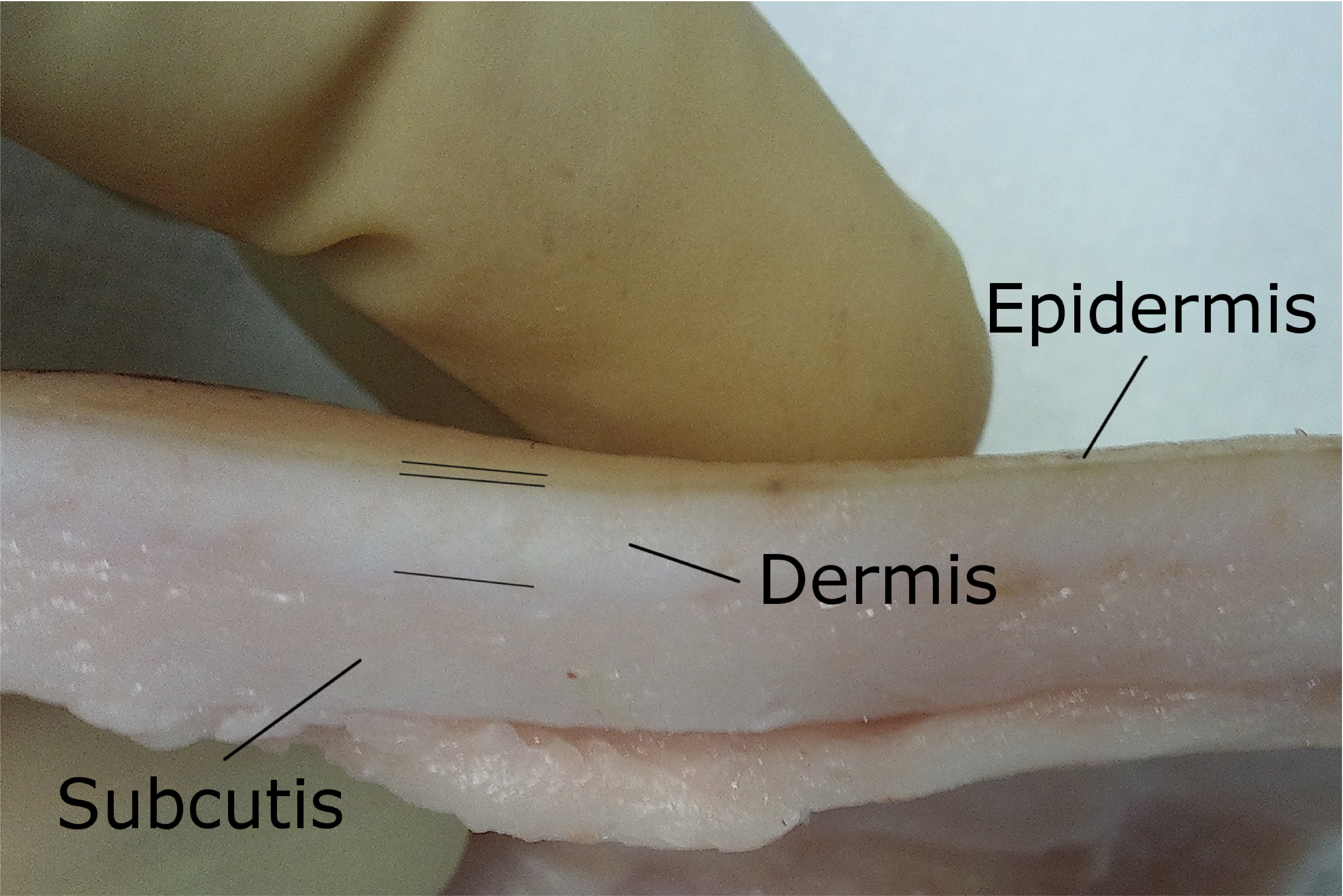}
\caption{\label{skin-picture}
Picture of the pig skin used in the experiment with the different skin layers as indicated. For the experiment we have cut out pieces of the dermis of the skin.}
\end{figure}

\vskip 0.3cm
{\bf 3.3 Crack propagation in the pig skin (dermis), with application to intradermal injections}

The delivery of a drug product through body tissue (muscle, skin, or organ tissue) is one of 
the most common routes of delivery for therapeutic drug products. Delivery into the skin layers includes injections in the sub-cutaneous space, which has growing interest, due to its applications to patient self-injection for chronic disease like diabetes. Intra-dermal injection, targeting the space between the outer skin layer (epidermis) and the sub-cutaneous space, is of particular interest for vaccines, due to its potential to elicit a stronger immune response. 
To understand fundamentals of injections, we need to consider the tissue as a media composed by
cells, extracellular matrices, and interstitial fluids\cite{derm1,derm2}.
The drug delivery depends on the
permeability of the tissue as well
as the compatibility between the
injecting fluid and the interstitial
fluids.

The majority of the studies
analyzing the fundamentals of
how the drug is delivered into
the tissue use porous
elastic models where the drug
product diffuses through the
tissue by permeability. The use of
Darcy's and Brinckman's
diffusion equations is a common practice\cite{addN}.

Here we consider the injection of fluid in the pig dermis.  
The dermis consist of a three-dimensional crosslinked network of elastic fibers (collagen, elastin) surrounded by
an amorphous gel-like (water rich) substance (containing mucopolysaccharides, chondroitin sulfates, and glycoproteins).
The gel-like substance provides lubrication for collagen fibers, which indicate that the bonding interaction between the elastic
fibers and the gel-like substance may be relative weak, and a likely region for crack to propagate during fluid injection.

We propose that the fluid injection in the dermis is similar to
hydraulic fracturing (also called fracking), used in oil and gas exploration,
involving the fracturing of an inhomogeneous material (here the dermis) by a pressurized liquid. 
In the present case, because of the low elastic modulus the skin dermis, the pressurized 
fluid will separate the cracked surfaces giving fluid filled cavities in the skin. Because of the
permeability of the dermis the fluid is diffusing away from the fluid filled cavities, but the speed of this
process will depend on the compatibility of the injected fluid and the interstitial fluids.

\vskip 0.2cm
{\bf A. Viscoelastic modulus of the dermis}

The experimental study uses pig skin where the dermis is relative thick, see Fig. \ref{skin-picture}.
The specimen used in the experiment was bought from the butcher and used one day after the pig was killed. 
The viscoelastic modulus was measured in tension mode using a DMA Q800 from TA Instruments.
From the skin specimen we have cut out a sample of the dermis removing first the subcutis and the epidermis. 
We used stripes of the dermis with rectangular cross section, $4.5 \times 2.5 \ {\rm mm}$, and  length of 7 to 12 mm. 
One problem is that the viscoelastic properties of the specimen changes with time, 
e.g., due to the loss of fluid evaporating to the atmosphere. 
As we can not control the humidity in the DMA Q800 this effect may have the largest influence on the results. 
In order to minimize this effect we have covered the sample with liquid pork fat prior to the experiment. 
The fat has been produced by heating up some of the pig skin in an oven for some time. 

To measure the viscoelastic modulus the specimen gets excited with different frequencies in tensile mode at different temperatures. 
The applied strain amplitude is chosen to be rather small (0.5\% strain) to avoid nonlinear effects. 
The temperature range covered start at $40\,^{\circ}\mathrm{C}$ down to $-50\,^{\circ}\mathrm{C}$. 
The results have then been shifted to obtain a smooth mastercurve. This is found to be rather 
complicated as their is a strong change when the temperature is around $-10\,^{\circ}\mathrm{C}$. 

The first measurement is done at a constant temperature of $40\,^{\circ}\mathrm{C}$ while the frequency is changed in steps from 25 to 0.25 Hz. 
After, the temperature is decreased by $10\,^{\circ}\mathrm{C}$ and the experiment repeated until $-50\,^{\circ}\mathrm{C}$ is reached. 
We have shifted the imaginary part of the viscoelastic modulus to obtain a smooth mastercurve. 
This time-temperature shifting procedure is often used for rubber-like materials, but may not hold accurately for the dermis.

\begin{figure}
\includegraphics[width=0.45\textwidth]{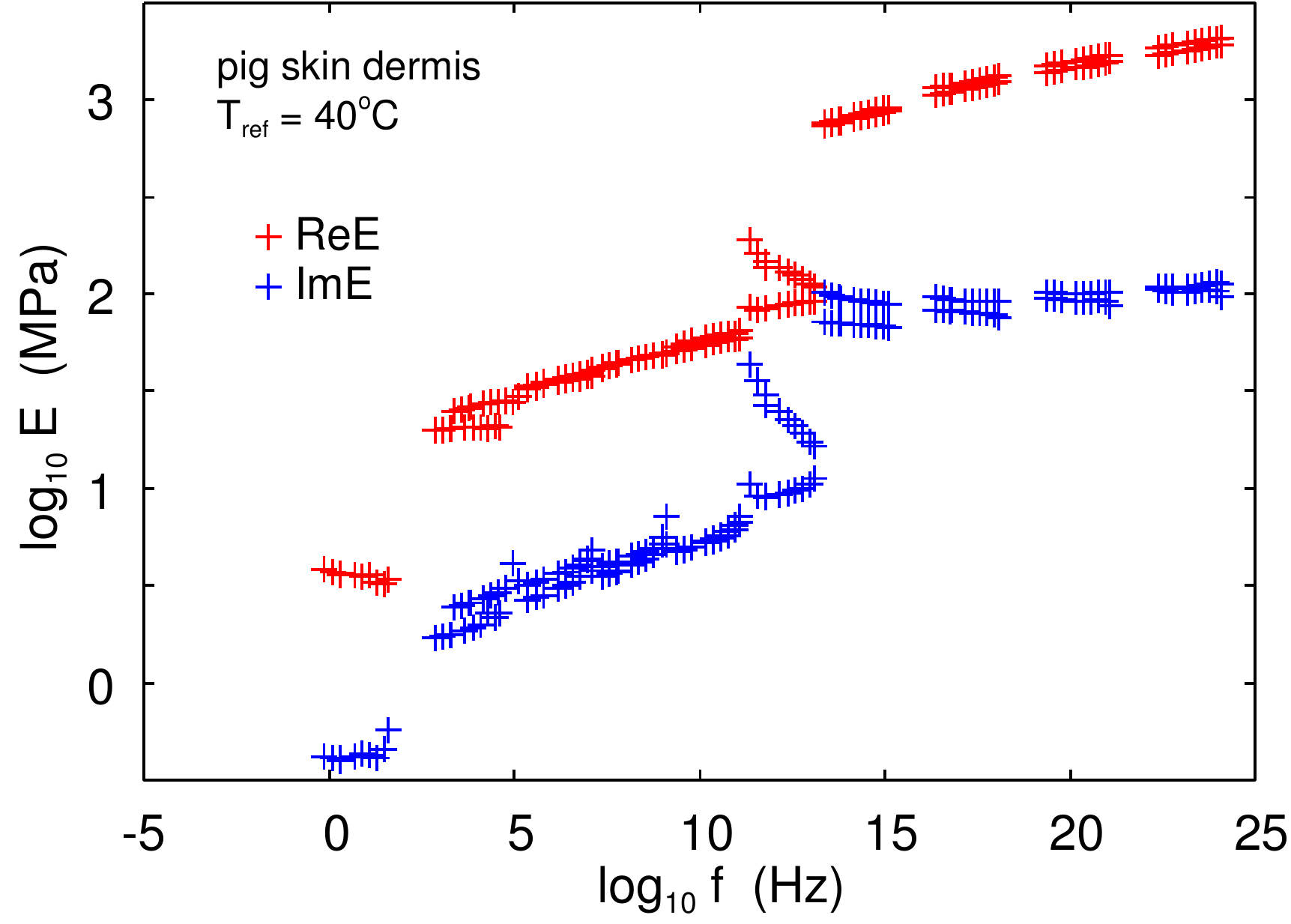}
\caption{\label{1logf.2logE.pork.pdf}
Real and imaginary part of the elastic modulus of the pig dermis for with fat covered pig skin samples.
The experiment started at $T=40^\circ {\rm C}$ and ended at $T=-50^\circ {\rm C}$ after $\sim 2 \ {\rm hours}$.
}
\end{figure}

\begin{figure}
\includegraphics[width=0.45\textwidth]{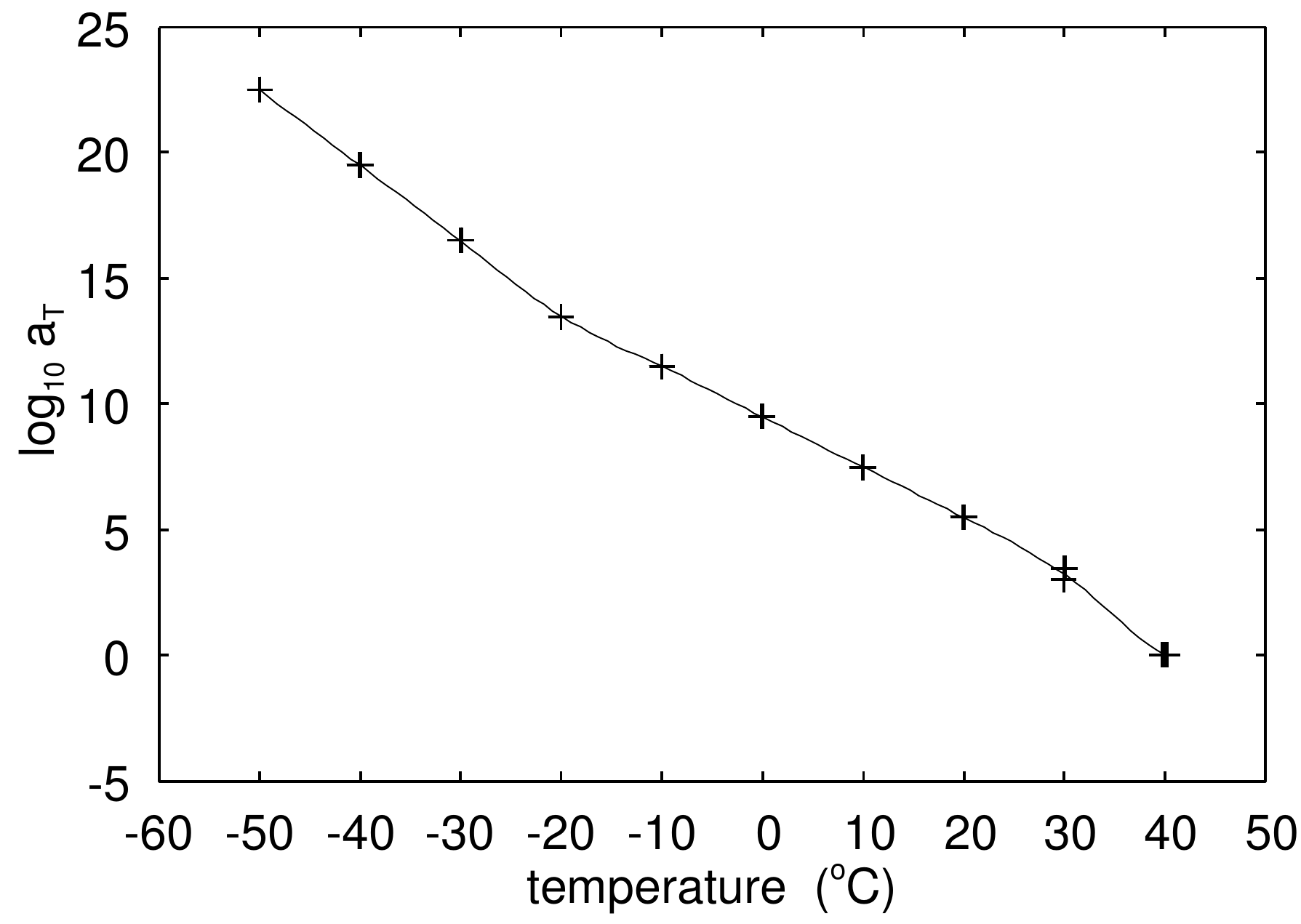}
\caption{\label{1temp.2aT.pork.pdf}
The shift factors obtained by shifting the imaginary part of the low-strain viscoelastic modulus so that a smooth master curve is observed
(shown in Fig. \ref{1logf.2logE.pork.pdf}. 
The reference temperature chosen is $40^\circ {\rm C}$.} 
\end{figure}

Fig. \ref{1logf.2logE.pork.pdf}
shows the real and imaginary part of the elastic modulus as a function of frequency (log-log scale) of the pig dermis.
Fig. \ref{1temp.2aT.pork.pdf}
shows the shift factors obtained by shifting the imaginary part of the viscoelastic modulus as to obtain
the (smooth) master curve shown in Fig. \ref{1logf.2logE.pork.pdf}. 
We have also plotted ${\rm ln} a_{\rm T}$ as a function of $1/T$, 
where $T$ is the absolute temperature (Kelvin) (not shown). We have found that, to a good approximation,
$a_{\rm T} = b{\rm e}^{-\epsilon/k_{\rm B}T}$, with the activation energy $\epsilon \approx 2.5 {\rm eV}$.

Preliminary analysis of pig skin dermis using deferential 
scanning calorimetry indicates several
thermal transitions within the physiological 
temperature range $0$ to $40^\circ {\rm C}$. Further analysis is ongoing to 
determine their relation to the viscoelastic behavior of the dermis.

\begin{figure}
\includegraphics[width=0.25\textwidth]{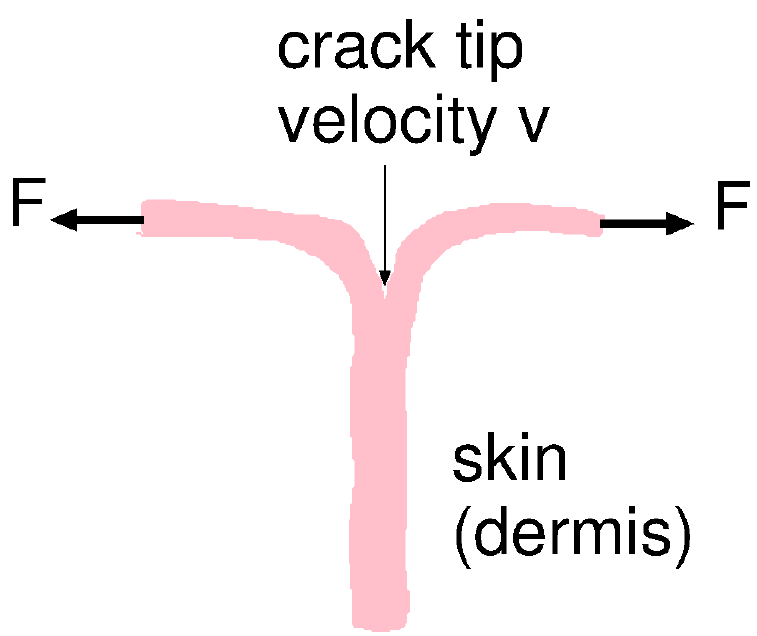}
\caption{\label{pullskin.ps}
The crack propagation energy $G\approx F/w$, where $w$ is the width of the strip.
}
\end{figure}

\begin{figure}
\includegraphics[width=0.45\textwidth]{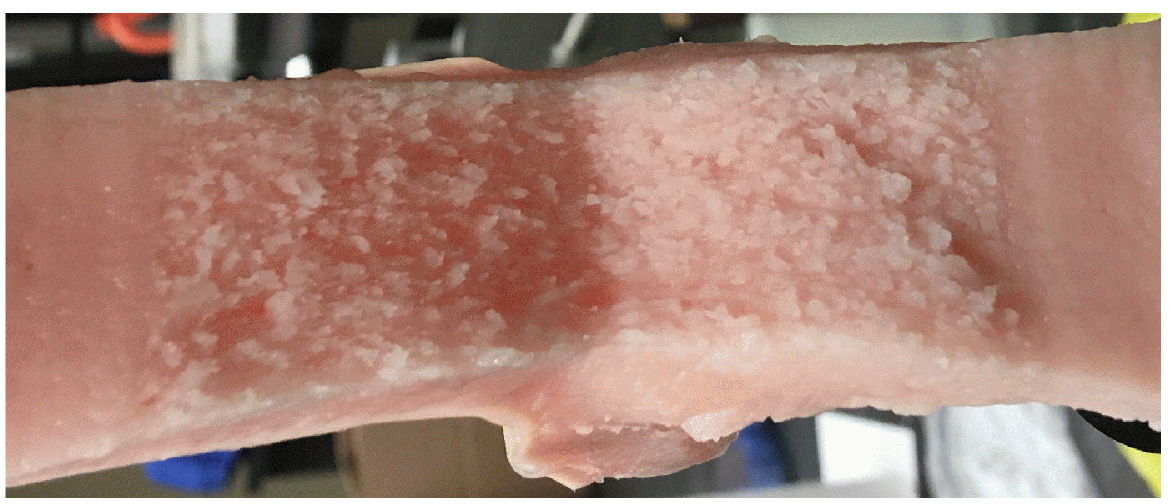}
\caption{\label{PulledDermisPig.ps}
Crack propagation in the pig dermis result in a rough surface.
}
\end{figure}

\begin{figure}
\includegraphics[width=0.45\textwidth]{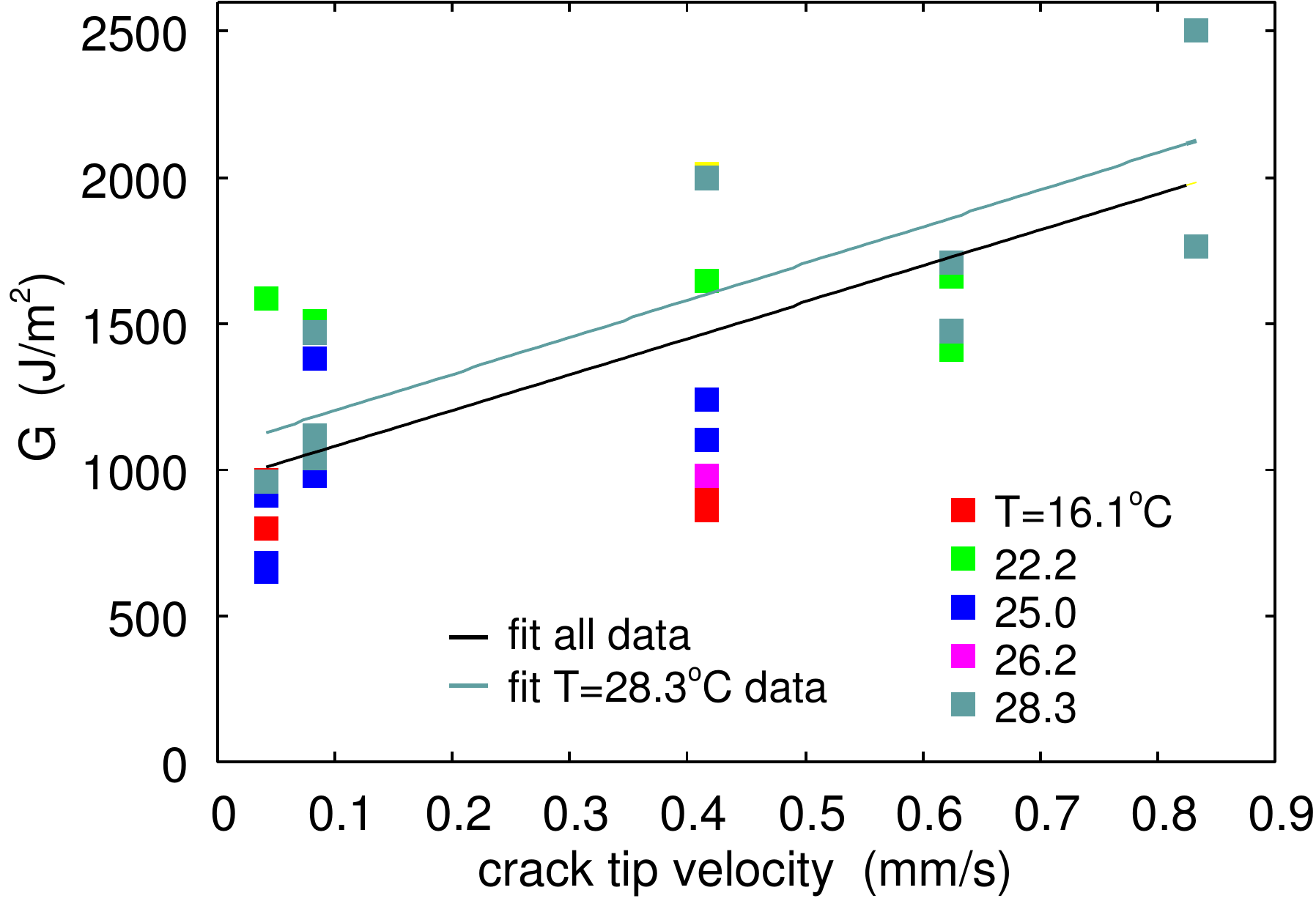}
\caption{\label{1velocity.2G.pork.skin.exp1.pdf}
Crack propagation energy $G(v)$ in the dermis of pig skin for different temperatures
and crack tip velocities.
}
\end{figure}

\begin{figure}
\includegraphics[width=0.45\textwidth]{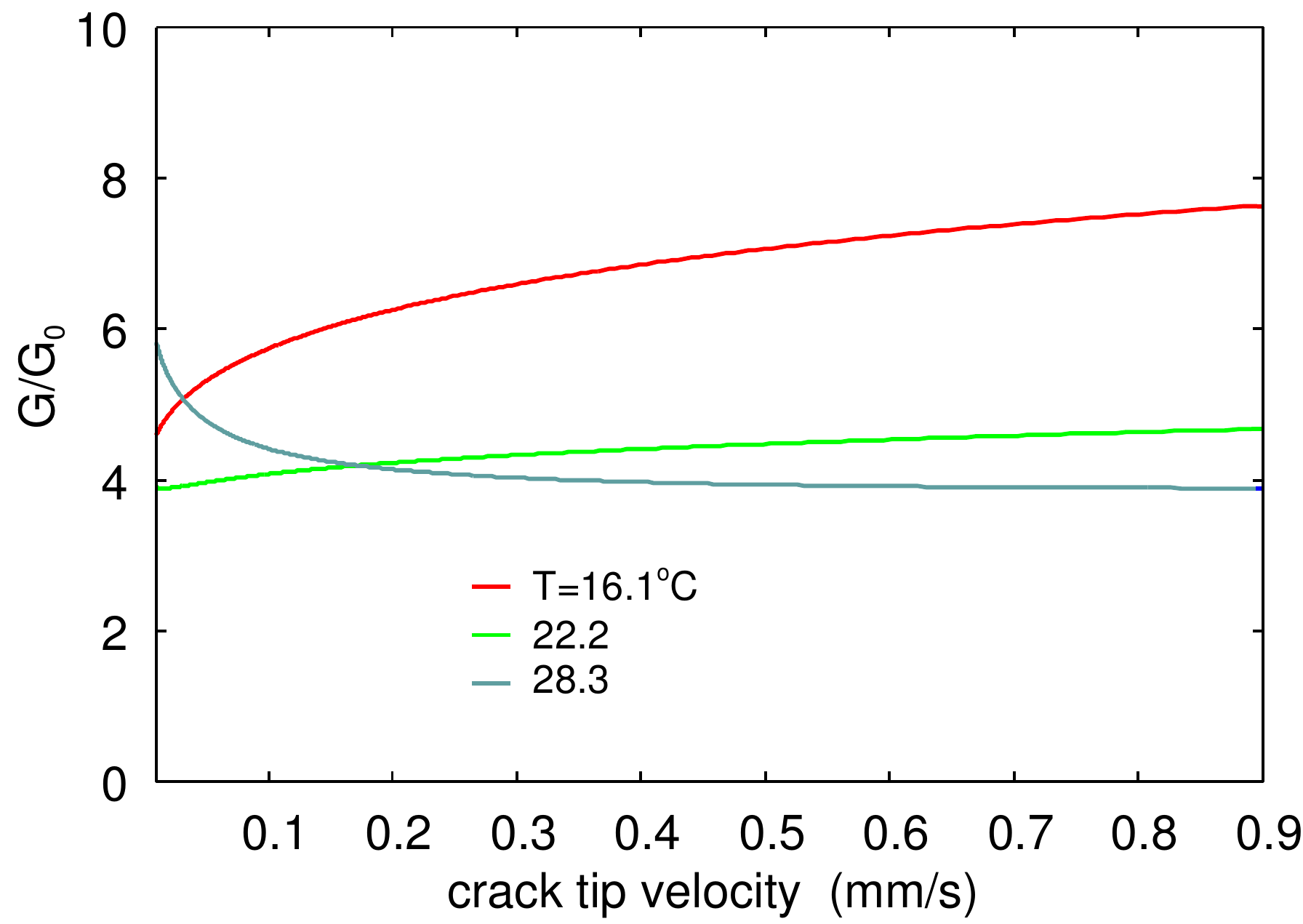}
\caption{\label{1logv.2G.pig.skin.crack.G.1.pdf}
The viscoelastic enhancement factor $G(v)/G_0 = 1+f(v,T)$ as a function of the logarithm of
the crack tip speed. In the calculation web have used the viscoelastic modulus of the pig
dermis shown in Fig. \ref{1logf.2logE.pork.pdf} and the shift factor from Fig. \ref{1temp.2aT.pork.pdf}.
Results are shown for $T=16.1$, 22.2 and $28.3^\circ \ {\rm C}$ and taking into account the finite-size effect with
$L=1 \ {\rm cm}$ of order the thickness of the dermis.
}
\end{figure}

\vskip 0.2cm
{\bf B. Crack propagation in the dermis}

We have measured the energy per unit area for crack propagation in the pig skin dermis. The experiments 
were conducted in an Instron tensile bench, model 5542 equipped with pneumatic grippers set to a pressure of
20 psi. The samples consisted of $w\approx 2.0 - 2.5 \ {\rm cm}$ wide strips of
skin where a crack was initiated by a razor blade cut in the dermis region of the skin,
see Fig. \ref{pullskin.ps}. We measured the $F(v)$ as a function
of the crack tip speed $v$. Neglecting the elastic energy stored in the skin the work $F S \approx G w S$ 
(where $S$ is the length of the crack). Using this equation in Fig. \ref{1velocity.2G.pork.skin.exp1.pdf}
we show $G$ as a function of the crack tip speed. Note that there is only a weak increase in $G$ with the speed $v$
and no systematic temperature dependency.

The crack propagation energy $G=G_0 (1+f(v,T))$ where $1+f(v,t)$ is the viscoelastic enhancement factor
and $G_0$ (which also depends on $v$ and $T$) the contribution from the crack-tip process zone to the crack propagation
energy. In Fig. \ref{1logv.2G.pig.skin.crack.G.1.pdf}
we show the calculated viscoelastic enhancement factor $G(v)/G_0 = 1+f(v,T)$ as a function of the logarithm of
the crack tip speed. In the calculation we have used the viscoelastic modulus of the pig
dermis shown in Fig. \ref{1logf.2logE.pork.pdf} and the shift factor from Fig. \ref{1temp.2aT.pork.pdf}.
Results are shown for $T=16.1$, 22.2 and $28.3^\circ \ {\rm C}$ and taking into account the finite-size effect with
$L=1 \ {\rm cm}$ of order the thickness of the dermis. For the crack speeds shown in Fig. \ref{1velocity.2G.pork.skin.exp1.pdf} the factor 
$1+f(v,t) $ is approximately temperature and velocity independent.
Taking into account the magnitude of $1+f(v,t)$ we conclude that $G_0(V)$
on the average increases from $\approx 200 \ {\rm J/m^2}$ for $v < 0.1 \ {\rm mm/s}$ 
to $\approx 400 \ {\rm J/m^2}$ for $v \approx 1 \ {\rm mm/s}$.

\begin{figure}
\includegraphics[width=0.45\textwidth]{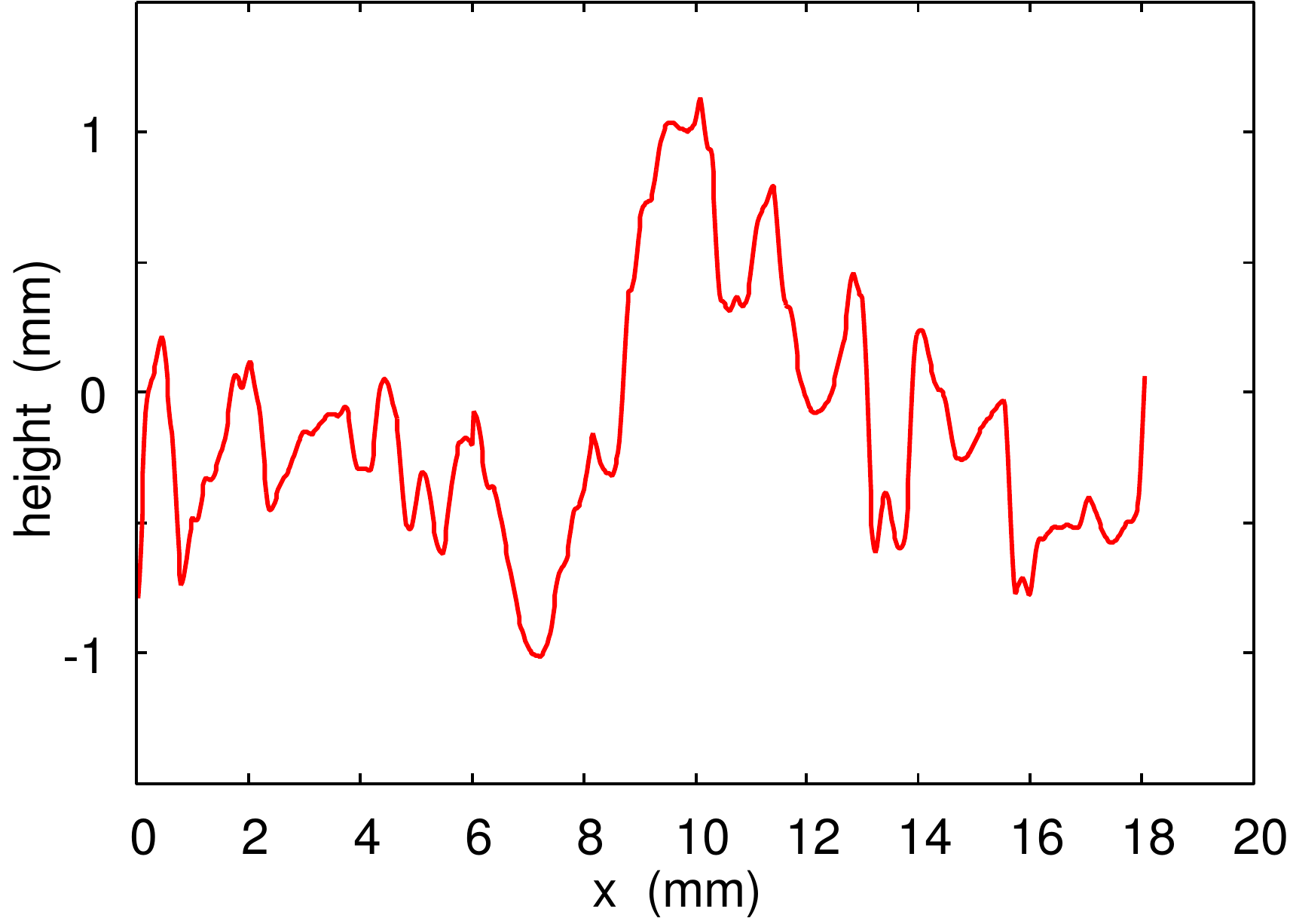}
\caption{\label{1x.2h.mm.pig.skin.pdf}
Topography (line scan) of pig skin dermis surface after crack propagation.
}
\end{figure}

\begin{figure}
\includegraphics[width=0.45\textwidth]{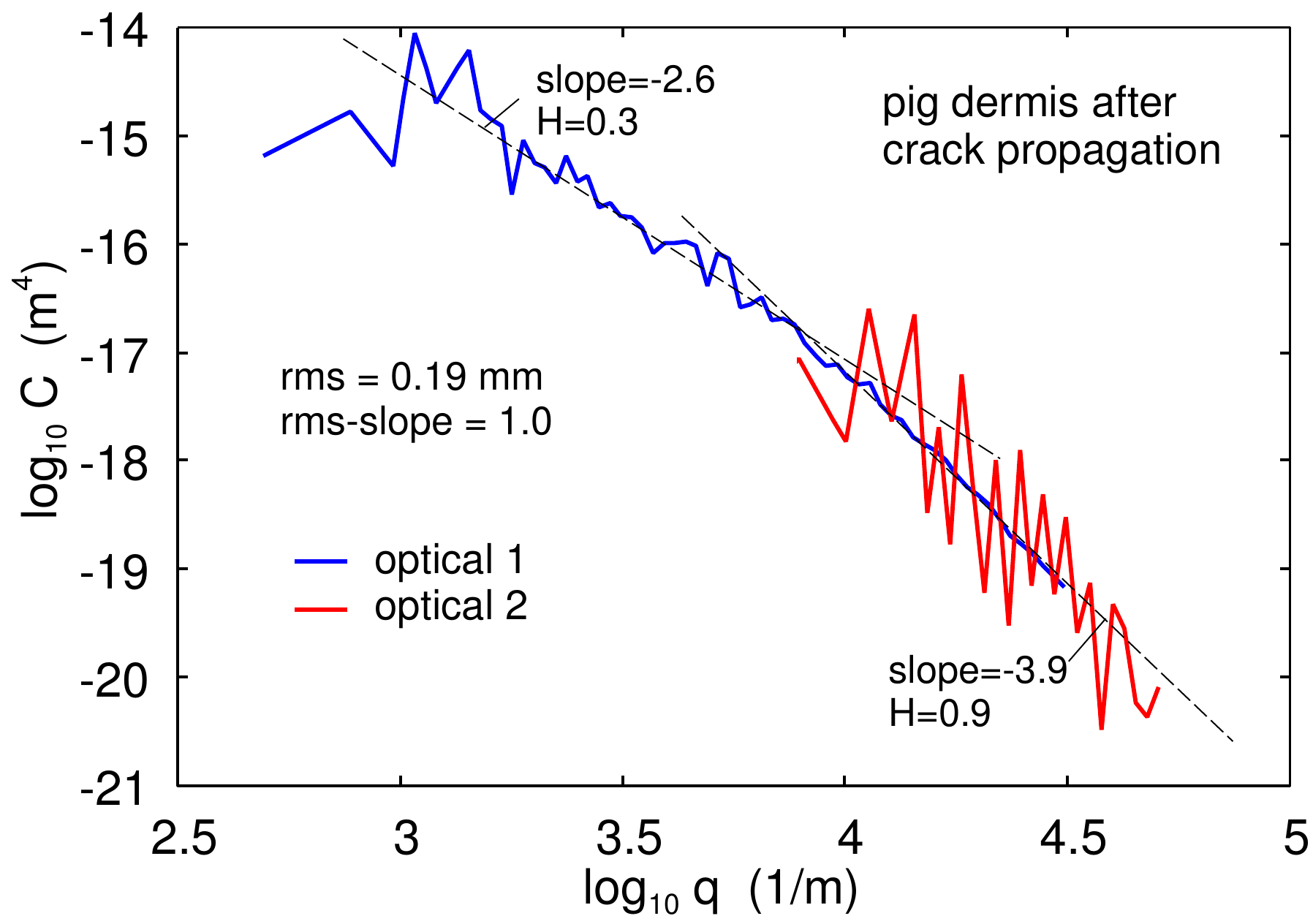}
\caption{\label{1logq.2logC1D.pig.skin.pdf}
The 2D surface roughness power spectrum of pig skin dermis surface after crack propagation.
}
\end{figure}

\vskip 0.2cm
{\bf C. Surface roughness of dermis crack surfaces}

The roughness resulted from the pig skin crack propagation was studied using a Keyence VR 5200 3D measurement system. 
The scanned area was $20 \ {\rm mm}\times 20 \ {\rm mm}$.
The dermis crack surfaces exhibit strong surface roughness as shown in Fig. \ref{PulledDermisPig.ps}.
Fig. \ref{1x.2h.mm.pig.skin.pdf} shows that the amplitude of the height fluctuations is about $1 \ {\rm mm}$.
In Fig. \ref{1logq.2logC1D.pig.skin.pdf} we show the calculated two-dimensional (2D) surface roughness power spectrum. 
Including only the wavenumber region shown in the figure, the surface has the root-mean-square (rms) roughness amplitude $0.19 \ {\rm mm}$
and the rms-slope 1.0.

Note that the slope of the power spectrum curve (on the log-log scale) change for $q= 2 \pi /\lambda$,
$\lambda \approx 1 \ {\rm mm}$, from -2.6 to -3.9.  We associate the region $q < 2 \pi /\lambda$ with 
the large (mm-sized) protruding structures which
can be seen in Fig. \ref{PulledDermisPig.ps}, 
which probably are domains of the soft extracellular matrix detached from the highly 
elongated fibril network (collagen and elastin) in the crack-tip process zone.
Thus, the morphology of the cracked surface (see Fig. \ref{PulledDermisPig.ps}) suggest that the (strong) fibril component of the dermis in not
homogeneously distributed, but are separated by relative large regions of the soft matrix. This will have important implications for the crack
propagation in the dermis during fluid injection (see Sec. 3.3E).

The roughness found on the cracked surfaces in Fig. \ref{PulledDermisPig.ps} is consistent with the crack propagation energy
shown in Fig. \ref{1velocity.2G.pork.skin.exp1.pdf}. Thus, experiments have shown that when a strip of dermis is 
elongated in tension, the maximum tensile stress before the 
dermis break is $\sigma_{\rm c} \approx 2 \ {\rm MPa}$ (see Ref. \cite{derm2}). 
The amplitude of the surface roughness created on the cracked surfaces in
Fig. \ref{PulledDermisPig.ps} is about $d\approx 1 \ {\rm mm}$ (typically $\approx 5$ times higher than the rms 
roughness amplitude). Thus one expect
the crack propagation energy to be $G_0 \approx \sigma_{\rm c} d /2 \approx 1000 \ {\rm J/m^3}$ which is consistent 
with Fig. \ref{1velocity.2G.pork.skin.exp1.pdf}.

\begin{figure}
\includegraphics[width=0.45\textwidth]{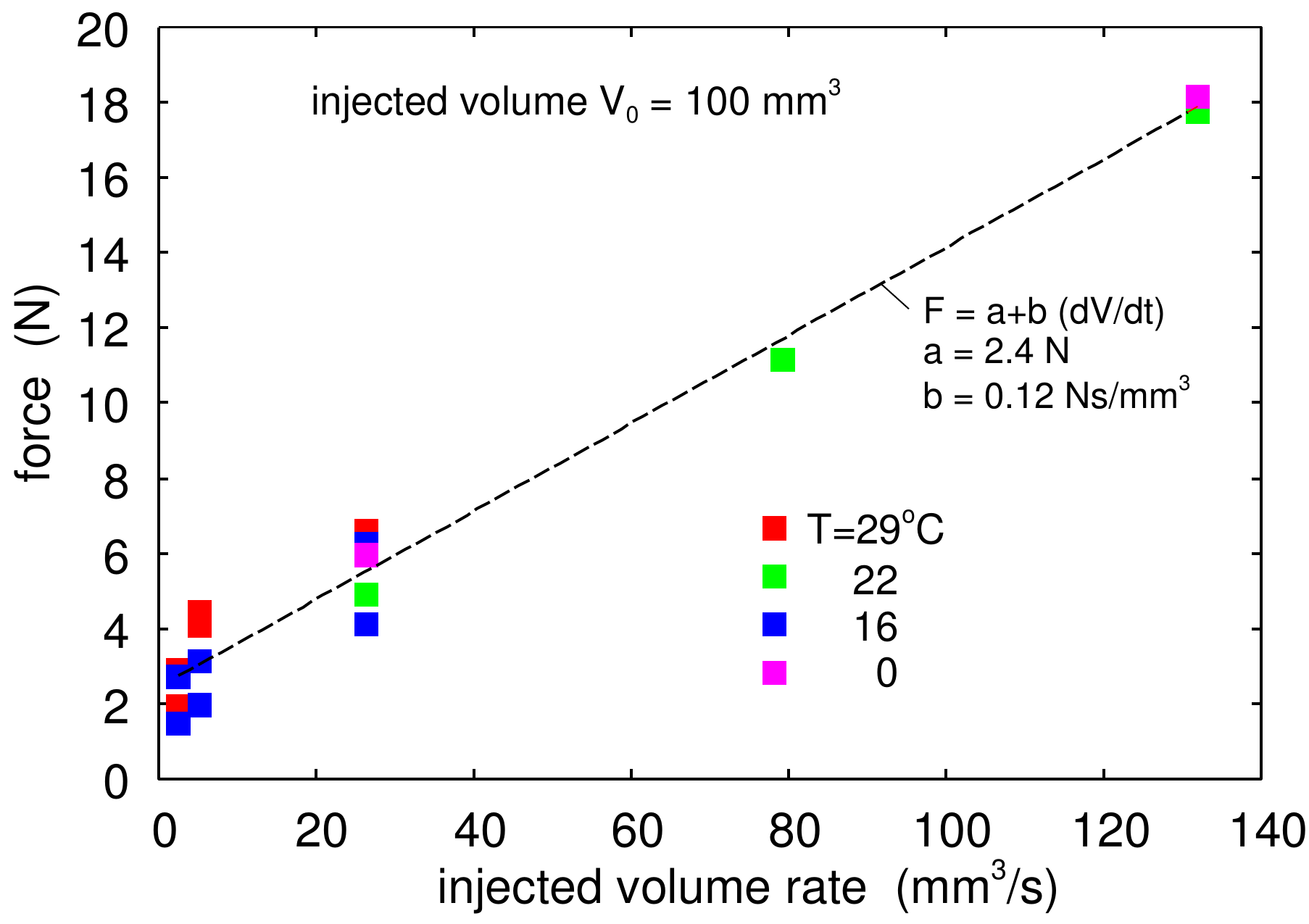}
\caption{\label{1dotV.2Fav.pdf}
The dependency on the force needed to inject a fluid in the pig dermis on the volume injection rate.
}
\end{figure}

\vskip 0.2cm
{\bf D. Intradermal fluid injection}

\begin{figure}
\includegraphics[width=0.40\textwidth]{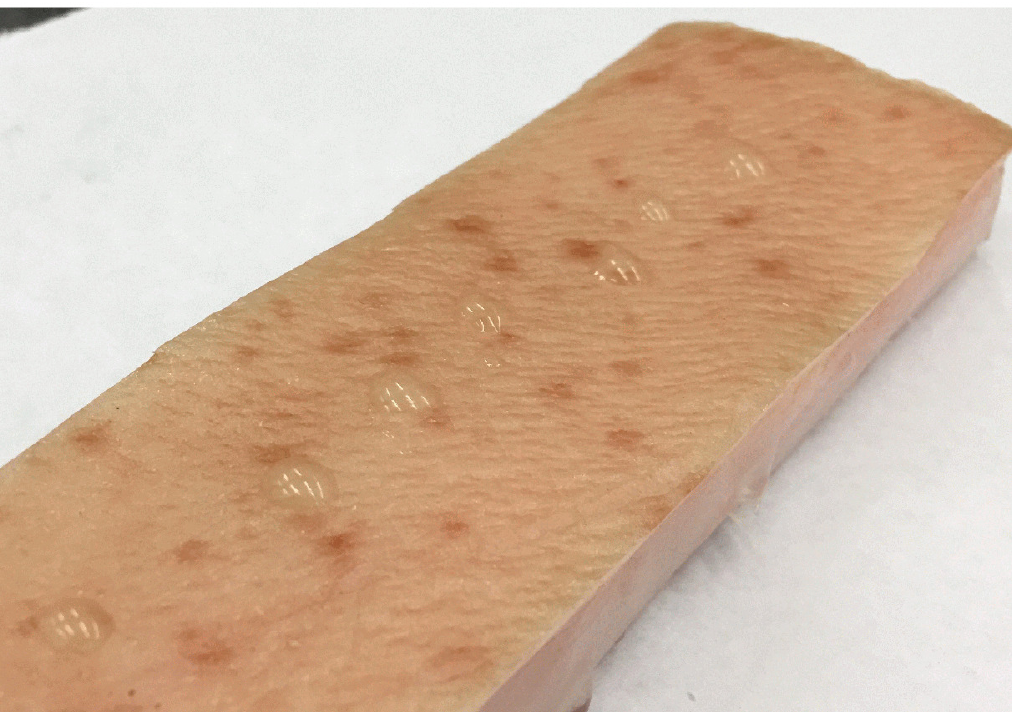}
\caption{\label{WaterDroppletPigSkin.ps}
After injecting the fluid and removing the needle the fluid pressure in the
cavity is high enough to allow some fluid to leak to the surface of the skin where it form
small water droplets.
}
\end{figure}

We have measured the injection of deionized water in the pig belly dermis (see Fig. \ref{WaterDroppletPigSkin.ps}). 
The water was injected using a $1 \ {\rm mL}$ glass syringe with
a barrel with inside cross section area $A\approx 30 \ {\rm mm}^2$. 
The syringe's needle was a 29 gauge with a $184 \ {\rm \mu m}$ inner diameter and $340 \ {\rm \mu m}$ outer diameter. The
injection depth was $6 \ {\rm mm}$. The injection setup was mounted in a Instron tensile bench model
5542 to control the speed and injected volume.

The total force applied to the rubber stopper is $F_{\rm tot} = F_{\rm f}+F_{\rm vis}+ Ap$ where
$F_{\rm f}$ is the friction force between the rubber stopper and the glass barrel, $F_{\rm vis}$ the force (due to the water viscosity)
needed to squeeze the fluid through the needle, and $Ap$ the force to inject the fluid in the dermis
($p$ is the water pressure in the dermis, and $A$ the barrel inner cross-section area).

The friction force between the rubber stopper and the glass barrel $F_{\rm f}$, and the viscous force needed 
to squeeze the fluid through the needle $F_{\rm vis}$, where measured before and after injecting fluid in the skin dermis. 
The so obtained force was subtracted from the total force $F_{\rm tot}$ needed to inject the fluid in the dermis.

Fig. \ref{1dotV.2Fav.pdf}
shows the dependency of the effective injection force $Ap$ on the volume injection rate $\dot V$.
The total injected volume is $V_0 = 100 \ {\rm mm}^3$. 
Note that the injection force tends to increase linearly with the injection rate,
and is nearly independent of the temperature.


\begin{figure}
\includegraphics[width=0.30\textwidth]{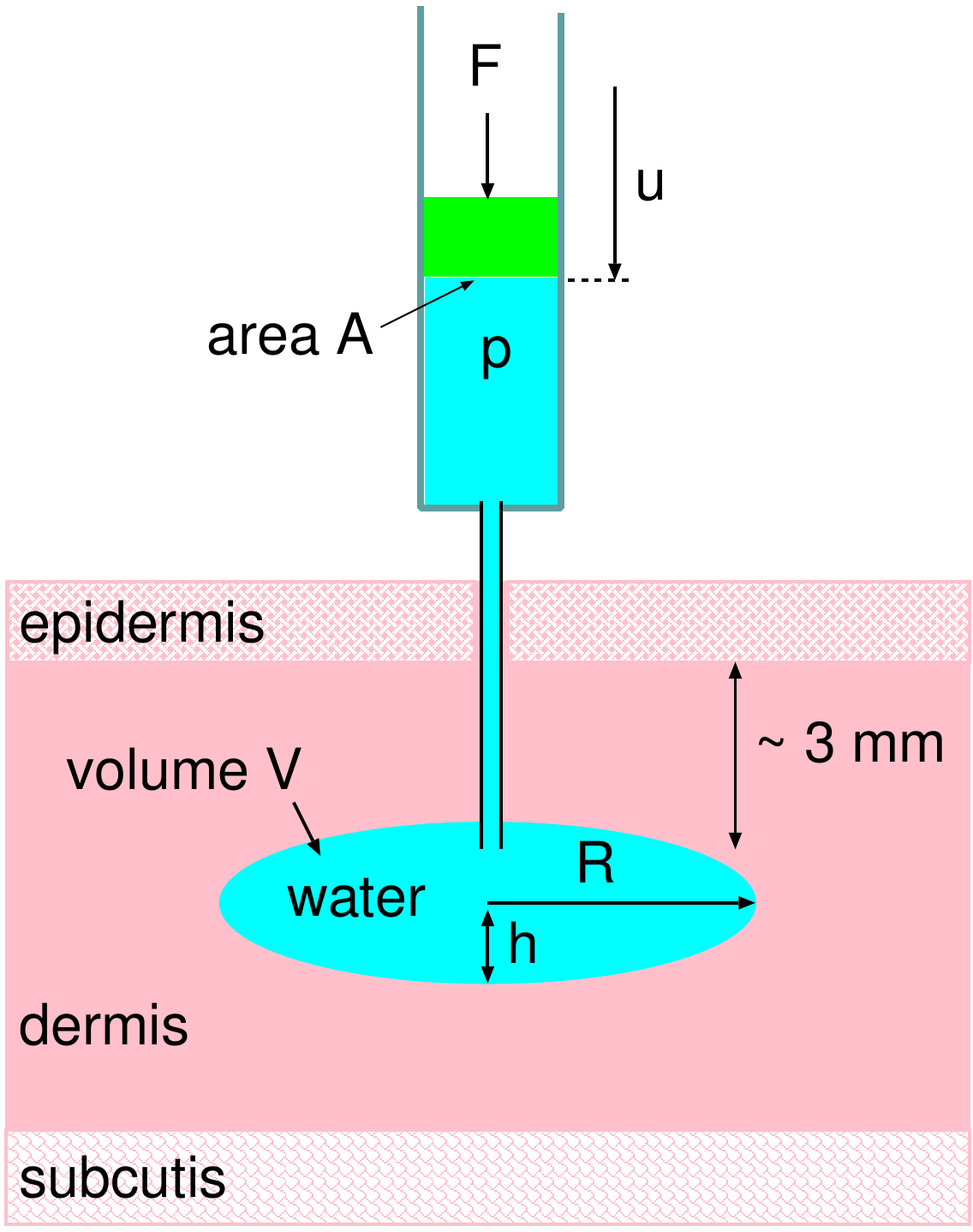}
\caption{\label{needlefluid}
Fluid injection into the skin.
}
\end{figure}

\begin{figure}
\includegraphics[width=0.45\textwidth]{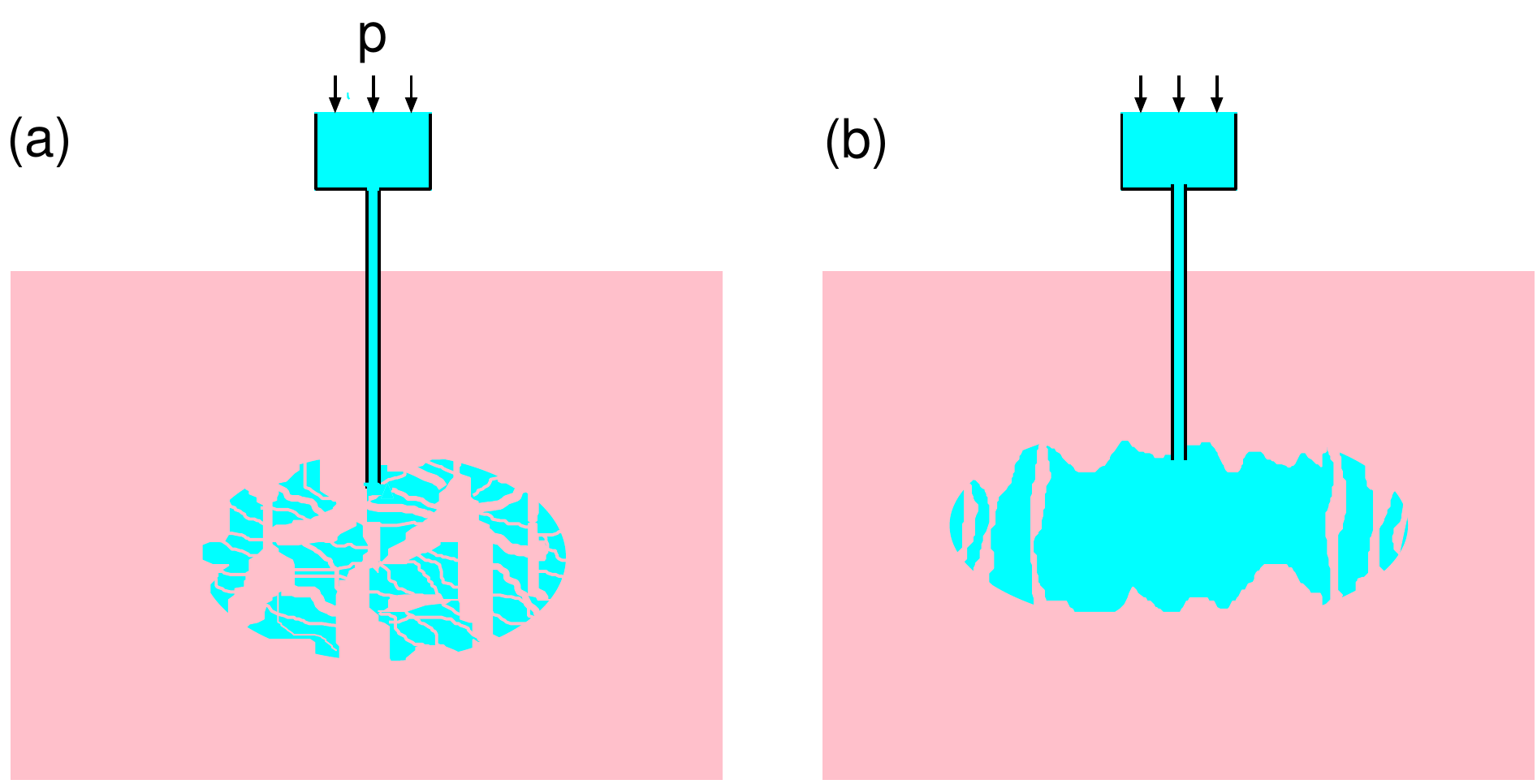}
\caption{\label{TwoStages.1.pdf}
Two limiting models of fluid injection. The dermis is assumed to consist of two components: a soft
(gel-like) matrix and a network of elastic fibers which break only at very high tension force. 
(a) If the pressure in the fluid is not too high it will generate cracks in the soft matrix, or at the interface between the
soft matrix and the fibers, forming a complex
network of fluid filled and connected regions, while the fiber network is intact but stretched.
(b) The fluid pressure is so high (e.g. as a result of very high injection rate) that also the
fiber network break, resulting in crack propagation in the dermis similar to in the model study reported
on in Sec.  3.2B. 
}
\end{figure}

\vskip 0.2cm
{\bf E. Theory and analysis of the experimental data}

Here we will study the fluid pressure $p$ necessary in order to inject fluid into the dermis of the skin.
We will calculate $p$ as a function of the size of the injected fluid volume and as a function of the speed of
fluid injection. We will assume that the fluid forms a roughly ellipsoidal shaped volume with height $h$ and radius $R$,
see Fig. \ref{needlefluid}. Thus the fluid volume $V\approx 4 \pi R^2 h/3$. 
We consider an equilibrium situation and determine $R$ and $h$ by minimizing the total potential energy $U_{\rm tot}$.
We have
$$U_{\rm tot} = -Fu +G \pi R^2 +{1\over 2} E \left ({h\over R}\right )^2 \kappa R^3$$
The first term is the potential energy of the loading mass acting on the stopper which is squeezing 
the fluid into the skin. The second term is the
energy to break the bonds on the surface area $\pi R^2$, and the last term is the elastic energy stored in the system
when bending the surfaces (by the amounts $h$) in the area $\pi R^2$ so it can occupy the fluid volume $V\approx 4 \pi R^2 h/3$.
This term can be understood as the strain energy density $\sigma \epsilon /2 = E\epsilon^2/2$ integrated over the volume where
the strain is finite. The strain is of order $h/R$ and the volume of order $R^3$. The factor $\kappa$ is a number of order unit which
depends on the location of the dermis relative to the skin surface, and also on the exact skin elastic properties such as the
elastic modulus and thickness of the epidermis (the skin top layer) and subcutis (see Fig. \ref{skin-picture}).

We assume that the fluid (water) is incompressible so that fluid volume conservation require
$$V=Au={4\pi \over 3} R^2 h\eqno(20)$$
Using this equation and $F=pA$ we can write
$$U_{\rm tot} = -{4\pi \over 3} R^2 h p +G \pi R^2 +{1\over 2} E \left ({h\over R}\right )^2 \kappa R^3\eqno(21)$$
Minimizing with respect to $h$ and $R$ gives
$$-{4\pi \over 3} R^2 p +E h \kappa R =0\eqno(22)$$
$$-{8\pi \over 3} R h p +G 2\pi R +{1\over 2} E h^2 \kappa =0\eqno(23)$$
which gives
$$h=\left ({4 \pi \over 3 \kappa} {GR\over E}\right )^{1/2}\eqno(24)$$
$$p=\left ({3\kappa \over 4 \pi} {EG\over R}\right )^{1/2}\eqno(25)$$
Using (20) and (24) we obtain the volume of injected fluid
$$V=\beta R^{5/2}\eqno(26)$$
where
$$\beta = {4 \pi \over 3} \left ({4 \pi \over 3 \kappa} {G\over E}\right )^{1/2}\eqno(27)$$

Assume that fluid is injected at a constant volume per unit time so that $V(t)=\dot V t$. If the injection occur during
the time period $0<t<t_0$ we get the injected volume $V(t_0) = V_0=\dot V t_0$. The time averaged force
$$\langle F \rangle = {1\over t_0}\int_0^{t_0} dt \ F(t) = {1\over V_0}\int_0^{V_0} dV \ F(V)\eqno(28)$$
We get
$$F= A p = \left ({3\kappa \over 4 \pi} EG\right )^{1/2} A R^{-1/2} = \alpha A R^{-1/2}\eqno(29)$$
where 
$$\alpha = \left ({3\kappa \over 4 \pi} EG\right )^{1/2}\eqno(30)$$ 
Combining (26) and (29) gives 
$$F=\alpha \beta^{1/5} A V^{-1/5}\eqno(31)$$

Assume now first that the $E$ and $G$ can be treated as constants independent of the injection rate $\dot V$, i.e.,
independent of the crack tip velocity $\dot R$. In this case, using (28) and (31) we get
$$\langle F \rangle = {1\over V_0}\int_0^{V_0} dV \ \alpha \beta^{1/5} A V^{-1/5} =  {5\over 4} \alpha \beta^{1/5} A V_0^{-1/5}\eqno(32)$$
Note that the force depends very weakly on the the injected volume (or injection time) e.g., doubling the injection time
result in a change in a reduction in the force with $\approx 13\%$. In reality $G(v)$ increases with increasing crack tip speed $v=\dot R$, 
and since $R=(V/\beta)^{2/5}$ we get $\dot R = (2/5)\beta^{-2/5} V^{-3/5} \dot V$. Thus the crack speed decreases with increasing time as
$t^{-3/5}$ which will reduce $G(v)$ with increasing time.   

Let us compare (32) with the experimental results shown in Fig. \ref{1dotV.2Fav.pdf}. 
The dermis has nonlinear viscoelastic properties, and the Young's modulus $E$ in the equations above must be considered as an
effective modulus obtained for the typical strain $\approx h/R$ involved in the cavity formation. If we use
$E=5 \ {\rm MPa}$ and $G=1000 \ {\rm J/m^2}$,
and assume $R$ a few mm using (24) we get a strain $h/R$ of order unity. For such large strain the effective 
modulus $E=5 \ {\rm MPa}$ appears resonable\cite{derm2}. Using this $E$ modulus 
for $V_0 = 100 \ {\rm mm}^3$ and $A=30 \ {\rm mm}^2$ we get $\langle F \rangle \approx 18 \ {\rm N}$. 
This is similar the observed injection force for the highest injection rate. However, 
the dependency of the calculated $\langle F \rangle$ on the injection rate $\dot V$ 
(via the dependency of $G$ and $E$ on the crack tip speed) is weaker than observed.
Thus, we note that the strain rate $\approx \dot R/R$ varies with the radius $R$, but is typically $\approx 0.1-10 \ {\rm s}^{-1}$
as $\dot V$ varies from the smallest to the highest value in Fig. \ref{1dotV.2Fav.pdf}. This correspond to an
increase in the $E$-modulus with only a factor of $\sim 2$. Thus, most of the dependency of $\langle F \rangle$
on the injection rate must be due to the crack propagation energy $G$. 
We conclude that at least for the lowest injection rates in Fig. \ref{1dotV.2Fav.pdf}
the fluid injection will not result in
a breaking of all components of the dermis material, and the crack propagation energy $G$
will be much smaller than used above, but may increase rapidly with increasing injection rate. Our present understanding 
of the fluid injection process is illustrated in Fig. \ref{TwoStages.1.pdf}.

Fig. \ref{TwoStages.1.pdf} shows two limiting models of fluid injection. The dermis is assumed to consist of two components: a soft
(gel-like) matrix and a network of elastic fibers which break only at very high tension force. 
If the pressure in the fluid is not too high it will generate cracks in the soft matrix, or at the interface between the fibers and the 
gel matrix, forming a complex
network of connected fluid filled regions, while the fiber network is intact but stretched (see Fig. \ref{TwoStages.1.pdf}(a)).  
In Fig. \ref{TwoStages.1.pdf}(b) it is instead assumed that the fluid pressure is so high 
(e.g. as a result of very high fluid injection rate) that also the
fiber network break, resulting in crack propagation in the dermis similar to in the model study reported
on in Sec.  3.3B. In practical applications the case (a) is likely to occur.

After injection of the fluid, resulting (from interfacial crack propagation) in a complex network 
of pressurized fluid filled regions, a slower process
will take place where the fluid diffuse into the skin dermis. This latter phase involves the skin permeability
and is governed by the (Darcy's and Brinkman's) diffusion equation and by poroelastic fluid dynamics.
This can be a slow process as manifested by the fact that 
after injecting the fluid and removing the needle the fluid pressure in the
cavity is high enough to allow some fluid to leak to the surface of the skin where it form
small water droplets (see Fig. \ref{WaterDroppletPigSkin.ps}).
We note that after the needle is removed the dermis and especially the epidermis (which is elastically
much stiffer then the dermis) will elastically rebound and tend to close the hole formed by the needle,
but Fig. \ref{WaterDroppletPigSkin.ps} shows that the fluid pressure is high enough to allow some fluid 
leakage to the skin surface.

\begin{figure}
\includegraphics[width=0.45\textwidth]{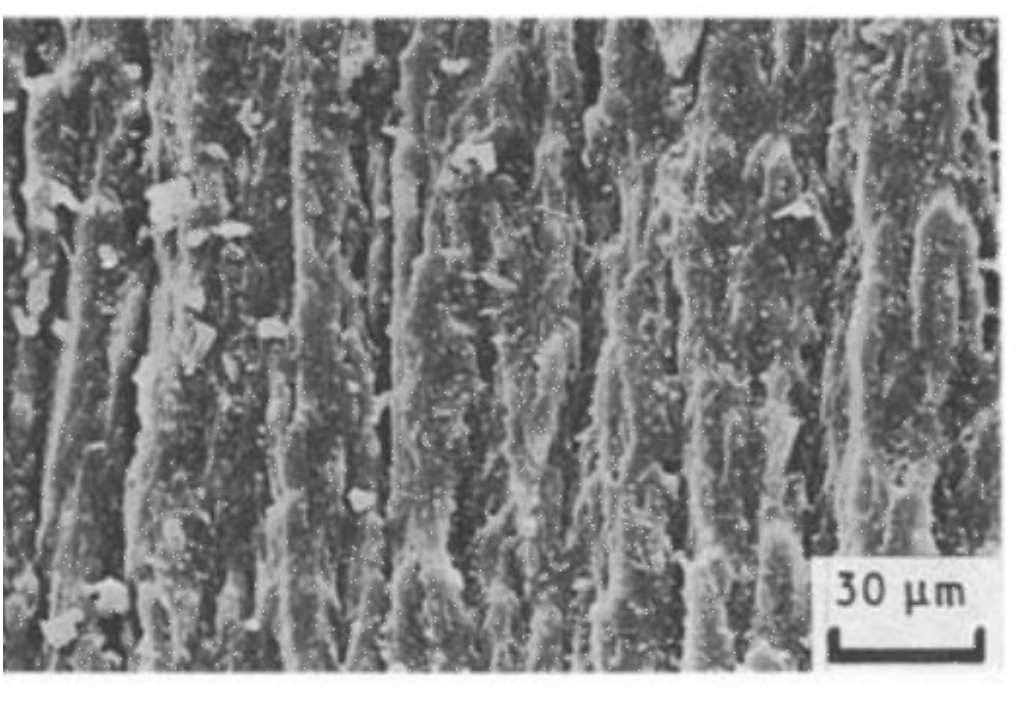}
\caption{\label{weartrack.pdf}
Wear tracks on rubber surface. 
}
\end{figure}

\vskip 0.2cm
{\bf 3.4 Rubber wear}

Rubber wear is a complex topic. Here we are interested in a rubber block sliding on a
a rigid substrate with surface roughness on different length scales, e.g., a tire tread block
on a concrete surface. There are several different limiting wear modes depending on the 
chemical composition of the rubber compound, the sharpness of the surface roughness, the composition of the atmospheric gas, 
the temperature, and the sliding speed.
Thus, at low temperature or for very sharp roughness (large rms slope and kurtosis) 
the wear may involve cutting 
the rubber surfaces forming linear wear tracks on the rubber surface (see Fig. \ref{weartrack.pdf}). This would result in 
rapid wear. Some rubber compounds tend to wear while forming a smear layer on the countersurface,
which may reduce the wear rate with increasing contact time (if slid repeatedly over
the same surface area). Other compounds may wear
at a constant rate by formation of small rubber wear particles (dry rubber dust). This latter wear mode
appear to be most important for tires and will be considered in the following.  

There is at present no accurate theory to predict the wear rate of rubber materials. Here we will discuss
several aspects which must be taken into account in any realistic model of rubber wear.

\begin{figure}
\includegraphics[width=0.35\textwidth]{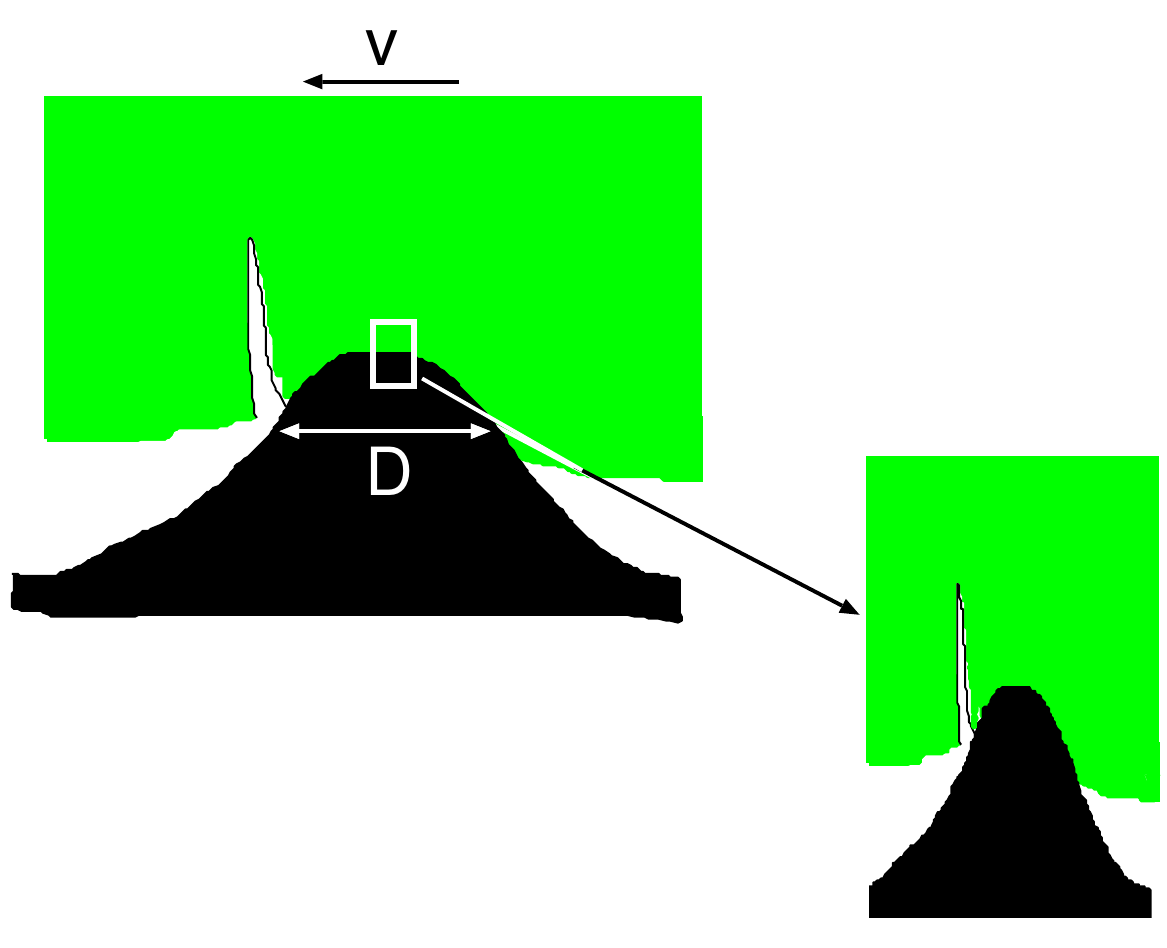}
\caption{\label{BigSmallCrackWear.pdf}
Big asperities drive big cracks in the rubber, and small asperities on top of bigger asperities
drive small cracks. The stress field of an asperity extend a distance into the rubber of order the width
of the contact region.
}
\end{figure}

\vskip 0.2cm
{\bf A. Multiscale crack propagation}

Rubber wear on road surfaces is a multiscale phenomena. 
All surfaces of solids have roughness on different length scales. The best picture of this is that a big asperity
has smaller asperities on top of it, and the smaller asperities have even smaller asperities on top of them, and so on.
When a rubber block is sliding on a surface with multiscale roughness the contact will in general not be complete,
but the contact area will decrease continuously as the magnification increases and new shorter wavelength surface roughness is observed.
If the rubber makes contact with an asperity, if $D$ is the width of the contact region then the deformation
(and stress) field will extend into the rubber block a distance of order $D$. The elastic energy (temporally) 
stored in the volume element $D^3$ can drive a crack in the surface region only if the crack extend into the rubber over a length less than
$D$. Thus different stages in the propagation of a crack will involve the surface roughness of different wavelength,
and a crack extending a distance $D$ into the rubber can only be driven further by road asperities of similar (or larger) size as
the crack length $D$. 

In order for a rubber wear particle to form, a crack cannot just propagate into the rubber surface but it must ``turn around''.
The distribution of sizes of wear particles has been studied experimentally by collecting the particles
generated from a tire in rolling contact (with some small slip) with a road surface\cite{Dannis,fiber1}. The probability distribution
of wear particle sizes (effective diameter $D$) was found to be exponential 
$$P(D)=l^{-1} {\rm e}^{-(D-D_1)/l}\eqno(33)$$ 
which is normalized so that
$$\int_{D_1}^\infty dD \ P(D) = 1$$
Here $D_1$ is the smallest wear particle diameter observed, which was found to be $\approx 4 \ {\rm \mu m}$.
Using (33) the average volume of a wear particle $\approx \pi l^3$.

In Ref. \cite{wear} a theory was developed which gives a probability distribution of the form (33) in which
the distance $l$ was interpreted as the crack mean free path. That is, it was assumed that after an average distance $l$
the crack abruptly changed direction e.g. by hitting into a filler particle cluster, or some other impenetrable
inhomogenity. For the rubber used in the experimental study of Ref. \cite{Dannis} $l \approx 16 \ {\rm \mu m}$.

\begin{figure}
\includegraphics[width=0.45\textwidth]{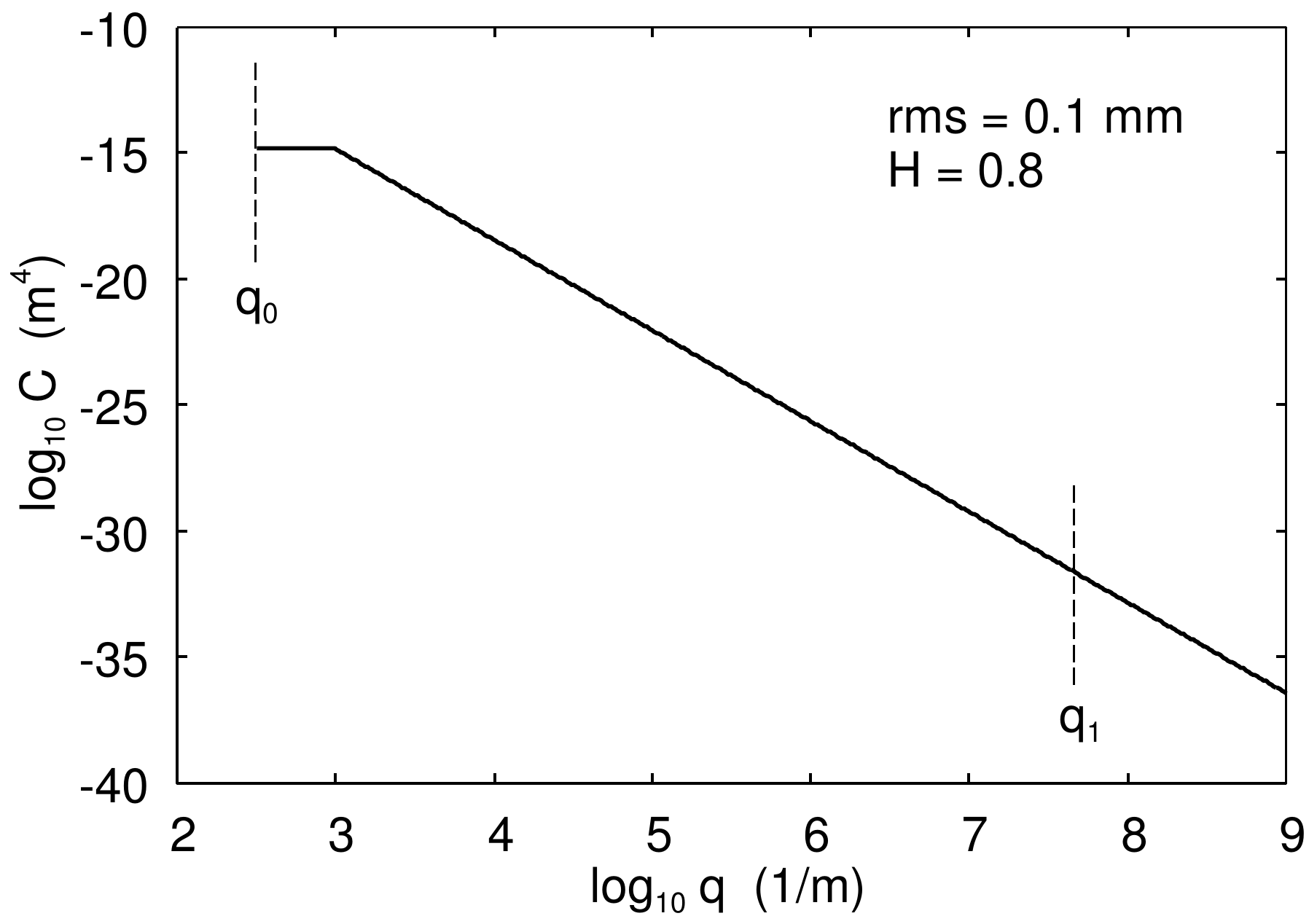}
\caption{\label{1logq.2logC.used.pdf}
The surface roughness power spectrum used in the friction calculations.
The surface is assumed to be self affine fractal with the Hurst exponent $H=0.8$
and the rms roughness $0.1 \ {\rm mm}$. The vertical dashed line is the large wavenumber
cut-off $q_1$ used in the friction calculations. Including all the roughness with wavenumber
$q<q_1$ result in a surface with the rms slope $1.3$.  
}
\end{figure}


\begin{figure}
\includegraphics[width=0.45\textwidth]{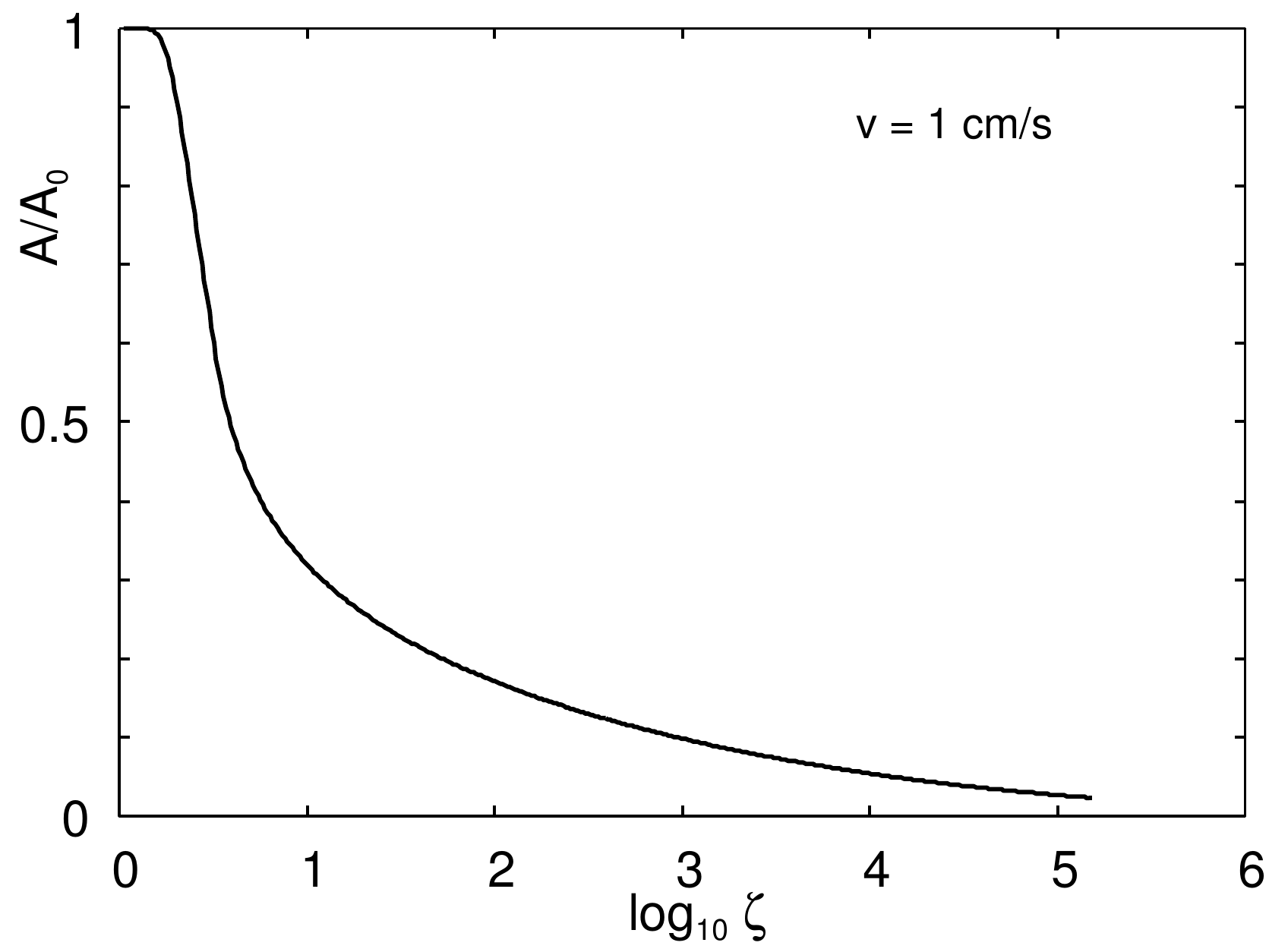}
\caption{\label{1logz.2A.for.1cm.per.s.Apollo4.pdf}
The relative area of contact $A/A_0$ (where $A_0$ is the nominal contact area) as a function
of the logarithm of the magnification for sliding speed $v=1 \ {\rm cm/s}$. 
The nominal contact pressure $P_0=0.3 \ {\rm MPa}$ and the temperature
$T=20^\circ {\rm C}$.
}
\end{figure}

\begin{figure}
\includegraphics[width=0.45\textwidth]{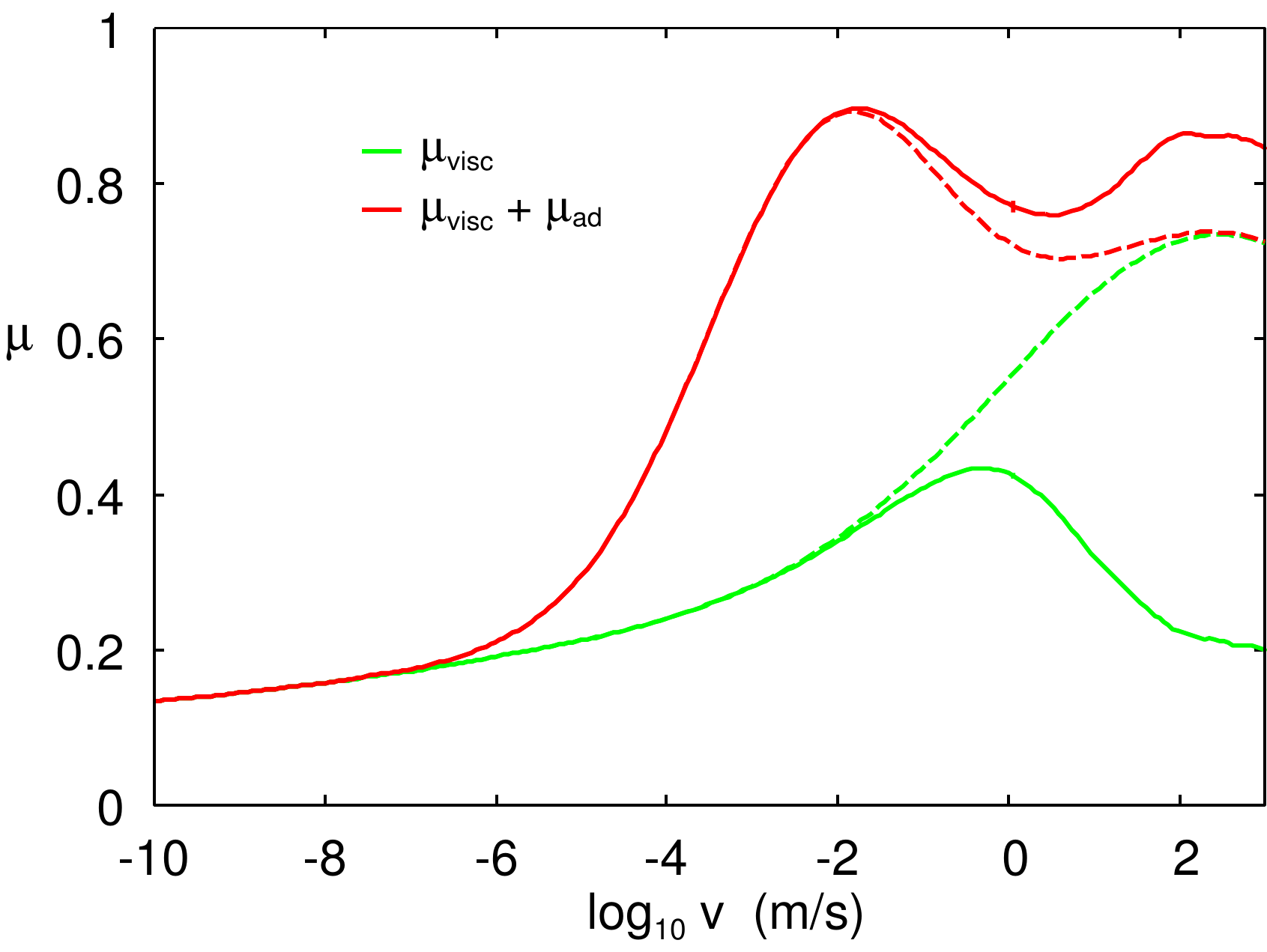}
\caption{\label{1logv.2mu.Apollo4.pdf}
The viscoelastic contribution to the friction coefficient $\mu_{\rm visc}$ (green lines) and the total
friction coefficient $\mu=\mu_{\rm visc}+\mu_{\rm ad}$ (where $\mu_{\rm ad}$ is the adhesive contribution,
i.e., the contribution from the area of real contact). The solid and dashed lines are with and without the
flash temperature. The nominal contact pressure $P_0=0.3 \ {\rm MPa}$ and the background temperature
$T=20^\circ {\rm C}$.
}
\end{figure}

\vskip 0.2cm
{\bf B. Frictional shear stress}

Rubber wear result from the frictional shear stress the rubber is exposed to when sliding on a
countersurface. At the exit side of asperity contact regions large tensile stress may develop which
can drive cracks in the rubber surface region (see Fig. \ref{BigSmallCrackWear.pdf}). 
Thus, rubber wear and rubber sliding friction are closely connected phenomena.

When a rubber block is sliding on hard countersurface with roughness on many length scales
the apparent rubber-substrate contact area will depend on the magnification. At the magnification
$\zeta$ only the roughness with wavenumber $q < q_0 \zeta$ can be observed. Here $q_0$ is the wavenumber of the most long
wavelength roughness component. Using the Persson rubber friction theory\cite{JCPP}, in Fig. \ref{1logz.2A.for.1cm.per.s.Apollo4.pdf}
we show the relative area of contact $A/A_0$ (where $A_0$ is the nominal contact area) as a function
of the logarithm of the magnification for the sliding speed $v=1 \ {\rm cm/s}$. 
The result is for a rubber tread compound used in a passenger car tire, assuming 
the nominal contact pressure $p_0=0.3 \ {\rm MPa}$ and the temperature $T=20^\circ {\rm C}$. In the calculation we have
used the surface roughness power spectrum shown in Fig. \ref{1logq.2logC.used.pdf}.

As the magnification increases the contact area decreases and the normal contact stress increases. Since the normal force
is constant we have $F_{\rm N} = p_0 A_0 = p(\zeta) A(\zeta)$ or $p(\zeta) = p_0 A_0/A(\zeta)$. As we increase
the magnification the contact pressure and the effective frictional shear stress $\tau (\zeta) = \mu F_{\rm N}/A(\zeta) =
\mu p (\zeta)$ will increase and finally, if the surface is rough enough, the shear stress becomes so high as to break the bonds in the
rubber. In the friction theory we assume this to occur at the point where including all the roughness with wavenumber $q=q_0\zeta <q_1$
result in a rms-slope of 1.3. This choice of cut-off is a condition obtained by analyzing a lot of experimental data,
but there is no rigorous theoretical argument for this cut-off. In fact, understanding how to determine the large wavenumber cut-off 
in the rubber friction calculation is
a very important but unsolved problem. However, we are convinced that on very rough surfaces, such as asphalt or concrete road
surfaces, the cut-off is related to the onset of strong wear as outlined above. We believe at the magnification
$\zeta_1 = q_1/q_0$ continuous cutting of the rubber surface by the road asperities occur, 
as observed at much larger length scale in Fig. \ref{weartrack.pdf} (which occur at low temperature or for much sharper roughness).

For the same system as used in Fig. \ref{1logz.2A.for.1cm.per.s.Apollo4.pdf} (rubber tread compound for a tire), 
in Fig. \ref{1logv.2mu.Apollo4.pdf} we
show the viscoelastic contribution to the friction coefficient $\mu_{\rm visc}$ (green lines) and the total
friction coefficient $\mu=\mu_{\rm visc}+\mu_{\rm ad}$ (where $\mu_{\rm ad}$ is the adhesive contribution,
i.e., the contribution from the area of real contact). The solid and dashed lines are with and without the
flash temperature.  

\begin{figure}
\includegraphics[width=0.45\textwidth]{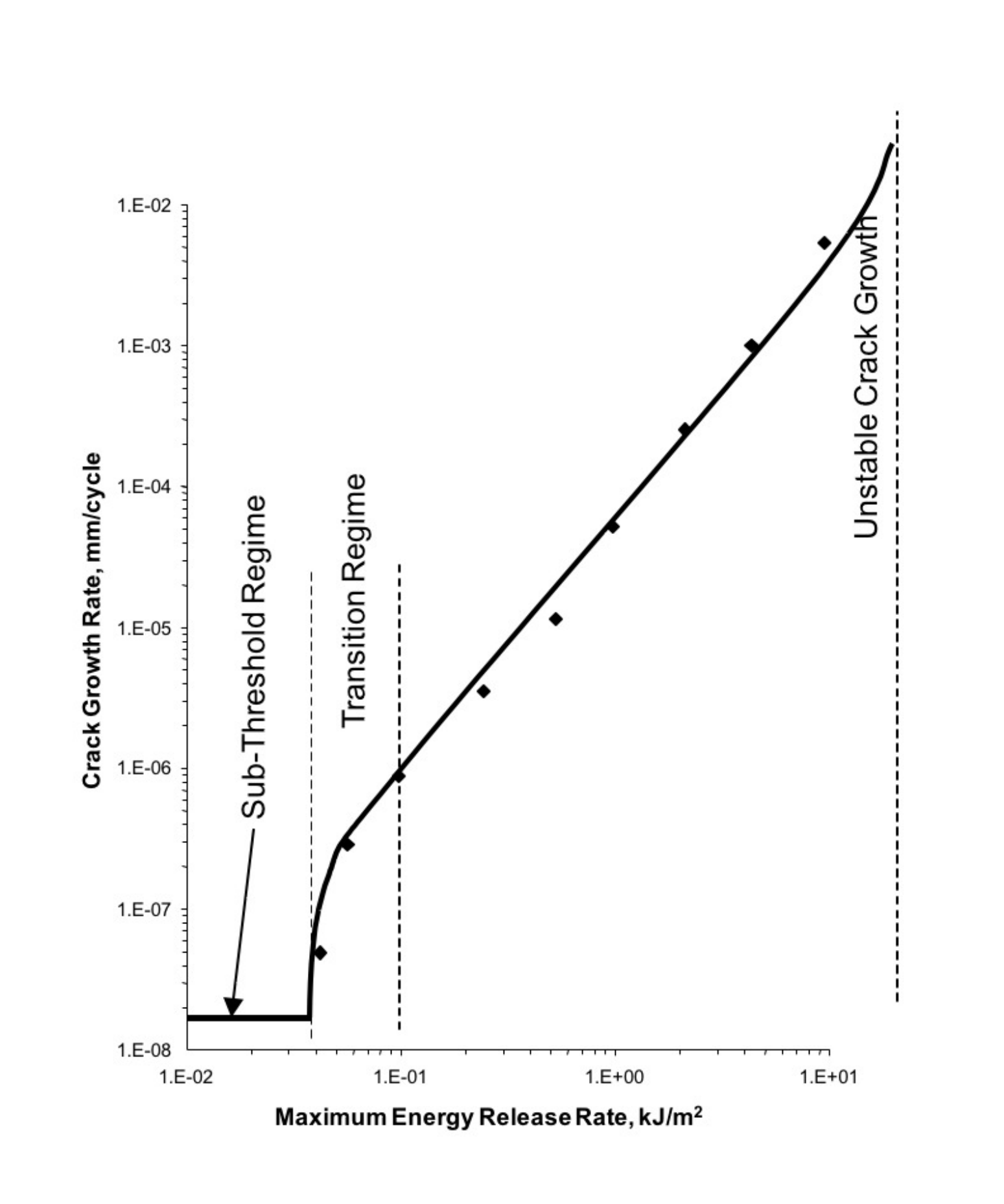}
\caption{\label{CrackGrowth.NR.pdf}
Crack growth in unfilled natural rubber. The rubber sample is exposed to an oscillating 
strain with the period $T=0.5 \ {\rm s}$. The minimum of the energy release rate is zero. 
In each stress cycle the crack extend by a distance $\Delta x$ so
the effective crack tip speed $v=\Delta x / T$. Adapted from Ref. \cite{Lake}.
}
\end{figure}

\begin{figure}
\includegraphics[width=0.4\textwidth]{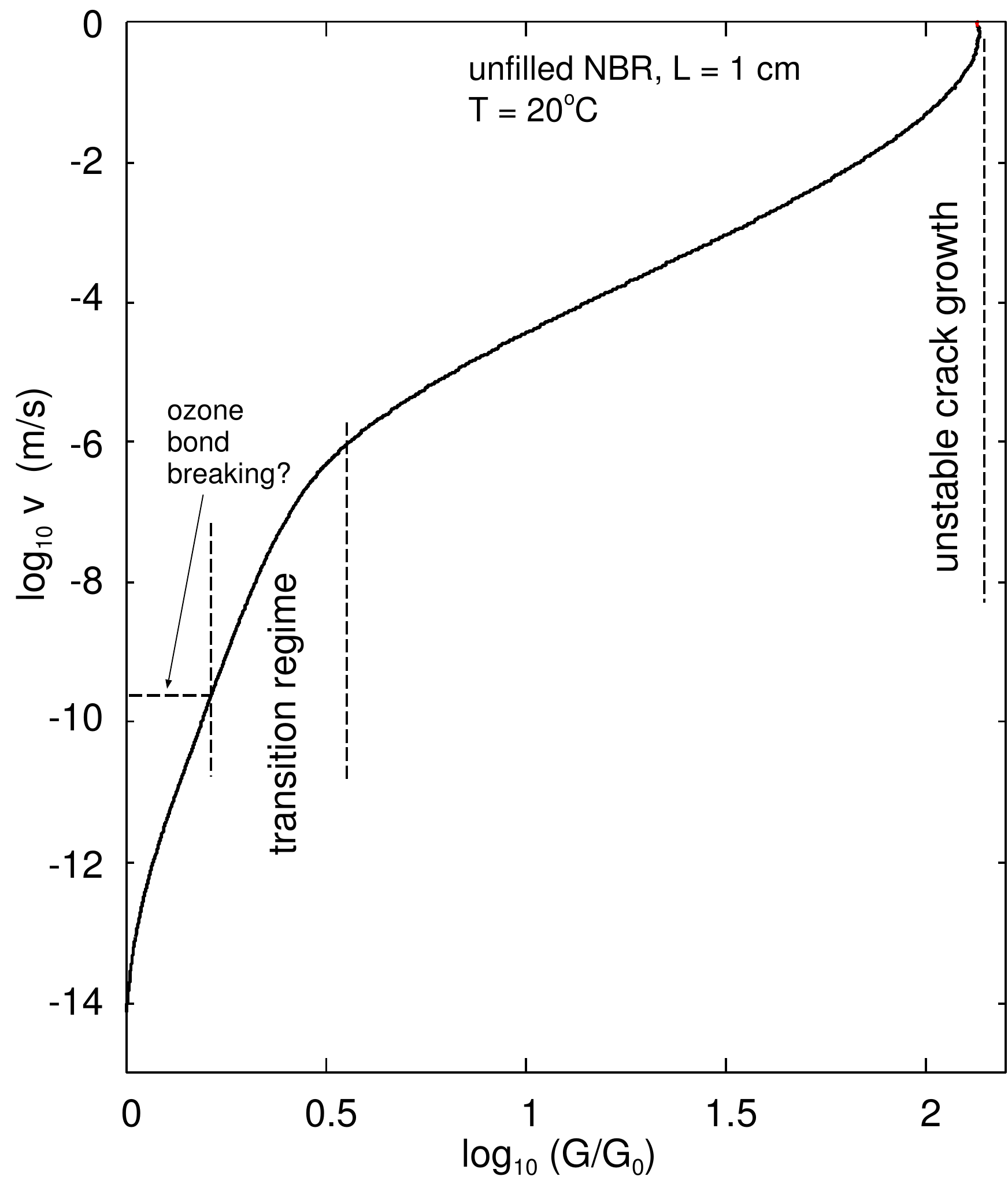}
\caption{\label{fatigueCURVE.NBR.theory.pdf}
Crack growth in unfilled NBR rubber.
}
\end{figure}

\begin{figure}
\includegraphics[width=0.45\textwidth]{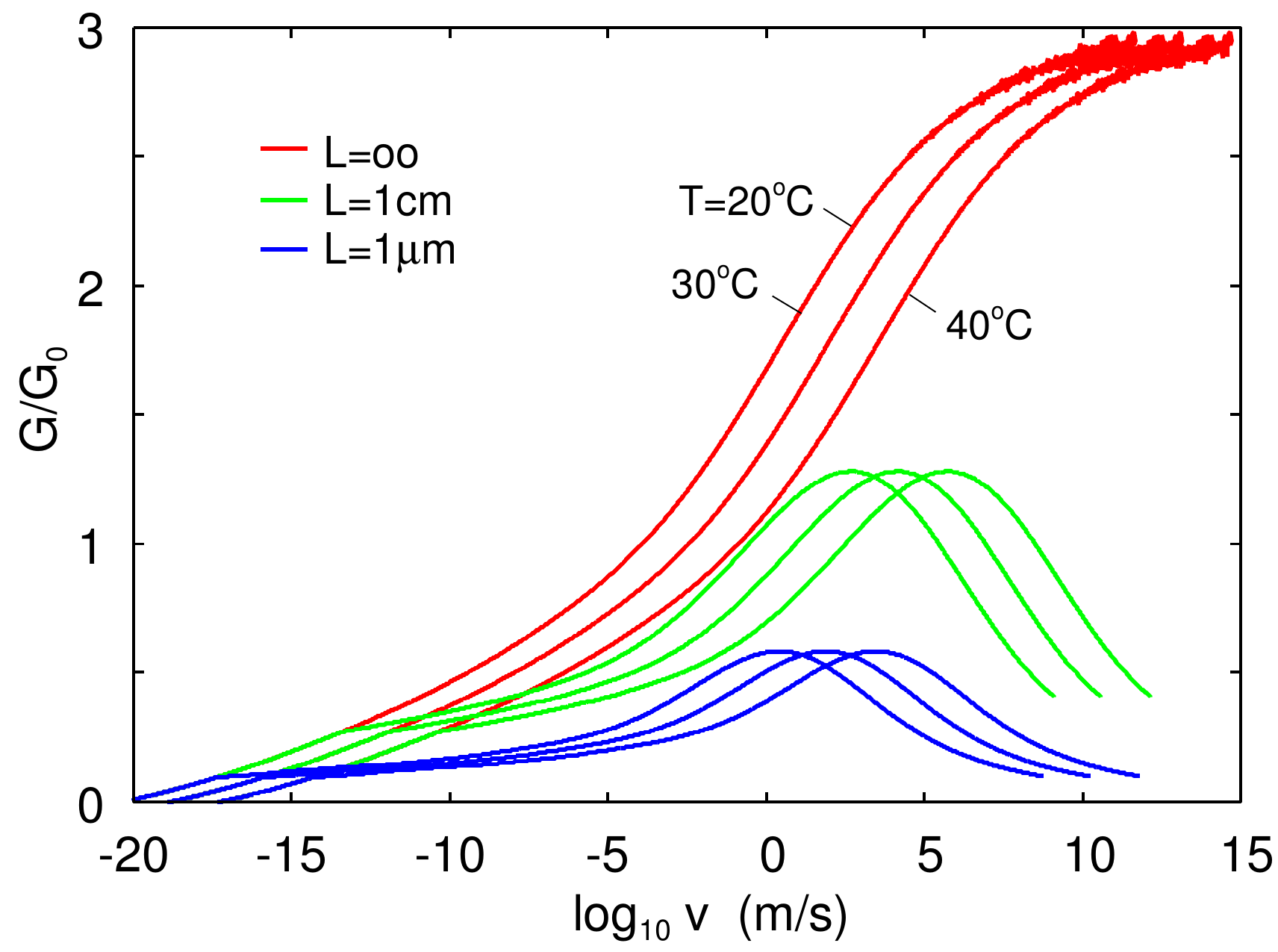}
\caption{\label{1logv.2Gcrack.Apollo4.1mu.1cm.infinite.1.pdf}
The crack propagation energy $G$ for an infinite system (red lines), and for finite size systems with
linear size $L=1 \ {\rm cm}$ (green) and $L=1 \ {\rm \mu m}$ (blue). Results are given for the temperatures
$T=20$, $30$ and $40^\circ {\rm C}$.
}
\end{figure}

\vskip 0.2cm
{\bf C. Size-dependent crack propagation energy}

We have argues that on surfaces with large roughness during slip 
strong rubber wear of the type seen in Fig. \ref{weartrack.pdf} will always occur  at short enough length scale,
and it will act as an effective cut-off in calculating the rubber friction (see Sec. 3.4B). 
This would typically result in rubber wear particles of micrometer size. 
However, larger rubber 
wear particles may also form but they result not from interaction with one road asperity,
but as a result of a cumulative influence of many accounts with road asperities. 
There are two reasons for this: as the length scale increases
(or the magnification decreases) the asperity stress decreases while simultaneously the crack propagation 
energy (at a given crack-tip velocity) increases. 
In this case, every time an asperity slide over a crack it will result in a small movement of the crack tip, and finally to the detachment
of a rubber particle. This process is denoted as fatigue wear. 
Given the time dependent stress acting on the rubber surface from the road asperities, one can estimate the fatigue 
wear rate from the knowledge how
the asperity stress field and the (velocity dependent) crack propagation energy change with the length scale (or magnification).
The former is given by the rubber friction theory and the latter by the theory of crack propagation, 
or both quantities can be obtained from experiments.

As an illustration, in Fig. \ref{CrackGrowth.NR.pdf} shows the measured crack tip displacement 
$\Delta x$ (in mm) per cycle, as a function of the 
amplitude of the oscillating energy release rate\cite{Lake} (for other similar measurements see Ref. \cite{KKLL}.
In the experiment the crack (in natural rubber) 
is exposed to an oscillating external stress field which simulate the oscillations in the stress observed by
a crack in the surface region of a rubber block as it is sliding over road asperities. 
If $T$ is the oscillation time period (in the present case of order 1 second)
one can define an average crack tip velocity $v=\Delta x/T$. In this way Fig. \ref{CrackGrowth.NR.pdf}  is closely related to the crack tip $G(v)$ function.
Indeed, if the crack tip velocity $v$ is plotted as a function of $G$, one obtain a curve very similar to that shown in Fig. \ref{CrackGrowth.NR.pdf}.
To illustrate this, in Fig. \ref{fatigueCURVE.NBR.theory.pdf}
we show the calculated relation between $v$ and $G$ for unfilled NBR rubber for a system of size $L=1 \ {\rm cm}$. 
We show the curve only up to the maximum of $G(v)$ as there is no stable solution for larger $G$.

Finally, as pointed out before, the crack propagation curve $G(v)$ depends on the system size so measurements of the $G(v)$ relation
(or the relation between the maximum energy release rate and the crack growth rate, Fig. \ref{CrackGrowth.NR.pdf}) for a macroscopic system
cannot be directly applied to the very small cracks which prevail at an early stage in the crack propagation phase. In Fig. 
\ref{fatigueCURVE.NBR.theory.pdf} we illustrate this with the calculated crack propagation energy for the tread rubber used in the friction study in Sec. 3.4B.
We show results for an infinite system (red lines), and for finite size systems with
linear size $L=1 \ {\rm cm}$ (green) and $L=1 \ {\rm \mu m}$ (blue), and for the temperatures
$T=20$, $30$ and $40^\circ {\rm C}$.

\vskip 0.2cm
{\bf D. Discussion}

Crack propagation in rubber involves several effects not discussed above but which are important
in practical applications. For very slowly moving cracks the bond-breaking at the crack tip is influenced by atmospheric gases such as
oxygen or ozone\cite{ozone}. This chemical bond breaking result in a cracks speed which is nearly independent of the driving
stress for small stress. Another effect is strain crystallization. Some types of rubber, like natural rubber, 
undergoes crystallization when exposed to large strain\cite{GertCrys}. The crystalline state is mechanically 
stronger than the amorphous state, which will increase the crack propagation energy.
However, strain crystallization require some time to occur, and if the driving stress is changing (fluctating) fast enough there may be no time
for crystallization to occur. But if the fluctuating stress never vanish, e.g., $\sigma(t)=\sigma_0+\sigma_1 {\rm cos} \omega t$
with $\sigma_0 >\sigma_1$, then some crystallization may always occur at the crack tip.  
Finally we note that if sliding occur in one preferable direction, 
a wear or abrasion pattern may form on the rubber surface\cite{pattern}, usually consisting of
periodic parallel ridges orthogonal to the sliding direction.
When a wear pattern form it will will influence (usually increase) the wear rate.

\vskip 0.3cm
{\bf 4 Summary and conclusion}

We have reviewed a theory for crack propagation in viscoelastic solids.
We have considered opening and closing cracks and finite-size effects. The 
theory was applied to pressure sensitive adhesives, the ball-flat adhesion problem,
intradermal fluid injection and rubber wear. 

\vskip 0.2cm
{\bf Acknowledgments:}

We thank Boris Lorenz (Continental, Singapore) for the DMA measurements of the pig skin dermis. 
We thank William Peabody (Keyence Corporation of America, NJ, USA) for the measurements of the 
roughness profile of the pig dermis crack surface.
We thank A. Tiwari for the pull-off force measurements on pressure sensitive adhesive film.

\end{document}